\documentclass[aps,superscriptaddress,showpacs,floatfix]{revtex4-1}

\usepackage{hyperref}
\usepackage{color}
\usepackage{graphicx}	
\graphicspath{%
  {./},%
}

\newcommand{\pAl}{p+{\rm Al}\,}
\newcommand{\pAu}{p+{\rm Au}\,}
\newcommand{\pPb}{p+{\rm Pb}\,}
\newcommand{\dPb}{d+{\rm Pb}\,}
\newcommand{\hPb}{$^3${\rm He}+{\rm Pb}\,}
\newcommand{\dAu}{d+{\rm Au}\,}
\newcommand{\hAu}{$^3${\rm He}+{\rm Au}\,}

\usepackage{xspace}	

\begin{document}

\title{Light-Heavy Ion Collisions: A window into pre-equilibrium QCD dynamics?}

\author{P. Romatschke} \affiliation{University of Colorado at Boulder}
\date{\today}

\begin{abstract}
Relativistic collisions of light on heavy ions (\pAu at $\sqrt{s}$=7.7 GeV, \pAu, \dAu,\hAu at $\sqrt{s}=$62.4 GeV and 200 GeV and \pPb,\hPb at $\sqrt{s}=5.02$ TeV) are simulated using ``superSONIC'', a model that includes pre-equilibrium flow, viscous hydrodynamics and a hadronic cascade afterburner. Even though these systems have strong gradients and only consist of at most a few tens of charged particles per unit rapidity, one finds evidence that a hydrodynamic description applies to these systems. Based on these simulations, the presence of a triangular flow component in \dAu collisions at $\sqrt{s}=200$ GeV is predicted to be similar in magnitude to that found in \hAu collisions. Furthermore, the $v_3(p_T)$ ratio of \hAu to \dAu is found to be sensitive to the presence of pre-equilibrium flow. This would imply an experimentally accessible window into pre-equilibrium QCD dynamics using light-heavy ion collisions.
\end{abstract}

\maketitle

\section{Introduction}

Experimental data on \dAu, \pPb and \hAu collisions seems to indicate the presence of collective flow in an (at least partially) equilibrated system \cite{Abelev:2012ola, Aad:2012gla, Adare:2013piz, Chatrchyan:2013nka, Shengli}. While alternative explanation have been put forward \cite{Dumitru:2014yza}, the result from measurements of multi-particle cumulants are strongly suggestive of a hydrodynamic origin of this collectivity \cite{CMSprelim}.

This study is motivated by the following, interrelated, questions:
\begin{enumerate}
\item
Does hydrodynamics apply to systems created in light-heavy ion collisions? 
\item
At which system size or collision energy does hydrodynamics break down?
\item
What experimental observables would confirm or rule out a theorist's answer to the above questions?
\end{enumerate}

In order to obtain answers for these questions, I use the new simulation package ``superSONIC'', which is an event-by-event generalization of the SONIC model \cite{Habich:2014jna}, including pre-equilibrium flow, relativistic viscous hydrodynamics as well as a hadronic cascade afterburner (details are given below). Using a Monte-Carlo Glauber model for generating event-by-event initial conditions, superSONIC can be used to simulate a range of different collision systems at different collision energies, such that results are directly comparable. Comparison to data will be performed wherever experimental information is available. 

Quite a number of theoretical studies on light-heavy ion collisions have appeared in the recent literature, and in the following similarities and notable differences of these works with respect to the present study are highlighted.

In Ref.~\cite{Bozek:2011if}, Piotr Bozek studied \pPb and \dPb collisions at $\sqrt{s}=4.4$ TeV and $\sqrt{s}=3.11$ TeV, respectively, using Monte-Carlo Glauber initial conditions followed by a 3+1d viscous hydrodynamics evolution and a hadronic cascade afterburner. He successfully predicted the large flow signal in $\pPb$ collisions at $\sqrt{s}=5.02$ TeV.

In Ref.~\cite{Nagle:2013lja}, Nagle et al. studied \pAu, \dAu and \hAu collisions at $\sqrt{s}=200$ GeV and \pPb collisions at $\sqrt{s}=5.02$ TeV using Monte-Carlo Glauber initial conditions followed by a 2+1 viscous hydrodynamics evolution and a hadronic cascade afterburner (``SONIC'', the predecessor of the package used in the present study). Nagle et al. found results consistent with Ref.~\cite{Bozek:2011if}, and first proposed \hAu collisions as an interesting handle on QCD transport properties. Elliptic and triangular flow in \hAu at $\sqrt{s}$ have since been measured experimentally \cite{Shenglipc}.

In Refs.~\cite{Schenke:2014zha,Schenke:2014gaa}, Schenke and Venugopalan studied light-heavy ion collisions using IP-Glasma initial conditions, followed by a 2+1d viscous hydrodynamic evolution, but no hadronic cascade afterburner. Their study is similar to the present one in that the sudden matching from the Glasma evolution to hydrodynamics includes non-vanishing pre-equilibrium flow. However, given that the Glasma evolution never drives the system to equilibrium, one can expect the results to be strongly dependent on the Glasma-hydro switching time. In their study, Schenke and Venugopalan found sizable flow components $v_2,v_3$ for \dAu,\hAu collisions (in agreement with experiment), but very little flow for \pPb collisions (in disagreement with experiment). 

In Ref.~\cite{Kozlov:2014fqa}, Kozlov et al. studied \pPb collisions at $\sqrt{s}=5.02$ TeV using 3+1d viscous hydrodynamics, but without any pre-equilibrium flow or hadronic afterburner. They implemented initial conditions based on a Monte-Carlo Glauber model supplemented with negative binomial energy fluctuations, which turn out to be very important for describing rare ``high-multiplicity'' events. Kozlov et al. found that they could successfully describe experimental flow data in \pPb using their model.

Compared to some of the studies listed above, the present study suffers from the drawback of only simulating boost-invariant (2+1d) dynamics, which clearly is not a good approximation to the longitudinal dynamics for light-heavy ion collisions. However, many of the conclusions in the present article will be based on ratios of flow observables in the transverse plane, so that one can expect conclusions to not be dominated by longitudinal artifacts. Nevertheless, it would be good if the present results could be checked by a full 3+1d calculation in the future. 
Also, the present study does not include negative binomial energy fluctuations, on the basis that these fluctuations are most relevant only for rare high-multiplicity events. In this article, the emphasis is on central (impact parameter $b<2$ fm/c), but not high-multiplicity selected events. Finally, the present study is based on a geometric Monte-Carlo Glauber model \cite{Miller:2007ri} rather than the IP-Glasma model \cite{Schenke:2012wb}, mostly to offer a baseline result using transparent ingredients. If it would turn out that some experimental result could not be described by the present approach based on Monte-Carlo Glauber, but can be described using IP-Glasma initial conditions, this could be regarded as experimental evidence for Color-Glass-Condensate dynamics in QCD.

On the other hand, the present simulation package for the first time combines pre-equilibrium flow, viscous hydrodynamics and hadron cascade dynamics in an event-by-event study that is applicable to different collision geometries. In this sense, it is the most realistic description currently available, and the direct comparison to experimental data may therefore offer interesting clues about the nature of hydrodynamics and transport in strongly coupled QCD.

The remainder of the article is structured as follows: in Sec.\ref{sec:me}, a brief description of the components of superSONIC is given, followed by results for \pAu,\dAu, \hAu and \pPb collisions at various energies in Sec.\ref{sec:res}, and the conclusions in Sec.\ref{sec:conc}.

\section{Methodology}
\label{sec:me}

\subsection{Early Stage: Initial Conditions and Pre-equlibrium}

In principle, obtaining information about real-time evolution in relativistic ion collisions is a problem that can be well defined in QCD. However, currently there are no known techniques to solve this problem, and thus obtain information about the pre-equilibrium stage in these systems. Nevertheless, a number of approximate approaches exits. Among these, there are weak-coupling inspired methods that rigorously apply to QCD in the asymptotically high energy limit (cf.~\cite{Mrowczynski:1993qm,Arnold:2004ti,Rebhan:2004ur,Romatschke:2005pm,Kurkela:2011ti,Gelis:2013rba,Berges:2013fga,Strickland:2013uga,Kurkela:2014tea}), and strong-coupling methods that rigorously apply to a certain class of non-abelian gauge theories (but not QCD) in the limit of large number of colors and large t'Hooft coupling (cf.~\cite{Nastase:2005rp,Kovchegov:2007pq,Amsel:2007cw,Grumiller:2008va,Chesler:2010bi,Wu:2011yd,vanderSchee:2013pia,Casalderrey-Solana:2013sxa,Bantilan:2014sra,Chesler:2015wra}). 

From a modeling perspective, what is needed from any of these methods is information about when the system behaves approximately hydro-dynamically (equilibration time) as well as the values of the hydrodynamic degrees of freedom (hydrodynamic initial conditions).  Other than the hydrodynamic starting time \cite{Gelis:2013rba,Kurkela:2014tea}, results for hydrodynamic initial conditions are presently available only from the strong-coupling method \cite{vanderSchee:2013pia,Bantilan:2014sra,Chesler:2015wra}, so this is what will be used in the present study. 

In more detail, from numerical relativity simulations of space-times modeling the relativistic collision of ``ions'' in ${\cal N}=4$ SYM it has been found that the radial fluid velocity is proportional to the gradient of the initial density distribution \cite{Romatschke:2013re,vanderSchee:2013pia,Chesler:2015wra}. In this study, this finding is promoted to the fluid velocity in all directions, such that
\begin{equation}
\label{pre-flow}
\vec{v}(\tau,{\bf x})=-\frac{\tau}{3.0}\vec \nabla \ln R^2({\bf x})
\end{equation}
where $\tau\equiv \sqrt{t^2-z^2}$ and $R^2({\bf x})$ is the product of the particle densities at the time of collision ($\tau=0$). Note that this result is consistent with Ref.~\cite{Vredevoogd:2008id}, but the pre-factor is non-trivial (cf.~\cite{Habich:2014jna}). The result (\ref{pre-flow}) has the added benefit that final particle momenta were found to be almost insensitive to the choice of the hydrodynamic starting time $\tau_0$ in heavy-ion collisions \cite{vanderSchee:2013pia}.

For the energy-density $\epsilon(\tau_0,{\bf x})$, initial conditions for each event are constructed as follows. Using Woods-Saxon distribution functions for the heavy ions such as ${\rm Au,Pb}$ \cite{DeJager1974479,DeVries1987495}, the Hulthen wavefunction for the deuteron (cf.~\cite{Adare:2013nff}) and realistic calculations for the $^{3}{\rm He}$ wavefunction \cite{Carlson:1997qn}, probability distributions of the nucleons within the nuclei of interest (cf.~\cite{Nagle:2013lja}) are obtained. Using a Monte-Carlo Glauber \cite{Miller:2007ri}, these probability distributions are mapped to positions of individual nucleons in the transverse ($x,y$) plane on an event-by-event basis implementing a hard-core repulsive potential of radius $0.4$ fm between nucleons. The positions of nucleons undergoing at least one inelastic collisions are recorded (``participants'') and converted into a density function $R^2({\bf x})$ by assuming that each participant contributes equally as a Gaussian with a width of $w=0.4$ fm (to match the RMS radius of a single nucleon). This fixes the initial fluid velocities via Eq.~(\ref{pre-flow}), and the energy density for hydrodynamics is assumed to be given as
\begin{equation}
\label{Eq:ed}
\epsilon(\tau_0,{\bf x})=E_0 R^2({\bf x})\,,
\end{equation}
with $E_0$ an overall constant (dependent on $\tau_0$, collision energy and collision system) that is related to the total multiplicity of the event. Typically 100 event initial conditions are generated for each collision system.

While the procedure to obtain the energy-density for hydrodynamics in Eq.~(\ref{Eq:ed}) is fairly standard in relativistic ion collision literature, the presence of the pre-equilibrium flow in Eq.~(\ref{pre-flow}) is relatively new. For this reason, also results without pre-equilibrium flow will be presented, and potential experimental ways to determine if pre-equilibrium flow is present in light-heavy ion collisions will be discussed. Other than the energy density and flow velocity field, relativistic hydrodynamic formulations also require the initial value for the shear and bulk stress tensors, which will be set to zero for simplicity. In heavy ion collisions, this assumption is harmless, as it has very little influence on final results cf.~\cite{Luzum:2008cw,vanderSchee:2013pia}. However, given the short life-time of systems created in light-heavy ion collisions compared to heavy ion collisions, this assumption should be carefully revisited in follow-up studies.

\subsection{Thermal Stage: Viscous Hydrodynamics}
In modern definitions, hydrodynamics is understood to be an effective theory of energy conservation at long wavelength \cite{Baier:2007ix,Bhattacharyya:2008jc}. Hydrodynamics is a good approximation of the bulk dynamics as long as higher order gradient corrections do not (strongly) change the leading order results. In standard nomenclature, viscous effects come into the energy-momentum tensor at first order in gradients (zeroth order would be ideal hydrodynamics). However, since all known consistent formulations for relativistic dissipative fluid dynamics are second order in gradients, this suggests a convenient handle on the effects of higher order gradients: the value of second-order transport coefficients, foremost the shear viscous relaxation time $\tau_\pi$. 

The view that is adopted here is the following: while first-order gradient effects in a system may be large (viscous effects sizable), a (viscous) hydrodynamic description of the system may still be quantitatively reliable as long as higher-order gradient corrections are small compared to first order gradients. This view is informed by direct simulations of strongly coupled quantum field theories out of equilibrium where it has been shown that viscous hydrodynamics offers a reliable description even in regions where first-order (viscous) corrections to the ideal energy-stress tensor are approaching 100 percent \cite{vanderSchee:2013pia}!

To test if second-order corrections are small compared to first-order gradients, the value of $\tau_\pi$ (more precisely the ratio $C_\eta=\frac{\tau_\pi (\epsilon+P)}{2\eta}$) is varied by 50 percent around the reference value $C_\eta=3$, thereby generating a ``systematic'' error estimate of the applicability of viscous hydrodynamics. If the hydrodynamic gradient approximation was breaking down, one would expect second-order gradient terms to be as important as first-order gradient terms. Thus, final results for particle spectra should vary considerably when changing the strength of second order terms (via the value of $C_\eta$) by 50 percent. Conversely, if final results showed very little dependence on the value for $C_\eta$ (as is the case for heavy ion collisions, cf.~\cite{Luzum:2008cw}), this could be considered evidence that a hydrodynamics was well applicable to such systems. In this sense, the reliability of hydrodynamics as a approximation to the system dynamics can be quantified and expressed as an error band generated by varying $C_\eta$, which is the strategy adopted in the following.

Besides the second-order transport coefficients, the hydrodynamic evolution will also depend on choices for the (temperature-dependent) shear and bulk viscosity coefficients as well as the speed of sound (via the equation of state). Again, for simplicity constant values are used for the ratios of shear viscosity over entropy density $\eta/s$ and the bulk viscosity is set to identically zero ($\zeta=0$). For the equation of state, a parametrization of lattice QCD data, given in Ref.~\cite{Borsanyi:2013bia}, is employed. All of these choices should be revisited in a more detailed study.

The hydrodynamic evolution is performed using the open-source code VH2+1 (version 1.9) \cite{Luzum:2008cw}, adapted by smearing to prohibit code instabilities in the strong-gradient regions frequently encountered in event-by-event viscous hydrodynamics. The smearing (optimized from a version used before in Ref.~\cite{Nagle:2013lja}) is performed by replacing low-energy density values by an average over nearest-neighbor cells, and it is only implemented at temperatures below the QCD phase transition (typically below $150$ MeV). I have tested that final results are not sensitive to the details of the smearing implementation. The hydrodynamic evolution solves the equations of motion on a square lattice with area $20^2\ {\rm fm}^2$ and $200^2$ grid points (lattice spacing $0.1$ fm) and a time step of one hundredth of the lattice spacing. I have tested that the results are unchanged when simulating the same volume with a lattice spacing of $0.05$ fm.

During the hydrodynamic evolution, cells that cool below a certain switching temperature $T_{\rm SW}=170$ MeV are monitored. This condition defines a switching hyper-surface on which information about temperature, flow velocity, dissipative stress tensors as well as the normal vector of the hyper-surface are recorded for each cell. This information will provided the initial condition for the late-stage hadron cascade simulation.

\subsection{Late Stage: Hadron Interactions in a Hadron Cascade}

For the late-stage hadron interactions,the hadronic cascade code B3D \cite{Novak:2013bqa} is used. Using the hyper-surface information to boost to the rest frame of each cell, the cascade is initialized with particles in the rest frame drawn from a Boltzmann distribution at a temperature $T_{\rm SW}$ with modifications of the momentum distribution to include deformations from viscous stress tensors (see \cite{Pratt:2010jt} for details). B3D includes hadron resonances in the particle data book up to masses of $2.2$ GeV, which interact via simple s-wave scattering with a constant cross-section of $10$ mb as well as scattering through resonances (modeled as a Breit-Wigner form). Once the resonances have stopped interacting, one can obtain final charged hadron multiplicities $\frac{dN_{\rm ch}}{d Y}$, mean charged particle momentum $\langle p_T\rangle$ and flow coefficients $v_n(p_T)$ for $n\geq 1$ from summing over individual particles with momenta ${\bf p}$. 
Specifically,
\begin{eqnarray}
&\frac{dN_{\rm ch}}{2\pi p_T dY dp_T}=\frac{\sum^{\rm ch.\ particles}_{\rm in\ p_T\ bin}}{2 \pi p_T \Delta_T \Delta_Y}\,,\quad
\frac{dN_{\rm ch}}{dY}=\int_0^\infty dp_T \frac{dN_{\rm ch}}{dY dp_T}\,,\quad
\langle p_T\rangle=\frac{\int_0^\infty dp_T p_T \frac{dN_{\rm ch}}{dY dp_T}}{\frac{dN_{\rm ch}}{dY}}
&\nonumber \\
&
|v_n|(p_T)=\sqrt{s_n(p_T)^2+c_n(p_T)^2}\,,\quad
\left(\begin{array}{cc}
s_n(p_T)\\
c_n(p_T)\end{array}\right)=
\frac{\sum^{\rm ch.\ particles}_{\rm in\ p_T\ bin}
\left(\begin{array}{cc}
\sin(n\phi))\\
\cos(n\phi)\end{array}\right)}
{\sum^{\rm ch.\ particles}_{\rm in\ p_T\ bin}}
\,,\quad
\phi\equiv \arctan\left(\frac{p_y}{p_x}\right)\,,&
\end{eqnarray}
where $\Delta_T=80$ MeV, $\Delta_Y=2$ are the width of bins for particle $p_T$ and rapidity $Y$, respectively. Note that since the cascade is applied to a boost-invariance case, the large $\Delta_Y$ value is of no significance. In practice, a sum over both particles and anti-particles and division of the spectra by two is performed, in order to increase statistics. For every hydrodynamic evolution event, 100,000 B3D events are run to increase statistics. In doing so, 
the sums in the definition of $v_n$ above are extended over all B3D events, thereby explicitly ignoring fluctuations arising from hadronic decays. After thus obtaining results $\frac{dN_{\rm ch}}{2\pi p_T dY dp_T}$ and $v_n(p_T)$ for each hydrodynamic event, an event average to obtain the event-by-event mean and event-by-event fluctuation is performed, the latter of which is recast into a statistical error bar on the mean. The results from this procedure are reported on in the following.

The combined simulation package of the early-stage, thermal stage and late stage evolution thus described above will be referred to as ``superSONIC'' in the remainder of this article.

\section{Results}
\label{sec:res}

\begin{table}[t]
\begin{tabular}{lcccccccccc}
System & $\sqrt{s}$ [GeV] & $\sigma_{NN}$[mb] &  $\langle N_{\rm part}\rangle$ &$\eta/s$ & preflow?&  $\frac{dN_{\rm ch}}{d\eta}$(th) &${\bf \frac{dN_{\rm ch}}{d\eta}}$(exp) & $\langle p_T\rangle$ [GeV] (th) & $\langle p_T\rangle$ [GeV] (exp) \\ \hline
\pAu & 7.7 &     32 &  7.4   & 0.08 & yes & 4.5 &  & 0.550(6) &\\
\pAu & 7.7 &     32 &  7.4   & 0.08 & no & 4.6 &  & 0.497(8) &\\
\pAu & 62.4  & 36 &  8.6   & 0.08 & yes & 8  & & 0.588(5)  & \\
\pAu & 62.4  & 36 &  8.6   & 0.08 & no & 8.4  & & 0.528(6)  & \\
\pAl & 200  & 42 &  4.7   & 0.08 & yes & 5.4  & & 0.575(6)  & \\
\pAl & 200  & 42 &  4.7   & 0.08 & no & 5.5  & & 0.515(7)  & \\
\pAu & 200  & 42 &  9.6   & 0.08 & yes & 10 & & 0.604(5) & \\
\pAu & 200  & 42 &  9.6   & 0.08 & no & 10 & & 0.537(5) & \\
\dAu & 62.4  & 36 &  14.9   & 0.08 & yes & 15  & & 0.570 & \\
\dAu & 200  & 42 &  17.5  & 0.08 & yes & 20 & ${\bf 20.3\pm 1.7}$ (0-5\%)\cite{Shenglipc} & 0.576(1) & {\bf 0.554} (0-20\%)\cite{Adare:2013esx}  \\
\dAu & 200  & 42 &  17.5  & 0.08 & no & 20 & ${\bf 20.3\pm 1.7}$ (0-5\%)\cite{Shenglipc} & 0.523(5) & {\bf 0.554} (0-20\%)\cite{Adare:2013esx}  \\
\hAu & 62.4  & 36 &  21   & 0.08 & yes & 21 &  & 0.557(1) &  \\
\hAu & 62.4  & 36 &  21   & 0.08 & no & 21 &  &   0.509(4) & \\
\hAu & 200  & 42 &  24    & 0.08 & yes & $27\pm 1$ &  & 0.567(2) &  \\
\hAu & 200  & 42 &  24    & 0.08 & no & $27\pm 1$ &  & 0.520(3) &  \\
\pPb & 5020 & 70 &  15.2  & 0.16 & yes & $39\pm 1$ & ${\bf 35\pm0.5}$ (0-20\%)\cite{Abelev:2012ola} & 0.716(6)\\
\pPb & 5020 & 70 &  15.2  & 0.08 & no & $38\pm 1$ & ${\bf 35\pm0.5}$ (0-20\%)\cite{Abelev:2012ola} & 0.623(3)\\
\hPb & 5020 & 70 &  32.4  & 0.16 & yes & 74 &  & 0.676\\
\end{tabular}
\caption{\label{tab:one}
Comparison of superSONIC runs (``th'') with available experimental data (``exp''). For all model runs $w=0.4$ fm, $\tau_0=0.25$ fm/c and $T_{\rm sw}=0.17$ GeV as well as ``central collisions'' with impact parameter $b<2$ fm/c are chosen. The effect of changing $C_\eta=2-3$ is contained in the reported theoretical error estimates. Mean particle $p_T$ is for unidentified charged hadrons. For the experimental data both the centrality and the reference are reported.}
\end{table}

A summary of systems that were simulated is given in Tab.~\ref{tab:one}. The first column in this table gives the system configuration, the second the collision energy, the third the inelastic cross-section (from Ref.~\cite{Aad:2011eu}) used in the Monte-Carlo Glauber. All Monte-Carlo Glauber events were generated for ``central'' collisions by imposing an impact parameter $b<2$ fm, loosely corresponding to the 0-5\% most central collisions. The fourth column in Tab.~\ref{tab:one} refers to the mean number of participants obtained by averaging over 100 random Monte-Carlo Glauber events. The fifth column specifies the constant value of $\eta/s$ chosen in the hydrodynamic simulations, which were all started at an initial time $\tau_0=0.25$ fm/c with or without pre-equilibrium flow according to the sixth column of Tab.~\ref{tab:one}.
For all runs, the energy scale factor $E_0$ that was  chosen in order to match measured or expected values (\cite{Alver:2010ck}) for the charged particle multiplicity, reported in column seven and eighth of Tab.~\ref{tab:one}. (Note that the calculated $\frac{dN_{\rm ch}}{dY}$ is converted to experimentally measured pseudo-rapidity distribution $\frac{dN_{\rm ch}}{d\eta}$ by dividing by $1.1$). For experimentally measured quantities, also the centrality class for the measurement as well as the corresponding reference is reported. Finally, the last two columns in Tab.~\ref{tab:one} give the mean charged particle transverse momentum in the superSONIC simulation compared to experimental data where available. 

As one can see from Tab.~\ref{tab:one}, simulated particle multiplicities range from $\frac{dN_{\rm ch}}{d\eta}\simeq 74$ down to $\frac{dN_{\rm ch}}{d\eta}\simeq 4.5$. This implies that systems with very few particles are being simulated and one generally expects hydrodynamics to be less applicable to these fewer-body systems. Also, note that there is a clear change in the mean charged particle momentum for systems with compared to without pre-equilibrium flow. This is not too surprising given that systems created in light-heavy ion collisions live comparatively shorter than those created in heavy ion collisions, thus making light-heavy ion collision systems more sensitive to pre-equilibrium conditions. While other parameters (viscosity, choice of switching temperature $T_{\rm sw}$) also affect the particle mean momentum, the strong difference between results with and without pre-equilibrium flow could serve as important discriminatory tool that is easy to implement experimentally.

In figure \ref{fig:all5000}, results for the flow coefficients $v_n$, n=2,3,4 minus $v_5$ are shown for \pPb and \hPb collisions at $\sqrt{s}=5.02$ TeV per nucleon pair (LHC energies). Results are reported as a difference with respect to calculated $v_5$ (the highest harmonic calculated) because finite statistics from the 100,000 B3D runs start to pollute the results at high $p_T$ and the true $v_5(p_T)$ can reasonably be expected to be consistent with zero. Hence the calculated $v_5(p_T)$ is a good measure of the statistical error, and can be used to subtract the statistical fluctuation for the other flow harmonics. In principle, this procedure could be made unnecessary by rerunning all the simulations with at least $10^6$ B3D runs per hydro event, which is left for future work.

For comparison, results with and without pre-equilibrium flow are shown. Boxes indicate uncertainty arising from both statistical fluctuations as well as systematic errors, the latter of which are quantified by performing simulations at different values of $C_\eta=2-3$. From these plots, the first finding is that the hydrodynamic uncertainty range for light-heavy ion collisions at $\sqrt{s}=5.02$ TeV is rather small compared to the overall magnitude of the flow for $v_2,v_3,v_4$, thus providing solid theoretical support to the notion of true hydrodynamic behavior in light-heavy ion collisions. Furthermore, one finds that simulated flow coefficients are in overall agreement with experimental data, where available, when suitably adjusting the constant simulated shear viscosity over entropy density. Note that, somewhat surprisingly, it is hard to distinguish the cases of $\eta/s=0.16$ with pre-equilibrium flow and $\eta/s=0.08$ without pre-equilibrium flow using experimental data for $v_2,v_3,v_4$. For this reason alone, it would be highly desirable to have an experimental handle on the presence of pre-equilibrium flow. Finally, comparing simulations of \hPb and \pPb collisions at LHC energies one finds that one could expect $v_2$ to be 50 percent higher in \hPb collisions than in \pPb collisions, while the overall magnitude for $v_3,v_4$ would be approximately the same. This presumably points to the fact that in \hPb collision there is a sizable geometric component (other than event-by-event fluctuations) that drives $v_2$.

In figure \ref{fig:all200}, results for the flow coefficients $v_n$, n=2,3,4 minus $v_5$ are shown for \pAl, \pAu, \dAu and \hAu collisions at $\sqrt{s}=200$ GeV per nucleon pair (highest RHIC energies). As before, results with and without pre-equilibrium flow are shown and boxes indicate theoretical uncertainty. Again, one finds that simulated flow coefficients are in overall agreement with experimental data, where available. Interestingly, the simulated results, both with and without pre-equilibrium flow predict a $v_3$ component in \dAu collisions that is almost as large as in \hAu. While $v_3$ in central \hAu has been measured by the PHENIX experiment (\cite{Shengli}), a non-vanishing $v_3$ in \dAu central collisions has not been seen in any experiment yet. However, given the unambiguous presence of a sizable $v_3$ component in the superSONIC simulation package, as well as in other theory (hydrodynamic and non-hydrodynamic) simulations \cite{Schenke:2014gaa,Koop:2015wea}, measuring or putting an upper bound on $v_3$ in \dAu collisions at $\sqrt{s}=200$ GeV could serve as a very useful experimental verification of theory ingredients to light-heavy ion collisions.

Similar to the finding for $\sqrt{s}=5.02$ TeV, for results at $\sqrt{s}=200$ shown in Fig.~\ref{fig:all200} one finds noticeably smaller $v_2$ component in \pAl and \pAu collisions compared to \dAu and \hAu. The comparison between different systems for $v_3$ is highly dependent on the presence or absence of pre-equilibrium flow. For instance, with pre-equilibrium flow, one finds that $v_3$ in \pAu collisions at $\sqrt{s}=200$ GeV is almost as large as for \hAu, while without the presence of pre-equilibrium flow $v_3$ in \pAu is about half as large as in \hAu. This suggests that $v_3$ in light-heavy ion collisions at $\sqrt{s}=200$ GeV could be a good experimental handle on pre-equilibrium QCD dynamics. 
Unlike the situation found for $\sqrt{s}=5.02$ TeV, simulation results predict $v_4$ to be consistent with zero except for \pAu,\dAu and \hAu where some small, non-vanishing $v_4$ is found. Since $v_4$ is found to be so small at $\sqrt{s}<200$ GeV, it is likely to be hard to measure, and hence it will be discounted  as a probe for pre-equilibrium dynamics in the following.

What is striking about the results shown in Fig.~\ref{fig:all200} is the magnitude of $v_2$ (and even $v_3$ in the case of pre-equilibrium flow) predicted in \pAl collisions at $\sqrt{s}=200$ GeV. The system created in these collisions consists of an average multiplicity of only $\frac{dN_{\rm ch}}{d\eta}\sim 5.4$, yet superSONIC results exhibit a clear flow response much larger than the estimated uncertainty band for the applicability of hydrodynamics. According to the criterion defined above, I predict that systems created in central \pAl collisions at $\sqrt{s}=200$ GeV can be quantitatively described using viscous hydrodynamics, and that a $v_2$ flow component in these systems can be expected to be similar to that for \pAu collisions at $\sqrt{s}=200$ GeV.

In Fig.~\ref{fig:all62} results for the flow coefficients $v_n$, n=2,3,4 minus $v_5$ are shown for \pAu, \dAu and \hAu collisions at $\sqrt{s}=62.4$ GeV per nucleon pair. Comparing results in Fig.~\ref{fig:all62} to
results at $\sqrt{s}=200$ GeV in Fig.~\ref{fig:all200}, it is hard to find clear differences in any of the flow observables shown. All $v_n$ at $\sqrt{s}=62.4$ GeV in all systems are broadly consistent with results at $\sqrt{s}=200$ GeV, only slightly lower. Nevertheless, one should point out that this in particular implies a sizable $v_2$ component in \pAu at $\sqrt{s}=62.4$, and hence hydrodynamic behavior at these collision energies.

In order to answer the question of a breakdown of hydrodynamic applicability, results for flow coefficients in proton-nucleus collisions (\pAu and \pPb) at energies ranging from $\sqrt{s}=7.7$ GeV to $\sqrt{s}=5.02$ TeV are compared in figure \ref{fig:allproton}. Most remarkable, the simulations predict a sizable $v_2(2.5\ {\rm p_T})\simeq 5-8$ percent even for \pAu collisions at $\sqrt{s}=7.7$ GeV, with a predicted multiplicity of only $\frac{dN_{\rm ch}}{d\eta}\simeq 4.5$. Again, according to the criterion of applicability of hydrodynamics from above, hydrodynamics still applies for systems created in \pAu collisions at $\sqrt{s}=7.7$ GeV. Failure of finding a break-down of hydrodynamics in \pAu collisions at $\sqrt{s}=7.7$ GeV begs the question of which value of $\sqrt{s}$, if any, one would have to study in order to truly see hydrodynamic no longer apply. 
Unfortunately, at collision energies below $\sqrt{s}=7.7$ GeV, the zero chemical potential lattice equation of state employed in superSONIC is clearly no longer applicable, so studying proton-nucleus collisions at energies below $\sqrt{s}=7.7$ GeV is not feasible with the current approach.

While a complete break-down of hydrodynamics is not observed in superSONIC simulations, one does observe a {\it gradual} break-down of hydrodynamics. That is, starting with \pPb collisions at $\sqrt{s}=5.02$ TeV and lowering the collision energy down to $\sqrt{s}=7.7$ GeV for \pAu collisions, first $v_4$ and then $v_3$ first drop and then collapse to values consistent with zero for all momenta considered. In the same fashion, one does also observe a clear drop in $v_2$ as a function of lowering $\sqrt{s}$, even though at $\sqrt{s}=7.7$ GeV $v_2$ results have not (yet) collapsed to zero. Hydrodynamics {\it is} breaking down, but is has not broken down completely for \pAu collisions at $\sqrt{s}=7.7$ GeV. 
My interpretation of this finding is that the question of applicability of hydrodynamics or collectivity in small systems does not have a yes/no answer, but rather should be thought of as a gradual process similar to the confinement/deconfinement cross-over transition in QCD, where the value of the critical temperature is also dependent on the observable one considers.

In Fig.~\ref{fig:ratios}, the flow response $v_2,v_3$ is compared for different collision systems in an attempt to quantify the presence of pre-equilibrium flow. Similar to the results shown in figures above, one finds that $v_2$ is fairly insensitive to pre-equilibrium flow, regardless of the collision energy. However, $v_3$ turns out to be a good indicator for the presence of pre-equilibrium flow in different collision systems at $\sqrt{s}\leq 200$ GeV. Specifically, while $v_3$ in \hAu collisions has been measured at $\sqrt{s}=200$ GeV, a measurement of $v_3$ in \dAu collisions at that same energy could serve as an experimental handle on the presence of pre-equilibrium flow. Also, note that lower collision energies (such as $\sqrt{s}=62.4$ GeV) would be even better suited for an experimental test of pre-equilibrium flow since the systems created in these collisions live comparatively shorter, and are thus more sensitive to out-of-equilibrium QCD transport dynamics. Conversely, it would be harder to probe these pre-equilibrium transport effects at $\sqrt{s}=5.04$ TeV, both because \hPb collisions at the LHC are not currently planned, and because \hPb to \pPb $v_3$ ratios are only sensitive to pre-equilibrium flow at $p_T\lesssim  0.5$ GeV. 

\begin{figure}[t]
\includegraphics[width=0.7\linewidth]{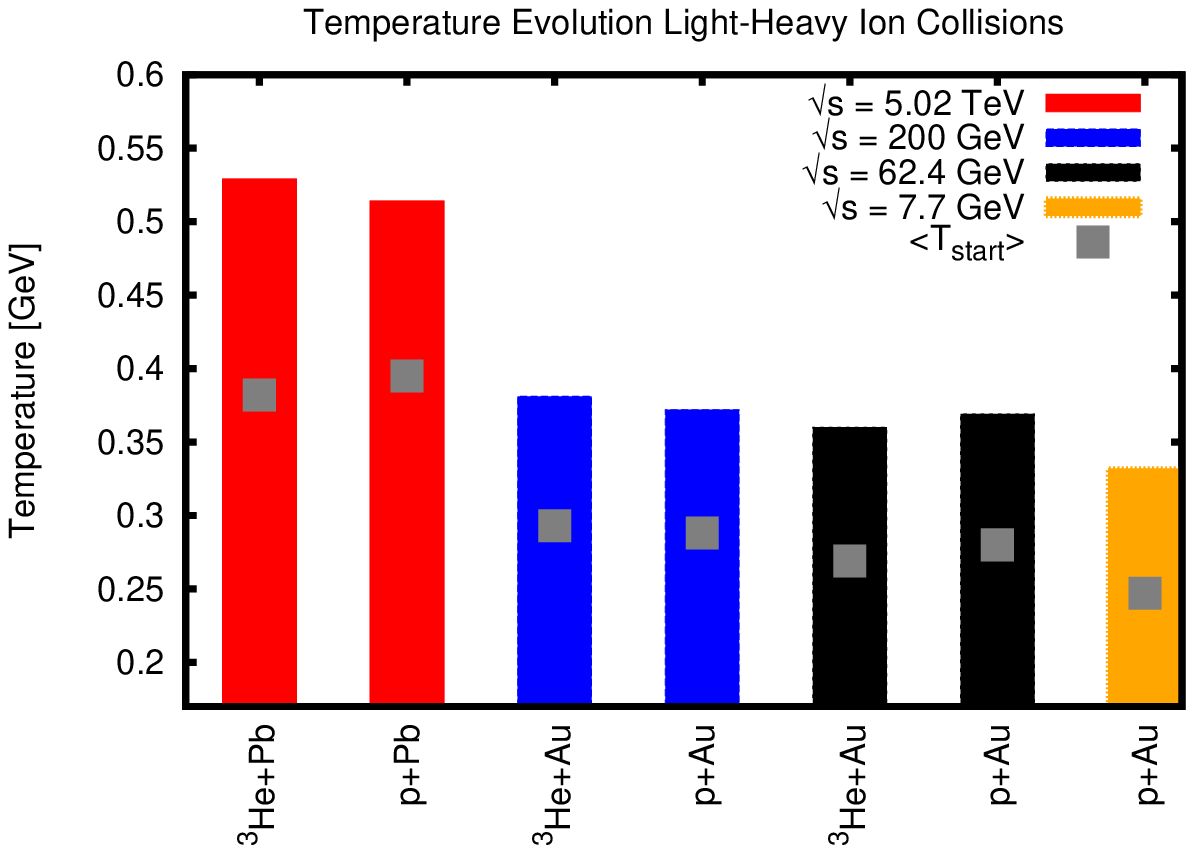}
\caption{\label{fig:Tdep} Temperature evolution in systems created in light-heavy ion collisions. Shown are temperature values encountered in event-by-event superSONIC simulations in \hAu,\hPb, \pAu and \pPb collisions at various energies (bars) and event-by-event mean starting (maximal) temperatures $<T_{\rm start}>$ at $\tau=0.25$ fm/c. Note that \pAu collisions at RHIC energies probe a smaller range of temperatures, thereby being insensitive to transport properties for $T>0.33$ GeV. }
\end{figure}

Another point worth noting is that the more limited temperature range encountered in light-heavy ion collisions as compared to heavy-ion collisions, especially at RHIC energies. Fig.~\ref{fig:Tdep} shows the maximal temperature encountered in various collision systems from $\sqrt{s}=7.7$ GeV to $\sqrt{s}=5.02$ TeV. Also shown in Fig.~\ref{fig:Tdep} is the typical (event-by-event mean) starting temperature, which is a measure of the maximal temperature the {\em average} system starts out with in these collisions (dots). One finds that hydrodynamic evolution in \pPb collisions probes and thus averages over transport properties (such as $\eta/s$ and $\zeta/s$) over a large temperature range $T\lesssim 0.53$ GeV ($<T>=0.4$ GeV, while by contrast \pAu collisions at $\sqrt{s}=7.7$ GeV only probe $T<0.33$ GeV ($<T>=0.25$ GeV). Hence \pAu collisions at $\sqrt{s}=7.7$ GeV are mostly sensitive to QCD transport properties at $T<0.25$ GeV, which could be a key experimental handle on probing the temperature dependence on e.g. $\eta/s$.

Finally, results on HBT Radii for selected collision systems are reported in Tab.~\ref{tab:one} in Figures \ref{fig:all5000hbt},\ref{fig:all200hbt} and \ref{fig:all62hbt}.

\section{Summary and conclusions}
\label{sec:conc}

In this work, central light on heavy ion collisions from $\sqrt{s}=7.7$ GeV to $\sqrt{s}=5.02$ TeV were simulated using superSONIC, a model that combines pre-equilibrium flow, relativistic viscous hydrodynamics and a hadron cascade afterburner. By varying the strength of the second order transport coefficients, one could quantify to which extent viscous hydrodynamics offers a reliable description of these systems . It was found that even in \pAu collisions at $\sqrt{s}=7.7$ GeV, a sizable collective flow component $v_2$ much larger than the systematic hydrodynamic uncertainty is present. Thus, viscous hydrodynamics is still applicable to describe $v_2$ in the systems created in these small, low-energy, few-body collisions. However, there is evidence that hydrodynamics is breaking down gradually as $\sqrt{s}$ is lowered from $5.02$ TeV to $\sqrt{s}=7.7$ GeV. Specifically, first $v_4$, then $v_3$ and finally $v_2$ start to decrease and eventually $v_3$ and $v_4$ become consistent with zero as the collision energy is lowered. The question of whether hydrodynamics applies to describe light-heavy ion collisions therefore cannot be answered by a simple yes or no, but depends on the quantity in question. In future work, it would be interesting to perform simulations at collision energies below $\sqrt{s}=7.7$ in order to potentially observe also $v_2$ collapse to zero in \pAu collisions.

For \dAu collisions at $\sqrt{s}=200$ GeV superSONIC simulations predicted $v_3$ results that were of the same order of magnitude as those measured in \hAu. Since similar results are observed in other theoretical models for \dAu collisions at $\sqrt{s}=200$ GeV, I predicted that $v_3$ should be observable in experiment. Furthermore, the sensitivity of flow results to the presence of pre-equilibrium flow was studied. It was found that low-energy collisions were most sensitive to the presence of pre-equilibrium flow and pointed out that the $v_3$ ratio of \hAu to \dAu could provide an experimental handle on pre-equilibrium QCD dynamics. 

Moreover, simulation results clearly show that light-heavy ion collision systems
probe a temperature window focused around the QCD phase transition temperature. Thus, a combination of simulation results and experimental data for these systems would offer a promising handle on the temperature dependence on QCD transport properties.

Many aspects of this study are amenable to improvements in future work, but despite the current set of limitations and approximations, the findings of this work could hopefully demonstrate the usefulness of studying light-heavy ion collisions to learn about QCD.

\begin{acknowledgments}
 
I would like to thank S.~Huang and J.L.~Nagle for many helpful discussions on this topic.
This work was supported by the Department of Energy, DOE award No. DE-SC0008027.
This work utilized the Janus supercomputer, which is supported by the National Science Foundation (award number CNS-0821794) and the University of Colorado Boulder. The Janus supercomputer is a joint effort of the University of Colorado Boulder, the University of Colorado Denver and the National Center for Atmospheric Research. Janus is operated by the University of Colorado Boulder.

\end{acknowledgments}

\bibliographystyle{apsrev} \bibliography{sphenix}

\begin{thebibliography}{53}
\expandafter\ifx\csname natexlab\endcsname\relax\def\natexlab#1{#1}\fi
\expandafter\ifx\csname bibnamefont\endcsname\relax
  \def\bibnamefont#1{#1}\fi
\expandafter\ifx\csname bibfnamefont\endcsname\relax
  \def\bibfnamefont#1{#1}\fi
\expandafter\ifx\csname citenamefont\endcsname\relax
  \def\citenamefont#1{#1}\fi
\expandafter\ifx\csname url\endcsname\relax
  \def\url#1{\texttt{#1}}\fi
\expandafter\ifx\csname urlprefix\endcsname\relax\def\urlprefix{URL }\fi
\providecommand{\bibinfo}[2]{#2}
\providecommand{\eprint}[2][]{\url{#2}}

\bibitem[{\citenamefont{Abelev et~al.}(2013)}]{Abelev:2012ola}
\bibinfo{author}{\bibfnamefont{B.}~\bibnamefont{Abelev}} \bibnamefont{et~al.}
  (\bibinfo{collaboration}{ALICE Collaboration}), \bibinfo{journal}{Phys.Lett.}
  \textbf{\bibinfo{volume}{B719}}, \bibinfo{pages}{29} (\bibinfo{year}{2013}),
  \eprint{1212.2001}.

\bibitem[{\citenamefont{Aad et~al.}(2013)}]{Aad:2012gla}
\bibinfo{author}{\bibfnamefont{G.}~\bibnamefont{Aad}} \bibnamefont{et~al.}
  (\bibinfo{collaboration}{ATLAS Collaboration}),
  \bibinfo{journal}{Phys.Rev.Lett.} \textbf{\bibinfo{volume}{110}},
  \bibinfo{pages}{182302} (\bibinfo{year}{2013}), \eprint{1212.5198}.

\bibitem[{\citenamefont{Adare et~al.}(2013{\natexlab{a}})}]{Adare:2013piz}
\bibinfo{author}{\bibfnamefont{A.}~\bibnamefont{Adare}} \bibnamefont{et~al.}
  (\bibinfo{collaboration}{PHENIX Collaboration}),
  \bibinfo{journal}{Phys.Rev.Lett.} \textbf{\bibinfo{volume}{111}},
  \bibinfo{pages}{212301} (\bibinfo{year}{2013}{\natexlab{a}}),
  \eprint{1303.1794}.

\bibitem[{\citenamefont{Chatrchyan et~al.}(2013)}]{Chatrchyan:2013nka}
\bibinfo{author}{\bibfnamefont{S.}~\bibnamefont{Chatrchyan}}
  \bibnamefont{et~al.} (\bibinfo{collaboration}{CMS Collaboration}),
  \bibinfo{journal}{Phys.Lett.} \textbf{\bibinfo{volume}{B724}},
  \bibinfo{pages}{213} (\bibinfo{year}{2013}), \eprint{1305.0609}.

\bibitem[{\citenamefont{Huang}(December 2014)}]{Shengli}
\bibinfo{author}{\bibfnamefont{S.}~\bibnamefont{Huang}}
  (\bibinfo{collaboration}{PHENIX Collaboration}), \bibinfo{journal}{Initial
  Stages, Napa Valley,}  (\bibinfo{year}{December 2014}),
  \urlprefix\url{https://indico.cern.ch/event/336283/session/16/contribution/6/material/slides/1.pdf}.

\bibitem[{\citenamefont{Dumitru et~al.}(2014)\citenamefont{Dumitru, McLerran,
  and Skokov}}]{Dumitru:2014yza}
\bibinfo{author}{\bibfnamefont{A.}~\bibnamefont{Dumitru}},
  \bibinfo{author}{\bibfnamefont{L.}~\bibnamefont{McLerran}}, \bibnamefont{and}
  \bibinfo{author}{\bibfnamefont{V.}~\bibnamefont{Skokov}}
  (\bibinfo{year}{2014}), \eprint{1410.4844}.

\bibitem[{CMS(2014)}]{CMSprelim}
\bibinfo{journal}{CMS-HIN-14-006}  (\bibinfo{year}{2014}),
  \urlprefix\url{https://cds.cern.ch/record/1705485/files/HIN-14-006-pas.pdf}.

\bibitem[{\citenamefont{Habich et~al.}(2015)\citenamefont{Habich, Nagle, and
  Romatschke}}]{Habich:2014jna}
\bibinfo{author}{\bibfnamefont{M.}~\bibnamefont{Habich}},
  \bibinfo{author}{\bibfnamefont{J.}~\bibnamefont{Nagle}}, \bibnamefont{and}
  \bibinfo{author}{\bibfnamefont{P.}~\bibnamefont{Romatschke}},
  \bibinfo{journal}{Eur.Phys.J.} \textbf{\bibinfo{volume}{C75}},
  \bibinfo{pages}{15} (\bibinfo{year}{2015}), \eprint{1409.0040}.

\bibitem[{\citenamefont{Bozek}(2012)}]{Bozek:2011if}
\bibinfo{author}{\bibfnamefont{P.}~\bibnamefont{Bozek}},
  \bibinfo{journal}{Phys.Rev.} \textbf{\bibinfo{volume}{C85}},
  \bibinfo{pages}{014911} (\bibinfo{year}{2012}), \eprint{1112.0915}.

\bibitem[{\citenamefont{Nagle et~al.}(2014)\citenamefont{Nagle, Adare, Beckman,
  Koblesky, Koop et~al.}}]{Nagle:2013lja}
\bibinfo{author}{\bibfnamefont{J.}~\bibnamefont{Nagle}},
  \bibinfo{author}{\bibfnamefont{A.}~\bibnamefont{Adare}},
  \bibinfo{author}{\bibfnamefont{S.}~\bibnamefont{Beckman}},
  \bibinfo{author}{\bibfnamefont{T.}~\bibnamefont{Koblesky}},
  \bibinfo{author}{\bibfnamefont{J.~O.} \bibnamefont{Koop}},
  \bibnamefont{et~al.}, \bibinfo{journal}{Phys.Rev.Lett.}
  \textbf{\bibinfo{volume}{113}}, \bibinfo{pages}{112301}
  (\bibinfo{year}{2014}), \eprint{1312.4565}.

\bibitem[{\citenamefont{Huang and Mitchell}(2014)}]{Shenglipc}
\bibinfo{author}{\bibfnamefont{S.}~\bibnamefont{Huang}} \bibnamefont{and}
  \bibinfo{author}{\bibfnamefont{J.}~\bibnamefont{Mitchell}},
  \bibinfo{journal}{private communication}  (\bibinfo{year}{2014}).

\bibitem[{\citenamefont{Schenke and
  Venugopalan}(2014{\natexlab{a}})}]{Schenke:2014zha}
\bibinfo{author}{\bibfnamefont{B.}~\bibnamefont{Schenke}} \bibnamefont{and}
  \bibinfo{author}{\bibfnamefont{R.}~\bibnamefont{Venugopalan}},
  \bibinfo{journal}{Phys.Rev.Lett.} \textbf{\bibinfo{volume}{113}},
  \bibinfo{pages}{102301} (\bibinfo{year}{2014}{\natexlab{a}}),
  \eprint{1405.3605}.

\bibitem[{\citenamefont{Schenke and
  Venugopalan}(2014{\natexlab{b}})}]{Schenke:2014gaa}
\bibinfo{author}{\bibfnamefont{B.}~\bibnamefont{Schenke}} \bibnamefont{and}
  \bibinfo{author}{\bibfnamefont{R.}~\bibnamefont{Venugopalan}},
  \bibinfo{journal}{Nucl.Phys.} \textbf{\bibinfo{volume}{A931}},
  \bibinfo{pages}{1039} (\bibinfo{year}{2014}{\natexlab{b}}),
  \eprint{1407.7557}.

\bibitem[{\citenamefont{Kozlov et~al.}(2014)\citenamefont{Kozlov, Luzum,
  Denicol, Jeon, and Gale}}]{Kozlov:2014fqa}
\bibinfo{author}{\bibfnamefont{I.}~\bibnamefont{Kozlov}},
  \bibinfo{author}{\bibfnamefont{M.}~\bibnamefont{Luzum}},
  \bibinfo{author}{\bibfnamefont{G.}~\bibnamefont{Denicol}},
  \bibinfo{author}{\bibfnamefont{S.}~\bibnamefont{Jeon}}, \bibnamefont{and}
  \bibinfo{author}{\bibfnamefont{C.}~\bibnamefont{Gale}}
  (\bibinfo{year}{2014}), \eprint{1405.3976}.

\bibitem[{\citenamefont{Miller et~al.}(2007)\citenamefont{Miller, Reygers,
  Sanders, and Steinberg}}]{Miller:2007ri}
\bibinfo{author}{\bibfnamefont{M.~L.} \bibnamefont{Miller}},
  \bibinfo{author}{\bibfnamefont{K.}~\bibnamefont{Reygers}},
  \bibinfo{author}{\bibfnamefont{S.~J.} \bibnamefont{Sanders}},
  \bibnamefont{and}
  \bibinfo{author}{\bibfnamefont{P.}~\bibnamefont{Steinberg}},
  \bibinfo{journal}{Ann.Rev.Nucl.Part.Sci.} \textbf{\bibinfo{volume}{57}},
  \bibinfo{pages}{205} (\bibinfo{year}{2007}), \eprint{nucl-ex/0701025}.

\bibitem[{\citenamefont{Schenke et~al.}(2012)\citenamefont{Schenke, Tribedy,
  and Venugopalan}}]{Schenke:2012wb}
\bibinfo{author}{\bibfnamefont{B.}~\bibnamefont{Schenke}},
  \bibinfo{author}{\bibfnamefont{P.}~\bibnamefont{Tribedy}}, \bibnamefont{and}
  \bibinfo{author}{\bibfnamefont{R.}~\bibnamefont{Venugopalan}},
  \bibinfo{journal}{Phys.Rev.Lett.} \textbf{\bibinfo{volume}{108}},
  \bibinfo{pages}{252301} (\bibinfo{year}{2012}), \eprint{1202.6646}.

\bibitem[{\citenamefont{Mrowczynski}(1993)}]{Mrowczynski:1993qm}
\bibinfo{author}{\bibfnamefont{S.}~\bibnamefont{Mrowczynski}},
  \bibinfo{journal}{Phys.Lett.} \textbf{\bibinfo{volume}{B314}},
  \bibinfo{pages}{118} (\bibinfo{year}{1993}).

\bibitem[{\citenamefont{Arnold et~al.}(2005)\citenamefont{Arnold, Lenaghan,
  Moore, and Yaffe}}]{Arnold:2004ti}
\bibinfo{author}{\bibfnamefont{P.~B.} \bibnamefont{Arnold}},
  \bibinfo{author}{\bibfnamefont{J.}~\bibnamefont{Lenaghan}},
  \bibinfo{author}{\bibfnamefont{G.~D.} \bibnamefont{Moore}}, \bibnamefont{and}
  \bibinfo{author}{\bibfnamefont{L.~G.} \bibnamefont{Yaffe}},
  \bibinfo{journal}{Phys.Rev.Lett.} \textbf{\bibinfo{volume}{94}},
  \bibinfo{pages}{072302} (\bibinfo{year}{2005}), \eprint{nucl-th/0409068}.

\bibitem[{\citenamefont{Rebhan et~al.}(2005)\citenamefont{Rebhan, Romatschke,
  and Strickland}}]{Rebhan:2004ur}
\bibinfo{author}{\bibfnamefont{A.}~\bibnamefont{Rebhan}},
  \bibinfo{author}{\bibfnamefont{P.}~\bibnamefont{Romatschke}},
  \bibnamefont{and}
  \bibinfo{author}{\bibfnamefont{M.}~\bibnamefont{Strickland}},
  \bibinfo{journal}{Phys.Rev.Lett.} \textbf{\bibinfo{volume}{94}},
  \bibinfo{pages}{102303} (\bibinfo{year}{2005}), \eprint{hep-ph/0412016}.

\bibitem[{\citenamefont{Romatschke and Venugopalan}(2006)}]{Romatschke:2005pm}
\bibinfo{author}{\bibfnamefont{P.}~\bibnamefont{Romatschke}} \bibnamefont{and}
  \bibinfo{author}{\bibfnamefont{R.}~\bibnamefont{Venugopalan}},
  \bibinfo{journal}{Phys.Rev.Lett.} \textbf{\bibinfo{volume}{96}},
  \bibinfo{pages}{062302} (\bibinfo{year}{2006}), \eprint{hep-ph/0510121}.

\bibitem[{\citenamefont{Kurkela and Moore}(2011)}]{Kurkela:2011ti}
\bibinfo{author}{\bibfnamefont{A.}~\bibnamefont{Kurkela}} \bibnamefont{and}
  \bibinfo{author}{\bibfnamefont{G.~D.} \bibnamefont{Moore}},
  \bibinfo{journal}{JHEP} \textbf{\bibinfo{volume}{1112}}, \bibinfo{pages}{044}
  (\bibinfo{year}{2011}), \eprint{1107.5050}.

\bibitem[{\citenamefont{Epelbaum and Gelis}(2013)}]{Gelis:2013rba}
\bibinfo{author}{\bibfnamefont{T.}~\bibnamefont{Epelbaum}} \bibnamefont{and}
  \bibinfo{author}{\bibfnamefont{F.}~\bibnamefont{Gelis}},
  \bibinfo{journal}{Phys.Rev.Lett.} \textbf{\bibinfo{volume}{111}},
  \bibinfo{pages}{232301} (\bibinfo{year}{2013}), \eprint{1307.2214}.

\bibitem[{\citenamefont{Berges et~al.}(2014)\citenamefont{Berges, Boguslavski,
  Schlichting, and Venugopalan}}]{Berges:2013fga}
\bibinfo{author}{\bibfnamefont{J.}~\bibnamefont{Berges}},
  \bibinfo{author}{\bibfnamefont{K.}~\bibnamefont{Boguslavski}},
  \bibinfo{author}{\bibfnamefont{S.}~\bibnamefont{Schlichting}},
  \bibnamefont{and}
  \bibinfo{author}{\bibfnamefont{R.}~\bibnamefont{Venugopalan}},
  \bibinfo{journal}{Phys.Rev.} \textbf{\bibinfo{volume}{D89}},
  \bibinfo{pages}{114007} (\bibinfo{year}{2014}), \eprint{1311.3005}.

\bibitem[{\citenamefont{Strickland}(2013)}]{Strickland:2013uga}
\bibinfo{author}{\bibfnamefont{M.}~\bibnamefont{Strickland}}
  (\bibinfo{year}{2013}), \eprint{1312.2285}.

\bibitem[{\citenamefont{Kurkela and Lu}(2014)}]{Kurkela:2014tea}
\bibinfo{author}{\bibfnamefont{A.}~\bibnamefont{Kurkela}} \bibnamefont{and}
  \bibinfo{author}{\bibfnamefont{E.}~\bibnamefont{Lu}},
  \bibinfo{journal}{Phys.Rev.Lett.} \textbf{\bibinfo{volume}{113}},
  \bibinfo{pages}{182301} (\bibinfo{year}{2014}), \eprint{1405.6318}.

\bibitem[{\citenamefont{Nastase}(2005)}]{Nastase:2005rp}
\bibinfo{author}{\bibfnamefont{H.}~\bibnamefont{Nastase}}
  (\bibinfo{year}{2005}), \eprint{hep-th/0501068}.

\bibitem[{\citenamefont{Kovchegov and Taliotis}(2007)}]{Kovchegov:2007pq}
\bibinfo{author}{\bibfnamefont{Y.~V.} \bibnamefont{Kovchegov}}
  \bibnamefont{and} \bibinfo{author}{\bibfnamefont{A.}~\bibnamefont{Taliotis}},
  \bibinfo{journal}{Phys.Rev.} \textbf{\bibinfo{volume}{C76}},
  \bibinfo{pages}{014905} (\bibinfo{year}{2007}), \eprint{0705.1234}.

\bibitem[{\citenamefont{Amsel et~al.}(2008)\citenamefont{Amsel, Marolf, and
  Virmani}}]{Amsel:2007cw}
\bibinfo{author}{\bibfnamefont{A.~J.} \bibnamefont{Amsel}},
  \bibinfo{author}{\bibfnamefont{D.}~\bibnamefont{Marolf}}, \bibnamefont{and}
  \bibinfo{author}{\bibfnamefont{A.}~\bibnamefont{Virmani}},
  \bibinfo{journal}{JHEP} \textbf{\bibinfo{volume}{0804}}, \bibinfo{pages}{025}
  (\bibinfo{year}{2008}), \eprint{0712.2221}.

\bibitem[{\citenamefont{Grumiller and Romatschke}(2008)}]{Grumiller:2008va}
\bibinfo{author}{\bibfnamefont{D.}~\bibnamefont{Grumiller}} \bibnamefont{and}
  \bibinfo{author}{\bibfnamefont{P.}~\bibnamefont{Romatschke}},
  \bibinfo{journal}{JHEP} \textbf{\bibinfo{volume}{0808}}, \bibinfo{pages}{027}
  (\bibinfo{year}{2008}), \eprint{0803.3226}.

\bibitem[{\citenamefont{Chesler and Yaffe}(2011)}]{Chesler:2010bi}
\bibinfo{author}{\bibfnamefont{P.~M.} \bibnamefont{Chesler}} \bibnamefont{and}
  \bibinfo{author}{\bibfnamefont{L.~G.} \bibnamefont{Yaffe}},
  \bibinfo{journal}{Phys.Rev.Lett.} \textbf{\bibinfo{volume}{106}},
  \bibinfo{pages}{021601} (\bibinfo{year}{2011}), \eprint{1011.3562}.

\bibitem[{\citenamefont{Wu and Romatschke}(2011)}]{Wu:2011yd}
\bibinfo{author}{\bibfnamefont{B.}~\bibnamefont{Wu}} \bibnamefont{and}
  \bibinfo{author}{\bibfnamefont{P.}~\bibnamefont{Romatschke}},
  \bibinfo{journal}{Int.J.Mod.Phys.} \textbf{\bibinfo{volume}{C22}},
  \bibinfo{pages}{1317} (\bibinfo{year}{2011}), \eprint{1108.3715}.

\bibitem[{\citenamefont{van~der Schee et~al.}(2013)\citenamefont{van~der Schee,
  Romatschke, and Pratt}}]{vanderSchee:2013pia}
\bibinfo{author}{\bibfnamefont{W.}~\bibnamefont{van~der Schee}},
  \bibinfo{author}{\bibfnamefont{P.}~\bibnamefont{Romatschke}},
  \bibnamefont{and} \bibinfo{author}{\bibfnamefont{S.}~\bibnamefont{Pratt}},
  \bibinfo{journal}{Phys.Rev.Lett.} \textbf{\bibinfo{volume}{111}},
  \bibinfo{pages}{222302} (\bibinfo{year}{2013}), \eprint{1307.2539}.

\bibitem[{\citenamefont{Casalderrey-Solana
  et~al.}(2014)\citenamefont{Casalderrey-Solana, Heller, Mateos, and van~der
  Schee}}]{Casalderrey-Solana:2013sxa}
\bibinfo{author}{\bibfnamefont{J.}~\bibnamefont{Casalderrey-Solana}},
  \bibinfo{author}{\bibfnamefont{M.~P.} \bibnamefont{Heller}},
  \bibinfo{author}{\bibfnamefont{D.}~\bibnamefont{Mateos}}, \bibnamefont{and}
  \bibinfo{author}{\bibfnamefont{W.}~\bibnamefont{van~der Schee}},
  \bibinfo{journal}{Phys.Rev.Lett.} \textbf{\bibinfo{volume}{112}},
  \bibinfo{pages}{221602} (\bibinfo{year}{2014}), \eprint{1312.2956}.

\bibitem[{\citenamefont{Bantilan and Romatschke}(2014)}]{Bantilan:2014sra}
\bibinfo{author}{\bibfnamefont{H.}~\bibnamefont{Bantilan}} \bibnamefont{and}
  \bibinfo{author}{\bibfnamefont{P.}~\bibnamefont{Romatschke}}
  (\bibinfo{year}{2014}), \eprint{1410.4799}.

\bibitem[{\citenamefont{Chesler and Yaffe}(2015)}]{Chesler:2015wra}
\bibinfo{author}{\bibfnamefont{P.~M.} \bibnamefont{Chesler}} \bibnamefont{and}
  \bibinfo{author}{\bibfnamefont{L.~G.} \bibnamefont{Yaffe}}
  (\bibinfo{year}{2015}), \eprint{1501.04644}.

\bibitem[{\citenamefont{Romatschke and Hogg}(2013)}]{Romatschke:2013re}
\bibinfo{author}{\bibfnamefont{P.}~\bibnamefont{Romatschke}} \bibnamefont{and}
  \bibinfo{author}{\bibfnamefont{J.~D.} \bibnamefont{Hogg}},
  \bibinfo{journal}{JHEP} \textbf{\bibinfo{volume}{1304}}, \bibinfo{pages}{048}
  (\bibinfo{year}{2013}), \eprint{1301.2635}.

\bibitem[{\citenamefont{Vredevoogd and Pratt}(2009)}]{Vredevoogd:2008id}
\bibinfo{author}{\bibfnamefont{J.}~\bibnamefont{Vredevoogd}} \bibnamefont{and}
  \bibinfo{author}{\bibfnamefont{S.}~\bibnamefont{Pratt}},
  \bibinfo{journal}{Phys.Rev.} \textbf{\bibinfo{volume}{C79}},
  \bibinfo{pages}{044915} (\bibinfo{year}{2009}), \eprint{0810.4325}.

\bibitem[{\citenamefont{Jager et~al.}(1974)\citenamefont{Jager, Vries, and
  Vries}}]{DeJager1974479}
\bibinfo{author}{\bibfnamefont{C.~D.} \bibnamefont{Jager}},
  \bibinfo{author}{\bibfnamefont{H.~D.} \bibnamefont{Vries}}, \bibnamefont{and}
  \bibinfo{author}{\bibfnamefont{C.~D.} \bibnamefont{Vries}},
  \bibinfo{journal}{Atomic Data and Nuclear Data Tables}
  \textbf{\bibinfo{volume}{14}}, \bibinfo{pages}{479 } (\bibinfo{year}{1974}),
  ISSN \bibinfo{issn}{0092-640X}, \bibinfo{note}{nuclear Charge and Moment
  Distributions},
  \urlprefix\url{http://www.sciencedirect.com/science/article/pii/S0092640X74800021}.

\bibitem[{\citenamefont{Vries et~al.}(1987)\citenamefont{Vries, Jager, and
  Vries}}]{DeVries1987495}
\bibinfo{author}{\bibfnamefont{H.~D.} \bibnamefont{Vries}},
  \bibinfo{author}{\bibfnamefont{C.~D.} \bibnamefont{Jager}}, \bibnamefont{and}
  \bibinfo{author}{\bibfnamefont{C.~D.} \bibnamefont{Vries}},
  \bibinfo{journal}{Atomic Data and Nuclear Data Tables}
  \textbf{\bibinfo{volume}{36}}, \bibinfo{pages}{495 } (\bibinfo{year}{1987}),
  ISSN \bibinfo{issn}{0092-640X},
  \urlprefix\url{http://www.sciencedirect.com/science/article/pii/0092640X87900131}.

\bibitem[{\citenamefont{Adare et~al.}(2014{\natexlab{a}})}]{Adare:2013nff}
\bibinfo{author}{\bibfnamefont{A.}~\bibnamefont{Adare}} \bibnamefont{et~al.}
  (\bibinfo{collaboration}{PHENIX Collaboration}), \bibinfo{journal}{Phys.Rev.}
  \textbf{\bibinfo{volume}{C90}}, \bibinfo{pages}{034902}
  (\bibinfo{year}{2014}{\natexlab{a}}), \eprint{1310.4793}.

\bibitem[{\citenamefont{Carlson and Schiavilla}(1998)}]{Carlson:1997qn}
\bibinfo{author}{\bibfnamefont{J.}~\bibnamefont{Carlson}} \bibnamefont{and}
  \bibinfo{author}{\bibfnamefont{R.}~\bibnamefont{Schiavilla}},
  \bibinfo{journal}{Rev.Mod.Phys.} \textbf{\bibinfo{volume}{70}},
  \bibinfo{pages}{743} (\bibinfo{year}{1998}).

\bibitem[{\citenamefont{Luzum and Romatschke}(2008)}]{Luzum:2008cw}
\bibinfo{author}{\bibfnamefont{M.}~\bibnamefont{Luzum}} \bibnamefont{and}
  \bibinfo{author}{\bibfnamefont{P.}~\bibnamefont{Romatschke}},
  \bibinfo{journal}{Phys.Rev.} \textbf{\bibinfo{volume}{C78}},
  \bibinfo{pages}{034915} (\bibinfo{year}{2008}), \eprint{0804.4015}.

\bibitem[{\citenamefont{Baier et~al.}(2008)\citenamefont{Baier, Romatschke,
  Son, Starinets, and Stephanov}}]{Baier:2007ix}
\bibinfo{author}{\bibfnamefont{R.}~\bibnamefont{Baier}},
  \bibinfo{author}{\bibfnamefont{P.}~\bibnamefont{Romatschke}},
  \bibinfo{author}{\bibfnamefont{D.~T.} \bibnamefont{Son}},
  \bibinfo{author}{\bibfnamefont{A.~O.} \bibnamefont{Starinets}},
  \bibnamefont{and} \bibinfo{author}{\bibfnamefont{M.~A.}
  \bibnamefont{Stephanov}}, \bibinfo{journal}{JHEP}
  \textbf{\bibinfo{volume}{0804}}, \bibinfo{pages}{100} (\bibinfo{year}{2008}),
  \eprint{0712.2451}.

\bibitem[{\citenamefont{Bhattacharyya et~al.}(2008)\citenamefont{Bhattacharyya,
  Hubeny, Minwalla, and Rangamani}}]{Bhattacharyya:2008jc}
\bibinfo{author}{\bibfnamefont{S.}~\bibnamefont{Bhattacharyya}},
  \bibinfo{author}{\bibfnamefont{V.~E.} \bibnamefont{Hubeny}},
  \bibinfo{author}{\bibfnamefont{S.}~\bibnamefont{Minwalla}}, \bibnamefont{and}
  \bibinfo{author}{\bibfnamefont{M.}~\bibnamefont{Rangamani}},
  \bibinfo{journal}{JHEP} \textbf{\bibinfo{volume}{0802}}, \bibinfo{pages}{045}
  (\bibinfo{year}{2008}), \eprint{0712.2456}.

\bibitem[{\citenamefont{Borsanyi et~al.}(2014)\citenamefont{Borsanyi, Fodor,
  Hoelbling, Katz, Krieg et~al.}}]{Borsanyi:2013bia}
\bibinfo{author}{\bibfnamefont{S.}~\bibnamefont{Borsanyi}},
  \bibinfo{author}{\bibfnamefont{Z.}~\bibnamefont{Fodor}},
  \bibinfo{author}{\bibfnamefont{C.}~\bibnamefont{Hoelbling}},
  \bibinfo{author}{\bibfnamefont{S.~D.} \bibnamefont{Katz}},
  \bibinfo{author}{\bibfnamefont{S.}~\bibnamefont{Krieg}},
  \bibnamefont{et~al.}, \bibinfo{journal}{Phys.Lett.}
  \textbf{\bibinfo{volume}{B730}}, \bibinfo{pages}{99} (\bibinfo{year}{2014}),
  \eprint{1309.5258}.

\bibitem[{\citenamefont{Novak et~al.}(2014)\citenamefont{Novak, Novak, Pratt,
  Vredevoogd, Coleman-Smith et~al.}}]{Novak:2013bqa}
\bibinfo{author}{\bibfnamefont{J.}~\bibnamefont{Novak}},
  \bibinfo{author}{\bibfnamefont{K.}~\bibnamefont{Novak}},
  \bibinfo{author}{\bibfnamefont{S.}~\bibnamefont{Pratt}},
  \bibinfo{author}{\bibfnamefont{J.}~\bibnamefont{Vredevoogd}},
  \bibinfo{author}{\bibfnamefont{C.}~\bibnamefont{Coleman-Smith}},
  \bibnamefont{et~al.}, \bibinfo{journal}{Phys.Rev.}
  \textbf{\bibinfo{volume}{C89}}, \bibinfo{pages}{034917}
  (\bibinfo{year}{2014}), \eprint{1303.5769}.

\bibitem[{\citenamefont{Pratt and Torrieri}(2010)}]{Pratt:2010jt}
\bibinfo{author}{\bibfnamefont{S.}~\bibnamefont{Pratt}} \bibnamefont{and}
  \bibinfo{author}{\bibfnamefont{G.}~\bibnamefont{Torrieri}},
  \bibinfo{journal}{Phys.Rev.} \textbf{\bibinfo{volume}{C82}},
  \bibinfo{pages}{044901} (\bibinfo{year}{2010}), \eprint{1003.0413}.

\bibitem[{\citenamefont{Adare et~al.}(2013{\natexlab{b}})}]{Adare:2013esx}
\bibinfo{author}{\bibfnamefont{A.}~\bibnamefont{Adare}} \bibnamefont{et~al.}
  (\bibinfo{collaboration}{PHENIX Collaboration}), \bibinfo{journal}{Phys.Rev.}
  \textbf{\bibinfo{volume}{C88}}, \bibinfo{pages}{024906}
  (\bibinfo{year}{2013}{\natexlab{b}}), \eprint{1304.3410}.

\bibitem[{\citenamefont{Aad et~al.}(2011)}]{Aad:2011eu}
\bibinfo{author}{\bibfnamefont{G.}~\bibnamefont{Aad}} \bibnamefont{et~al.}
  (\bibinfo{collaboration}{ATLAS Collaboration}), \bibinfo{journal}{Nature
  Commun.} \textbf{\bibinfo{volume}{2}}, \bibinfo{pages}{463}
  (\bibinfo{year}{2011}), \eprint{1104.0326}.

\bibitem[{\citenamefont{Alver et~al.}(2011)}]{Alver:2010ck}
\bibinfo{author}{\bibfnamefont{B.}~\bibnamefont{Alver}} \bibnamefont{et~al.}
  (\bibinfo{collaboration}{PHOBOS Collaboration}), \bibinfo{journal}{Phys.Rev.}
  \textbf{\bibinfo{volume}{C83}}, \bibinfo{pages}{024913}
  (\bibinfo{year}{2011}), \eprint{1011.1940}.

\bibitem[{\citenamefont{Koop et~al.}(2015)\citenamefont{Koop, Adare,
  McGlinchey, and Nagle}}]{Koop:2015wea}
\bibinfo{author}{\bibfnamefont{J.~D.~O.} \bibnamefont{Koop}},
  \bibinfo{author}{\bibfnamefont{A.}~\bibnamefont{Adare}},
  \bibinfo{author}{\bibfnamefont{D.}~\bibnamefont{McGlinchey}},
  \bibnamefont{and} \bibinfo{author}{\bibfnamefont{J.}~\bibnamefont{Nagle}}
  (\bibinfo{year}{2015}), \eprint{1501.06880}.

\bibitem[{\citenamefont{Aad et~al.}(2014)}]{Aad:2014lta}
\bibinfo{author}{\bibfnamefont{G.}~\bibnamefont{Aad}} \bibnamefont{et~al.}
  (\bibinfo{collaboration}{ATLAS Collaboration}), \bibinfo{journal}{Phys.Rev.}
  \textbf{\bibinfo{volume}{C90}}, \bibinfo{pages}{044906}
  (\bibinfo{year}{2014}), \eprint{1409.1792}.

\bibitem[{\citenamefont{Adare et~al.}(2014{\natexlab{b}})}]{Adare:2014keg}
\bibinfo{author}{\bibfnamefont{A.}~\bibnamefont{Adare}} \bibnamefont{et~al.}
  (\bibinfo{collaboration}{PHENIX Collaboration})
  (\bibinfo{year}{2014}{\natexlab{b}}), \eprint{1404.7461}.

\end{thebibliography}

\newpage
\begin{appendix}
\section{Plot Collection}
For better readability of the text, many of the plots for this article are collected in this appendix. 

\begin{figure}[h]
\begin{center}
\begin{Large}
LHC, $\sqrt{s}=5.02$ TeV
\end{Large}
\end{center}
\includegraphics[width=0.47\linewidth]{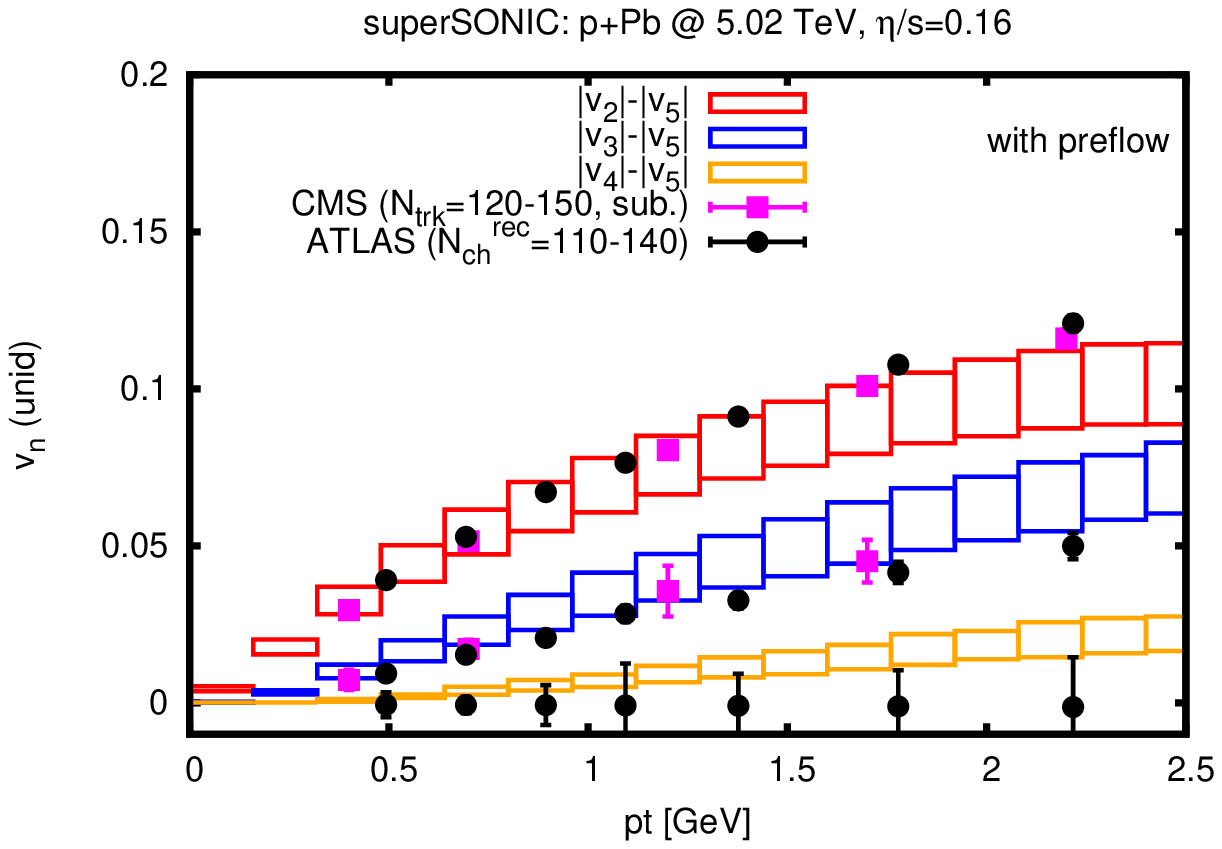}\hfill
\includegraphics[width=0.47\linewidth]{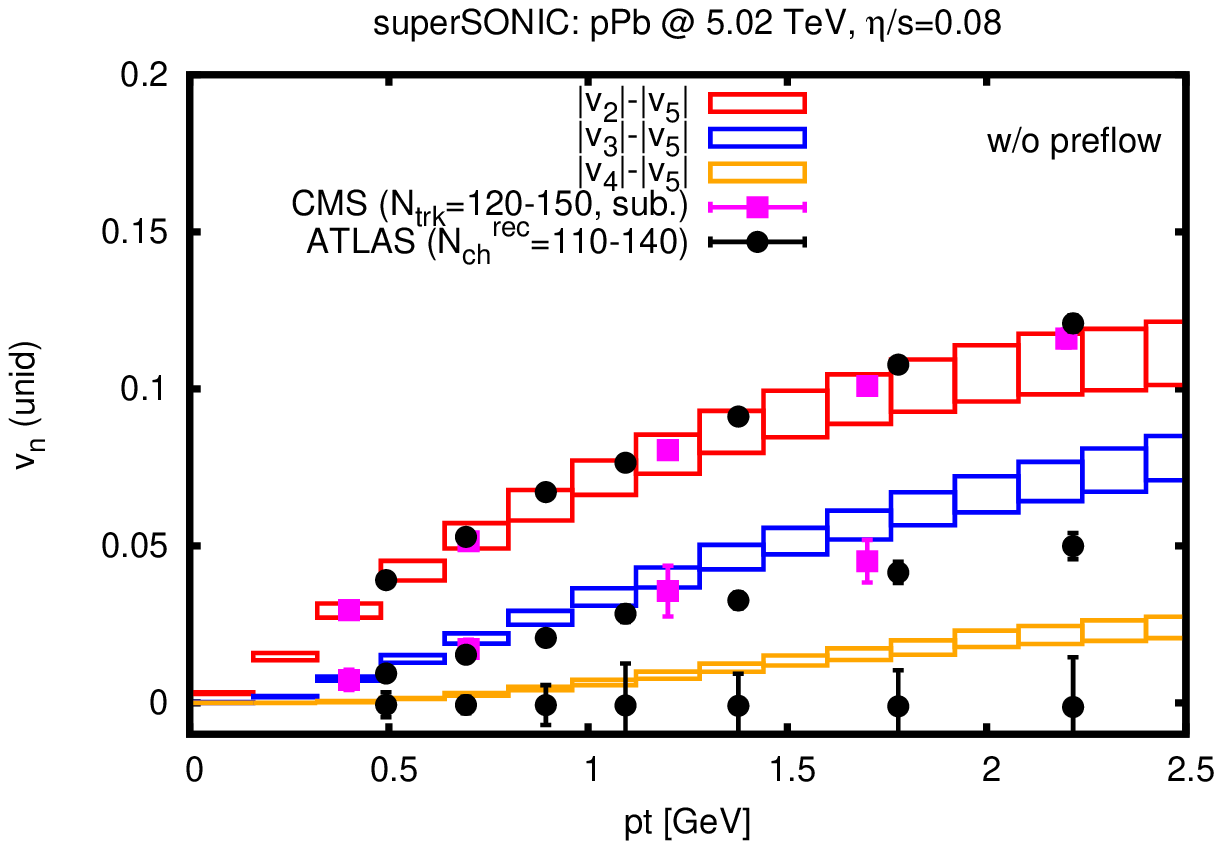}
\includegraphics[width=0.47\linewidth]{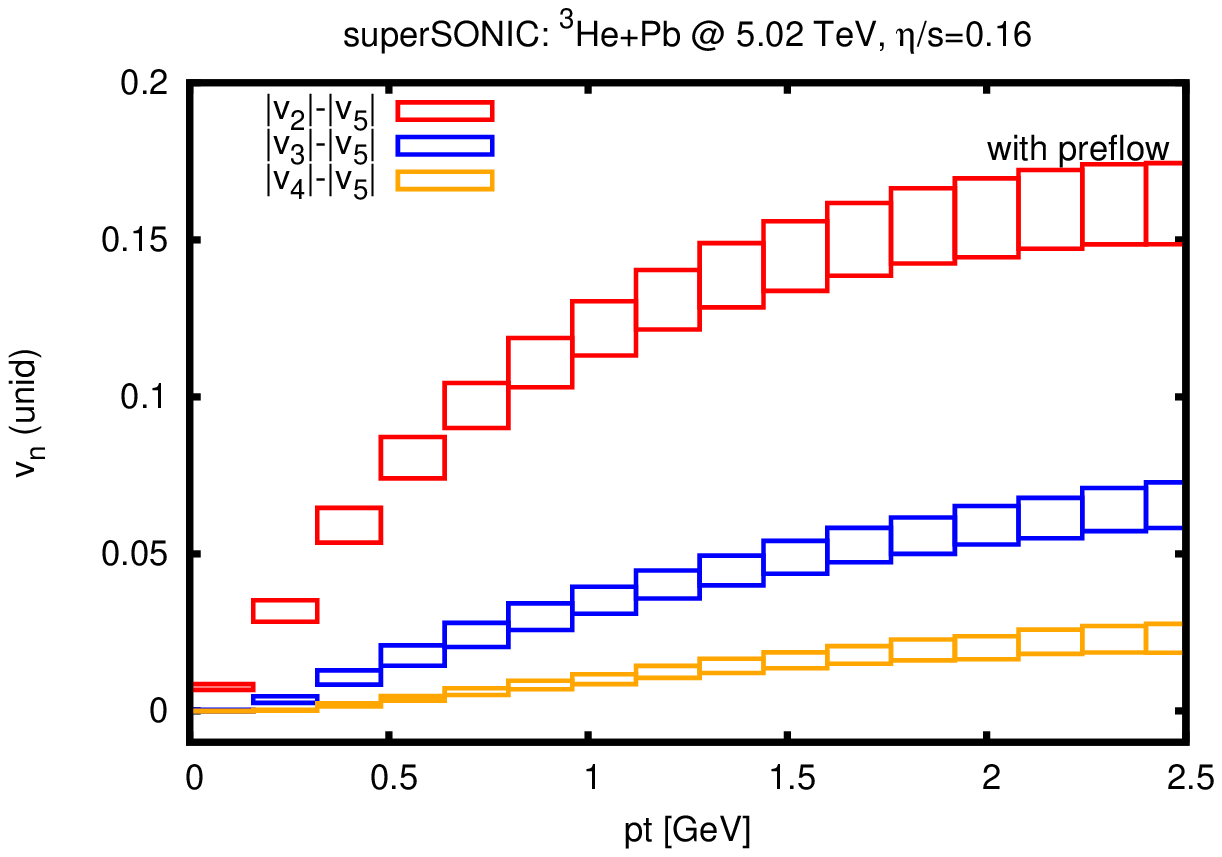}\hfill
\includegraphics[width=0.47\linewidth]{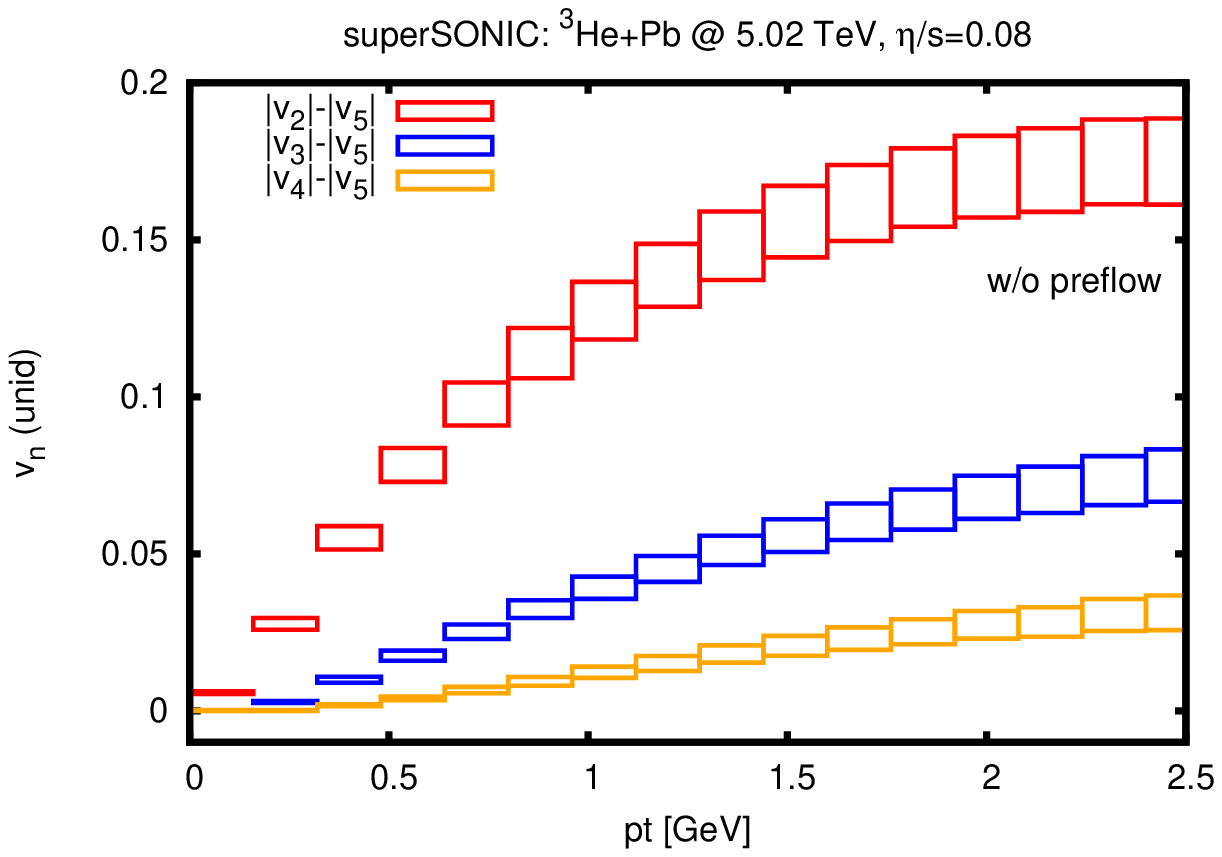}
\caption{\label{fig:all5000}
Flow harmonics $v_n(p_T)-v_5(p_T)$ for unidentified charged particles (unid) for $n=2,3,4$ from superSONIC, with and without pre-equilibrium flow. Boxes indicate combined statistic and estimated systematic error for hydrodynamics (latter from varying $C_\eta=2-3$). For reference, experimental data is shown where available \cite{Chatrchyan:2013nka,Aad:2014lta}. }
\end{figure}

\begin{figure}[htp]
\begin{center}
\begin{Large}
RHIC, $\sqrt{s}=200$ GeV
\end{Large}
\end{center}
\includegraphics[width=0.45\linewidth]{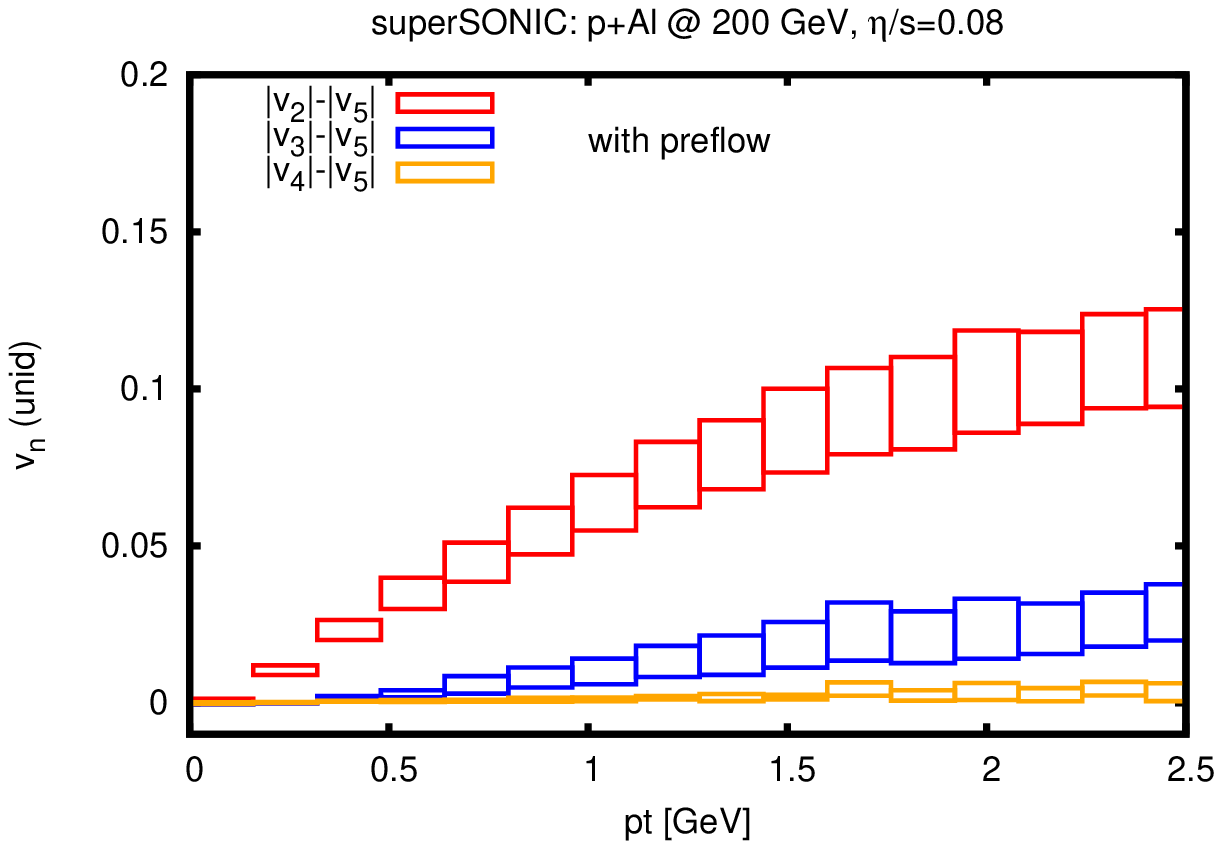}\hfill
\includegraphics[width=0.45\linewidth]{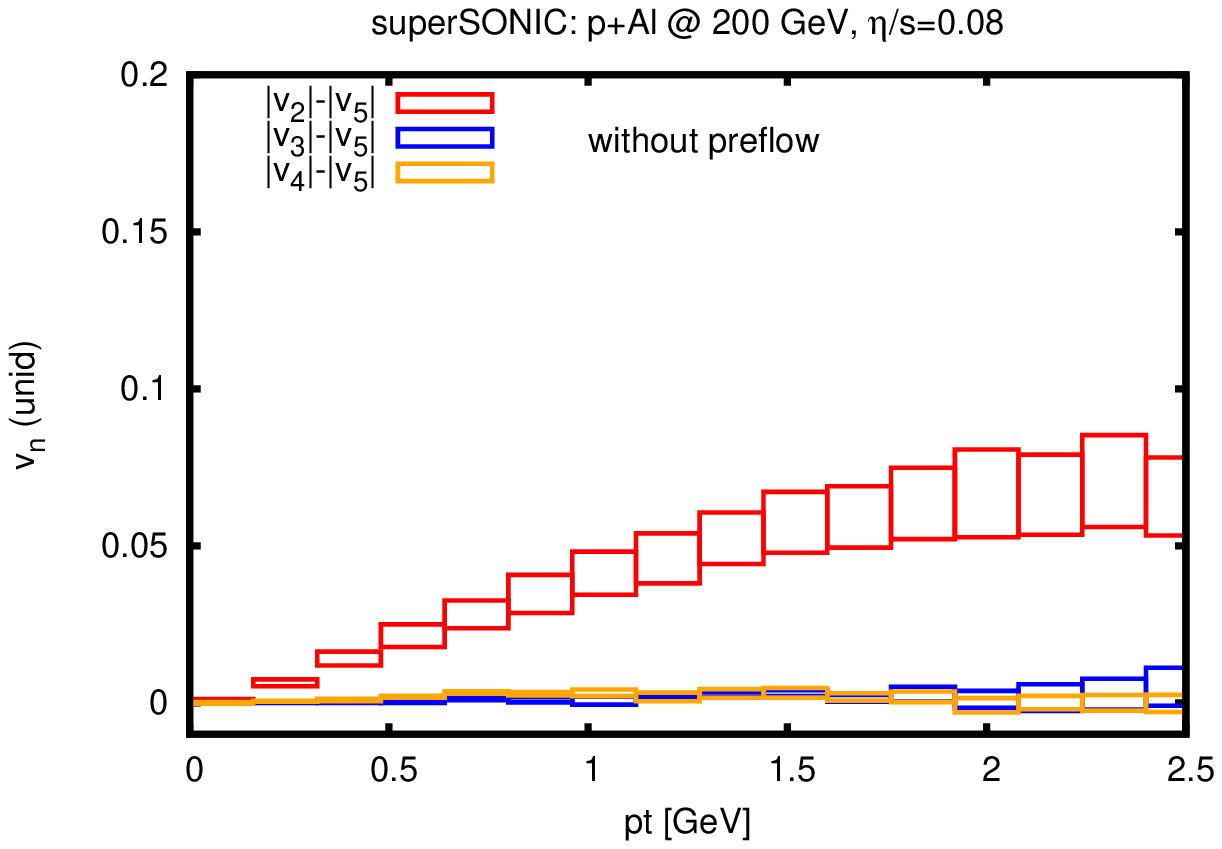}\hfill
\includegraphics[width=0.45\linewidth]{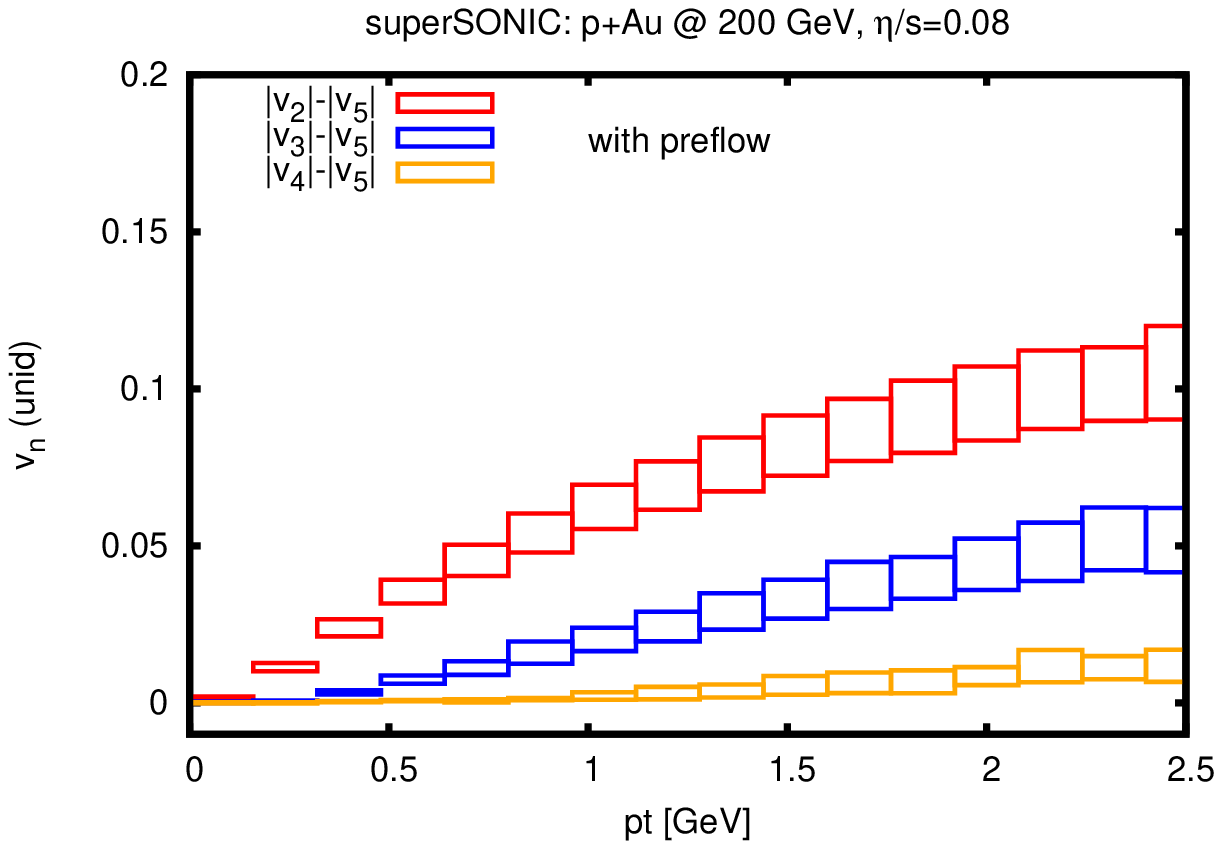}\hfill
\includegraphics[width=0.45\linewidth]{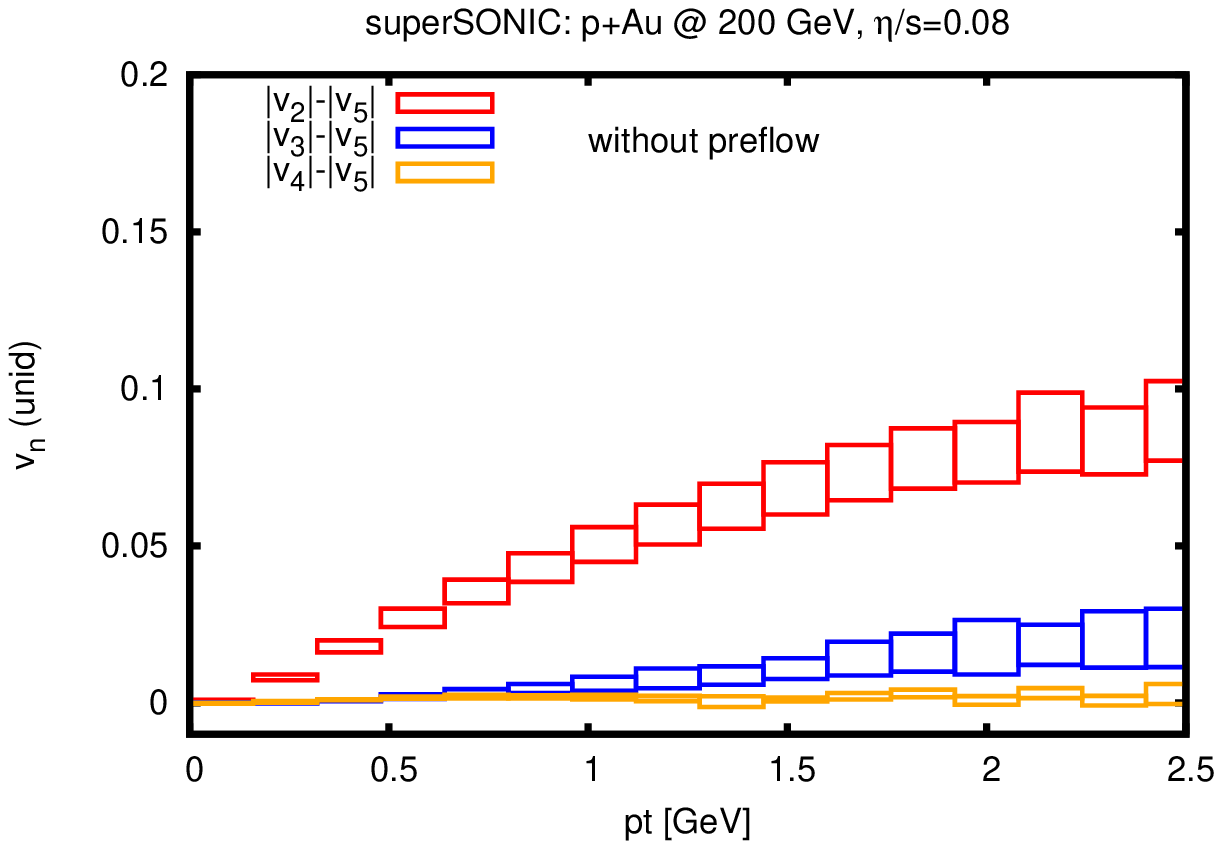}\hfill
\includegraphics[width=0.45\linewidth]{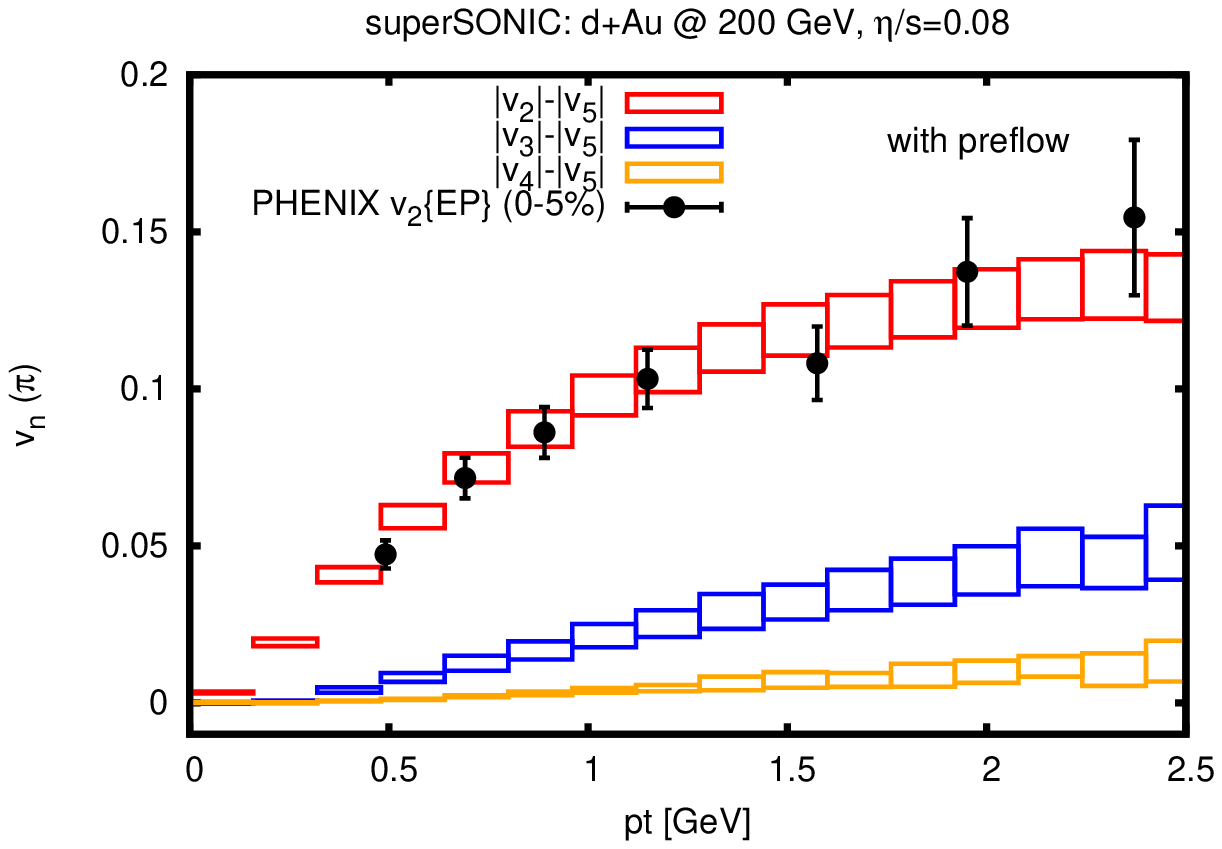}\hfill
\includegraphics[width=0.45\linewidth]{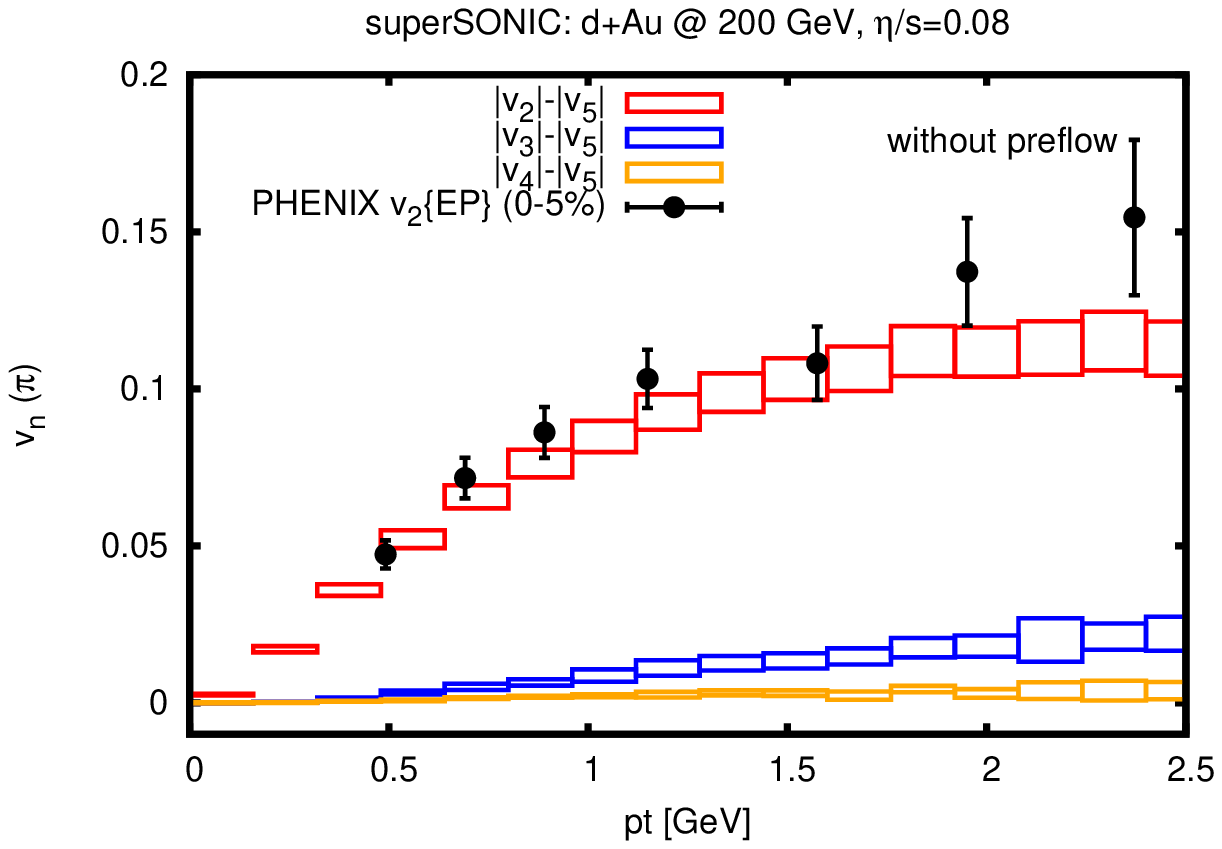}\hfill
\includegraphics[width=0.45\linewidth]{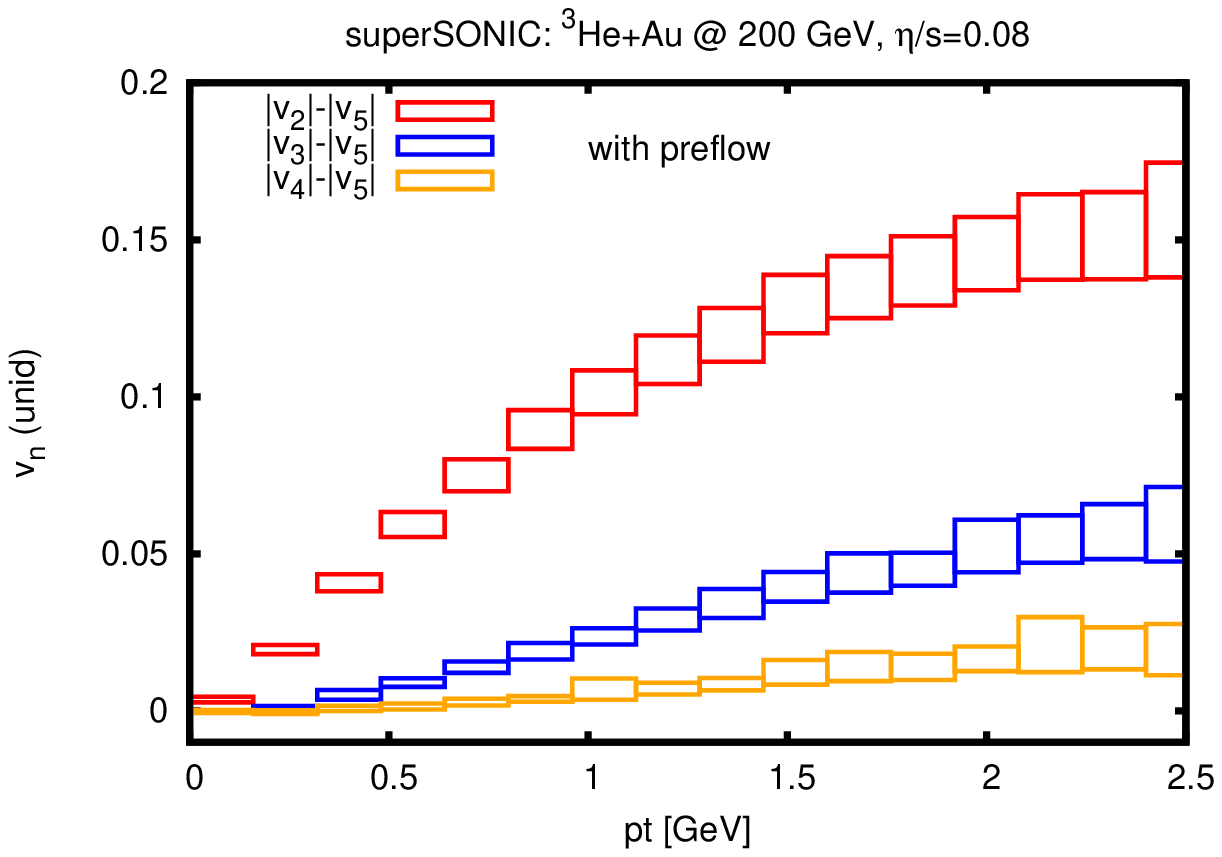}\hfill
\includegraphics[width=0.45\linewidth]{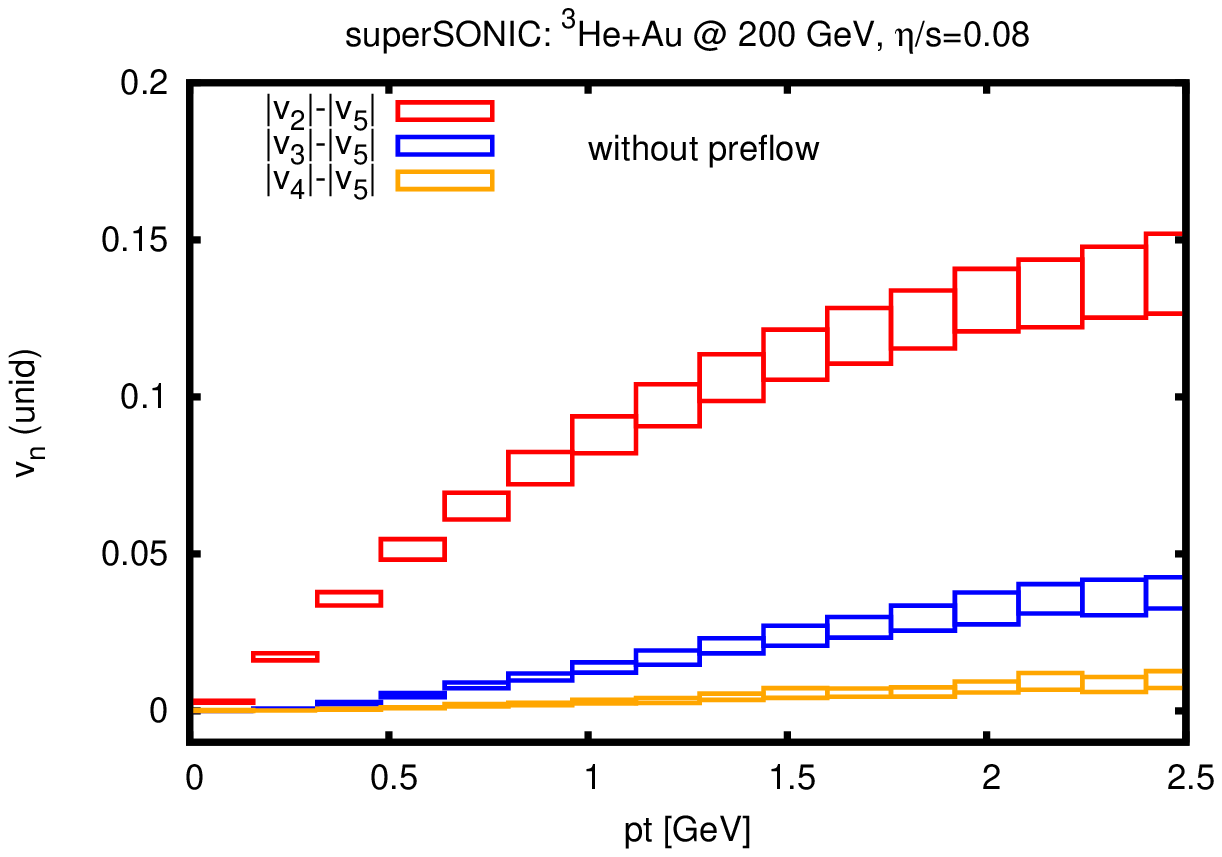}
\caption{\label{fig:all200}
Flow harmonics $v_n(p_T)-v_5(p_T)$ for identified pions ($\pi$) and unidentified charged particles (unid) for $n=2,3,4$ from superSONIC, with and without pre-equilibrium flow. Boxes indicate combined statistic and estimated systematic error for hydrodynamics (latter from varying $C_\eta=2-3$). For reference, experimental data is shown where available \cite{Adare:2014keg,Shengli}. }
\end{figure}

\begin{figure}[t]
\begin{center}
\begin{Large}
RHIC, $\sqrt{s}=62.4$ GeV
\end{Large}
\end{center}
\includegraphics[width=0.47\linewidth]{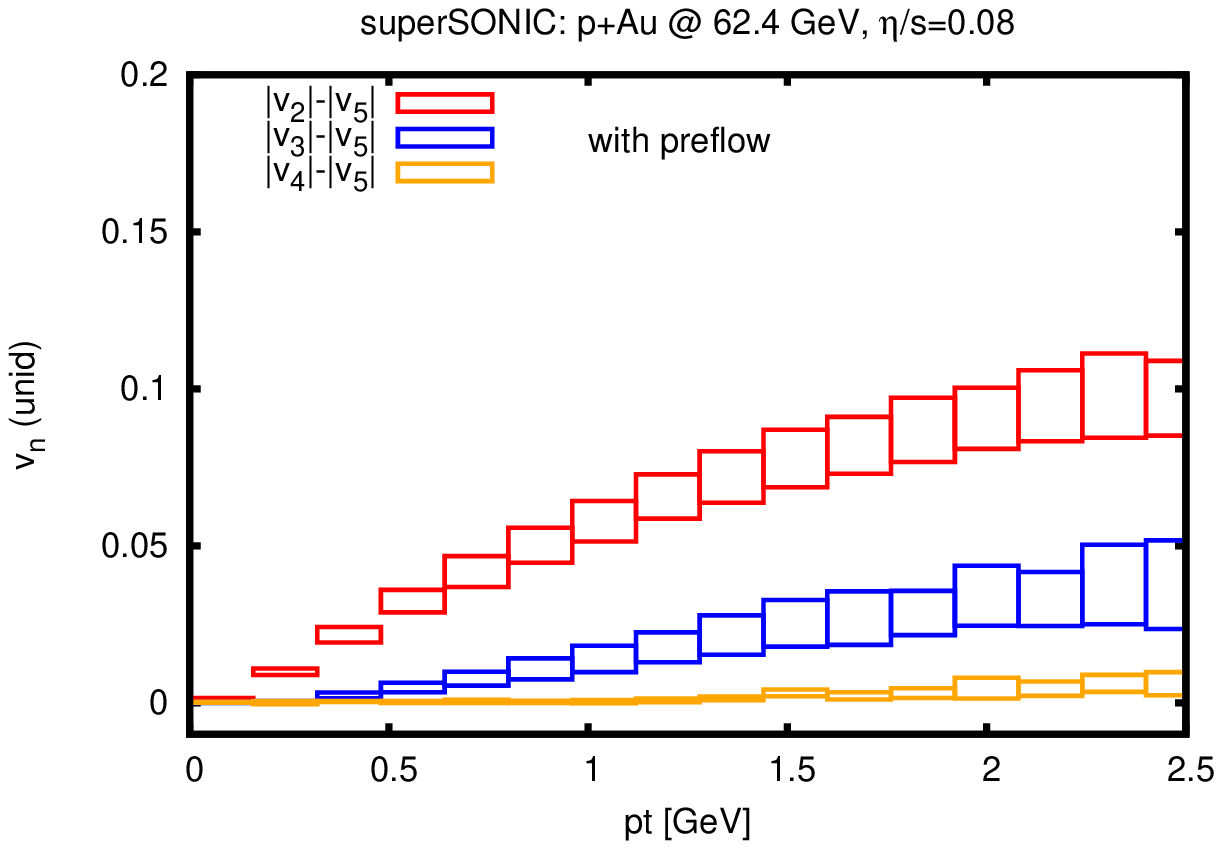}\hfill
\includegraphics[width=0.47\linewidth]{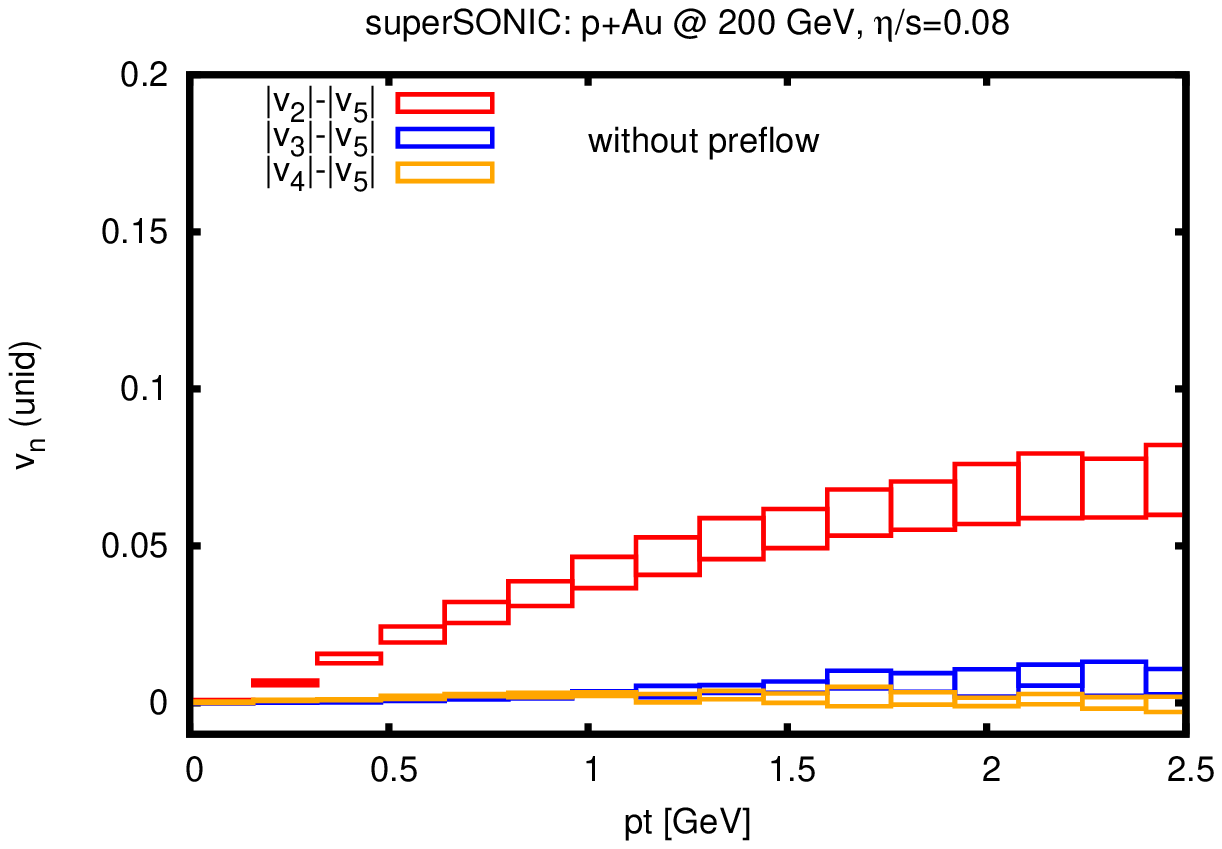}\hfill
\includegraphics[width=0.47\linewidth]{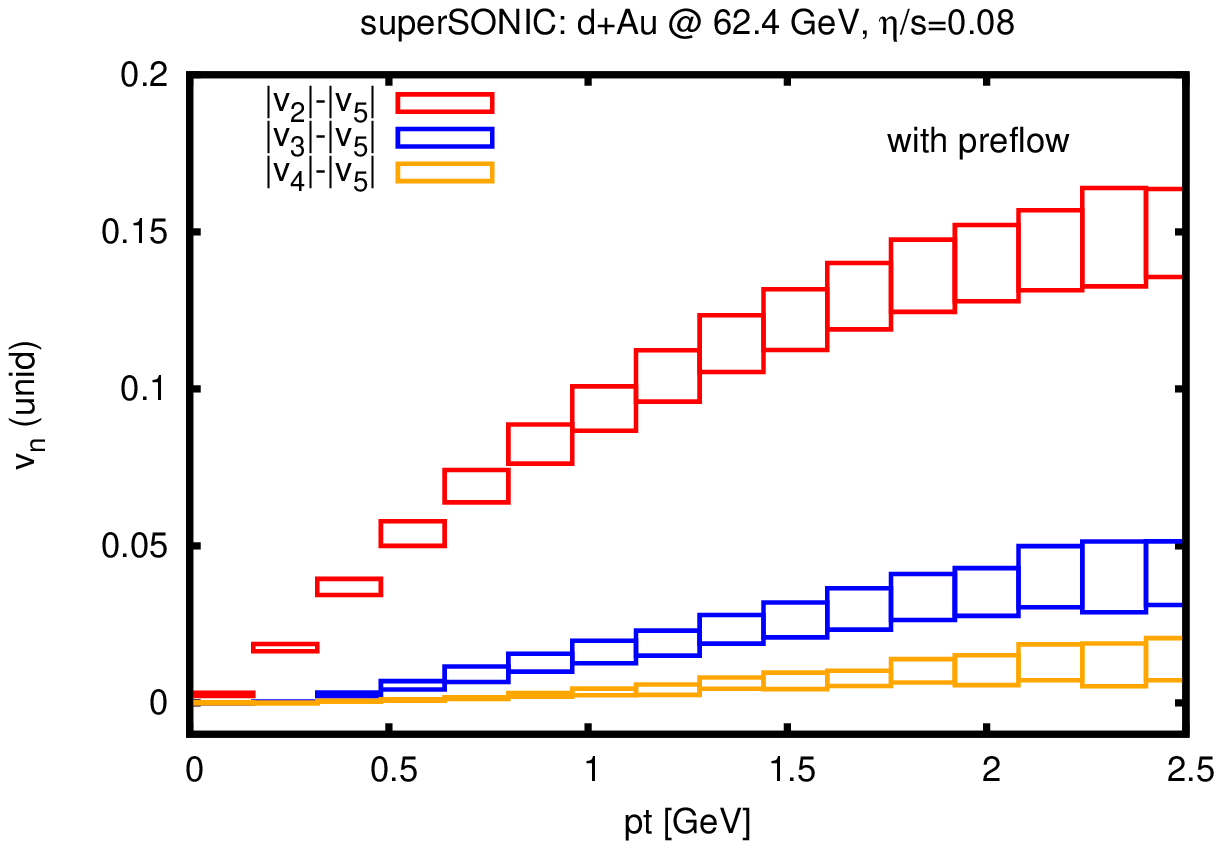}\hfill
\includegraphics[width=0.47\linewidth]{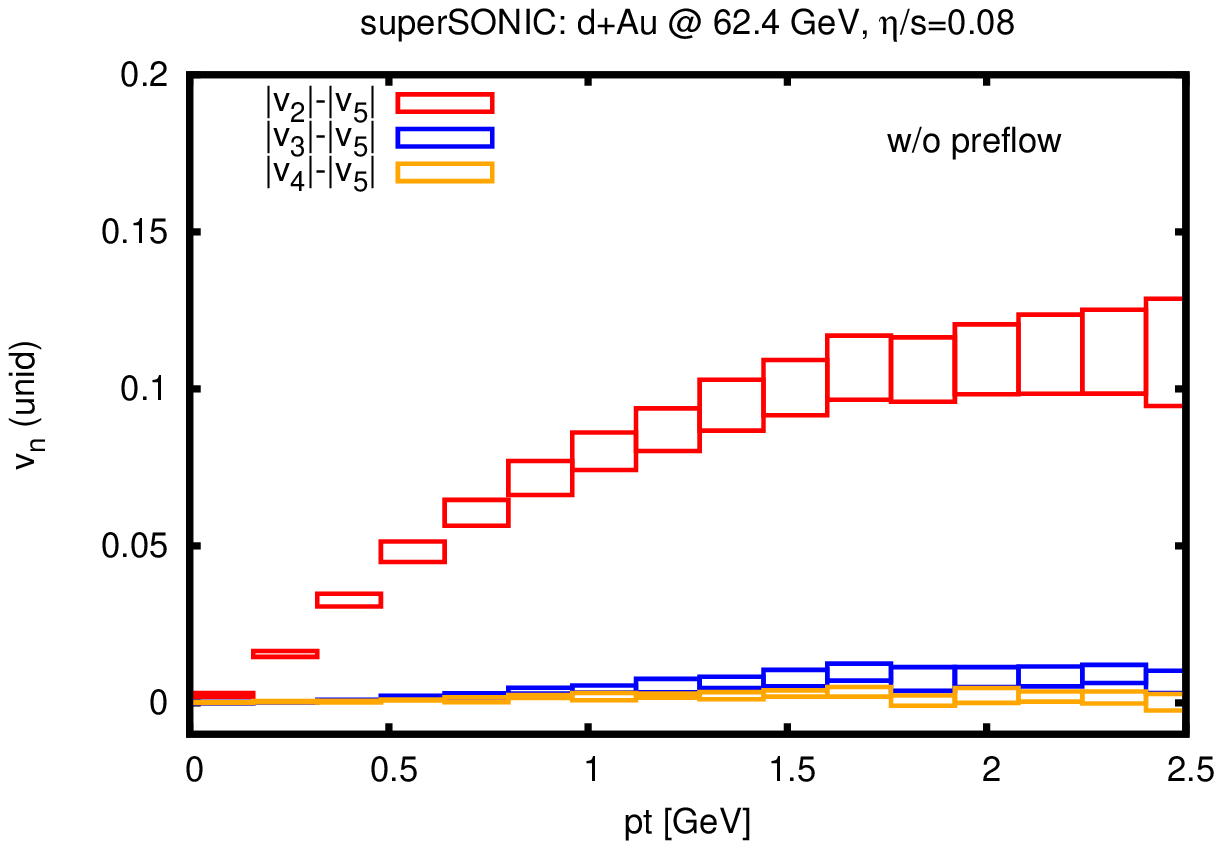}\hfill
\includegraphics[width=0.47\linewidth]{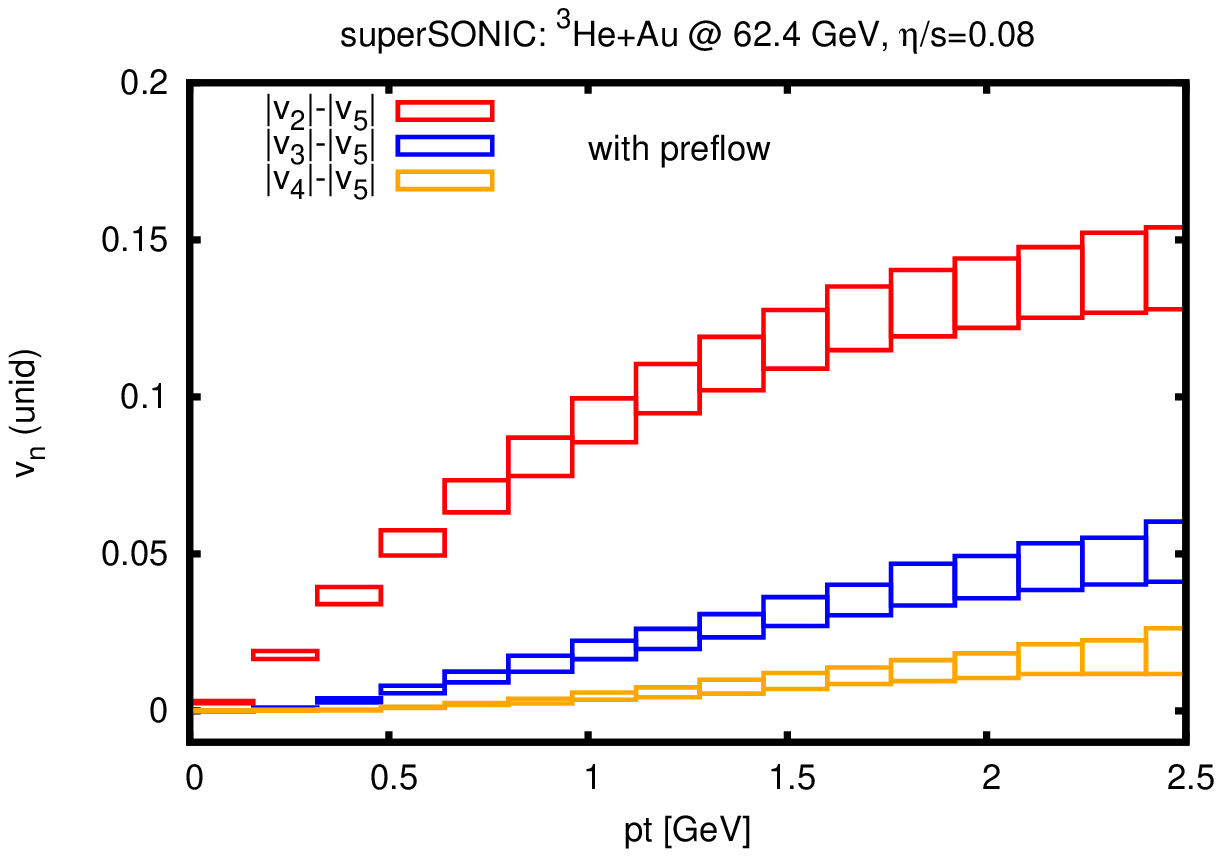}\hfill
\includegraphics[width=0.47\linewidth]{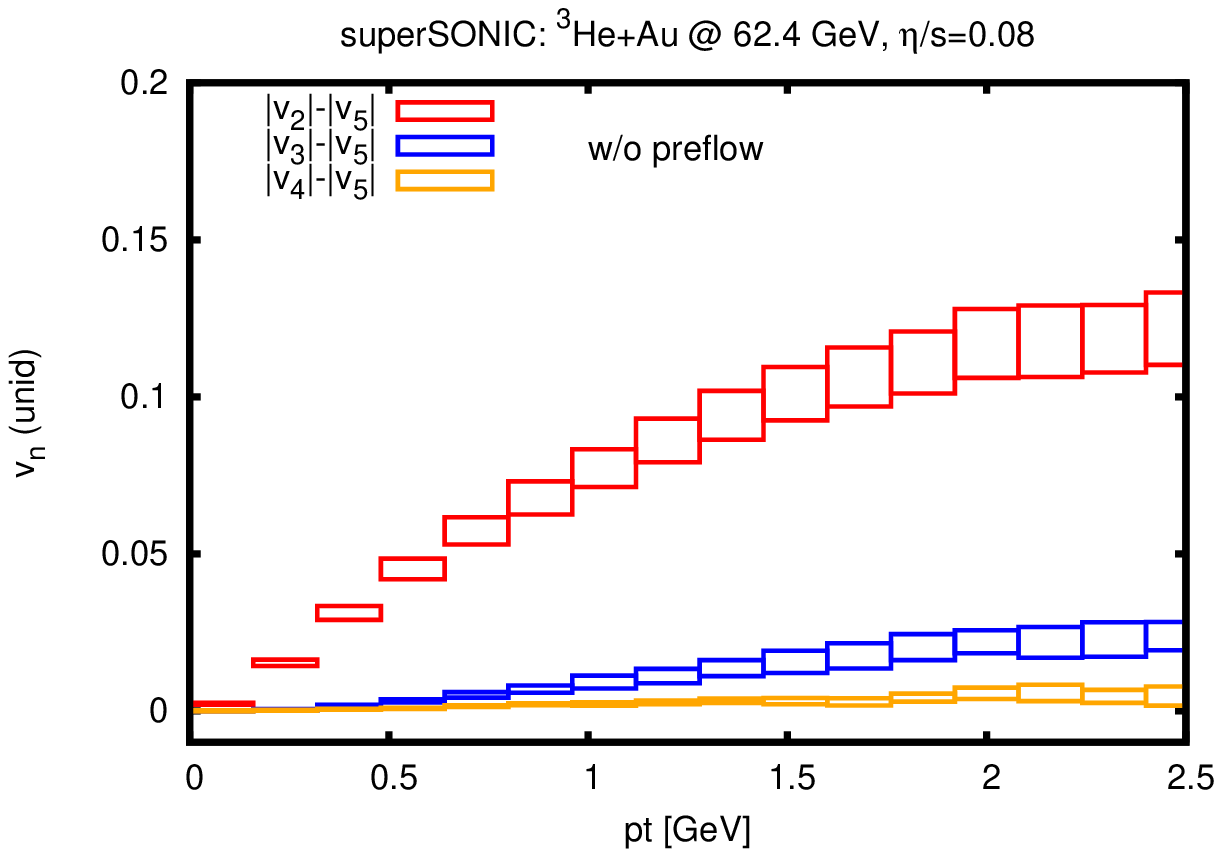}
\caption{\label{fig:all62}
Flow harmonics $v_n(p_T)-v_5(p_T)$ for identified pions ($\pi$) and unidentified charged particles (unid) for $n=2,3,4$ from superSONIC, with and without pre-equilibrium flow. Boxes indicate combined statistic and estimated systematic error for hydrodynamics (latter from varying $C_\eta=2-3$).
}
\end{figure}

\begin{figure}[t]
\begin{center}
\begin{Large}
Proton-Nucleus collisions, $\sqrt{s}$=7.7-5000 GeV
\end{Large}
\end{center}
\includegraphics[width=0.47\linewidth]{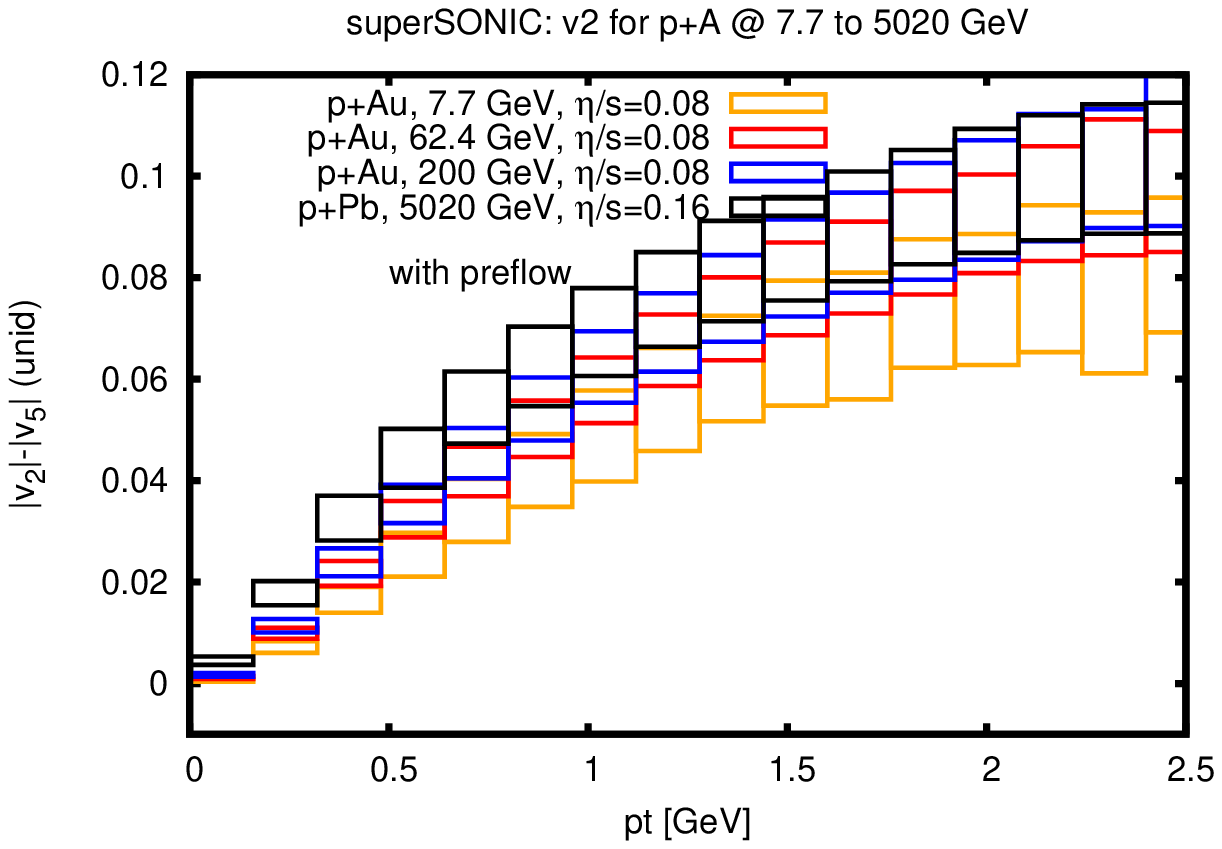}\hfill
\includegraphics[width=0.47\linewidth]{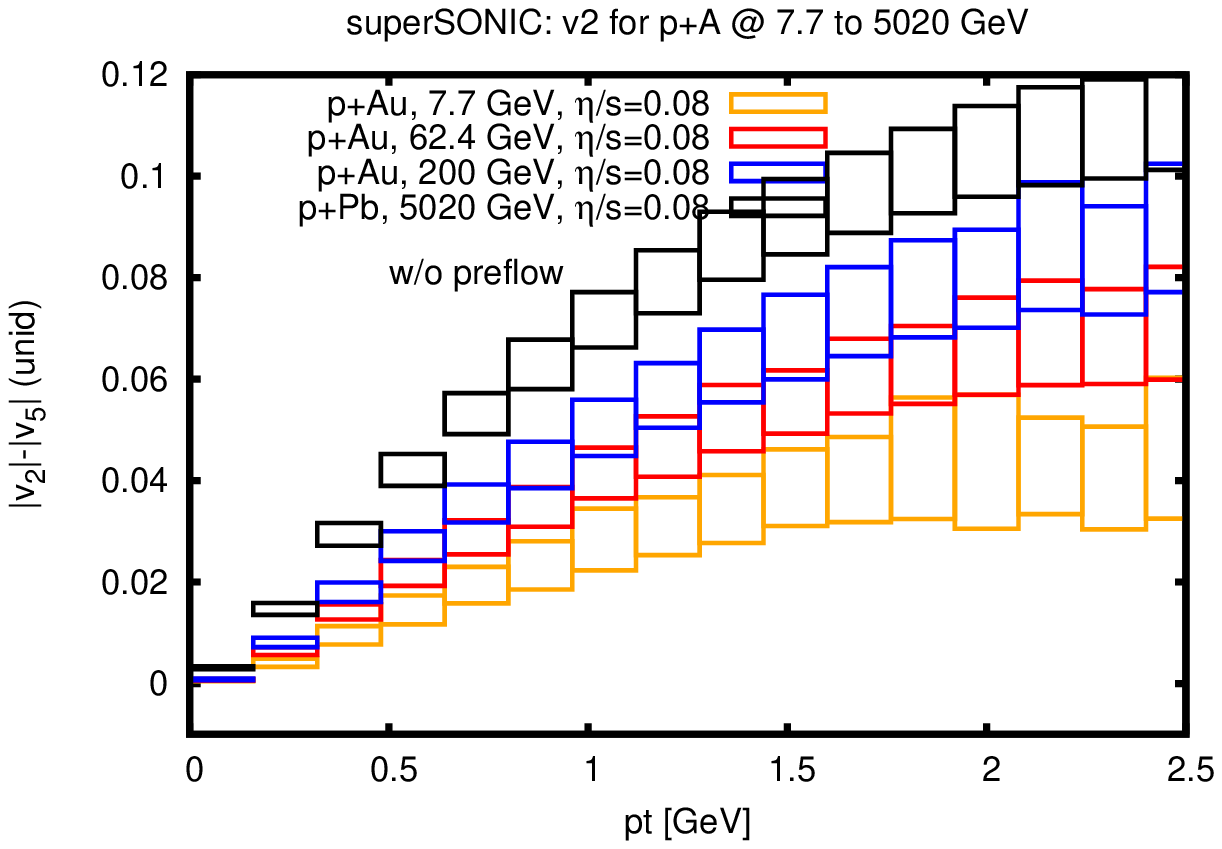}\hfill
\includegraphics[width=0.47\linewidth]{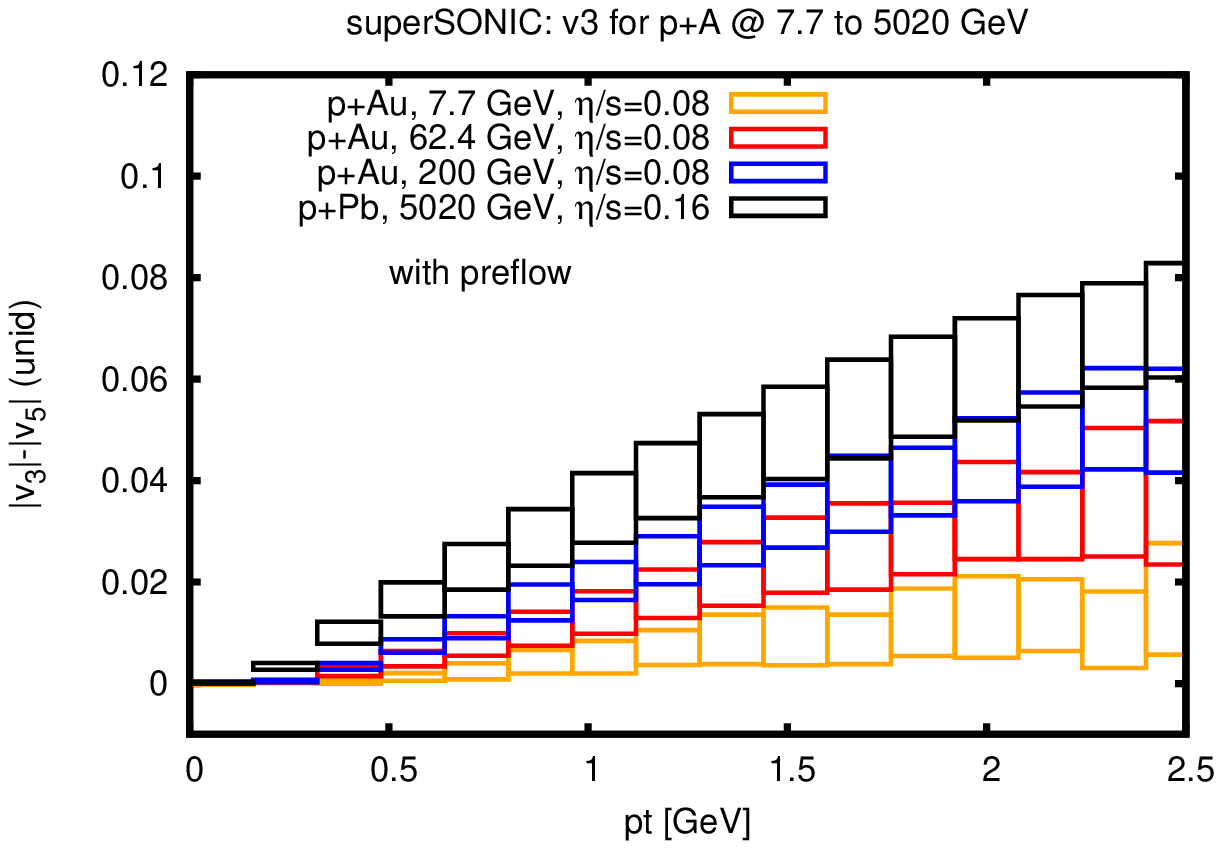}\hfill
\includegraphics[width=0.47\linewidth]{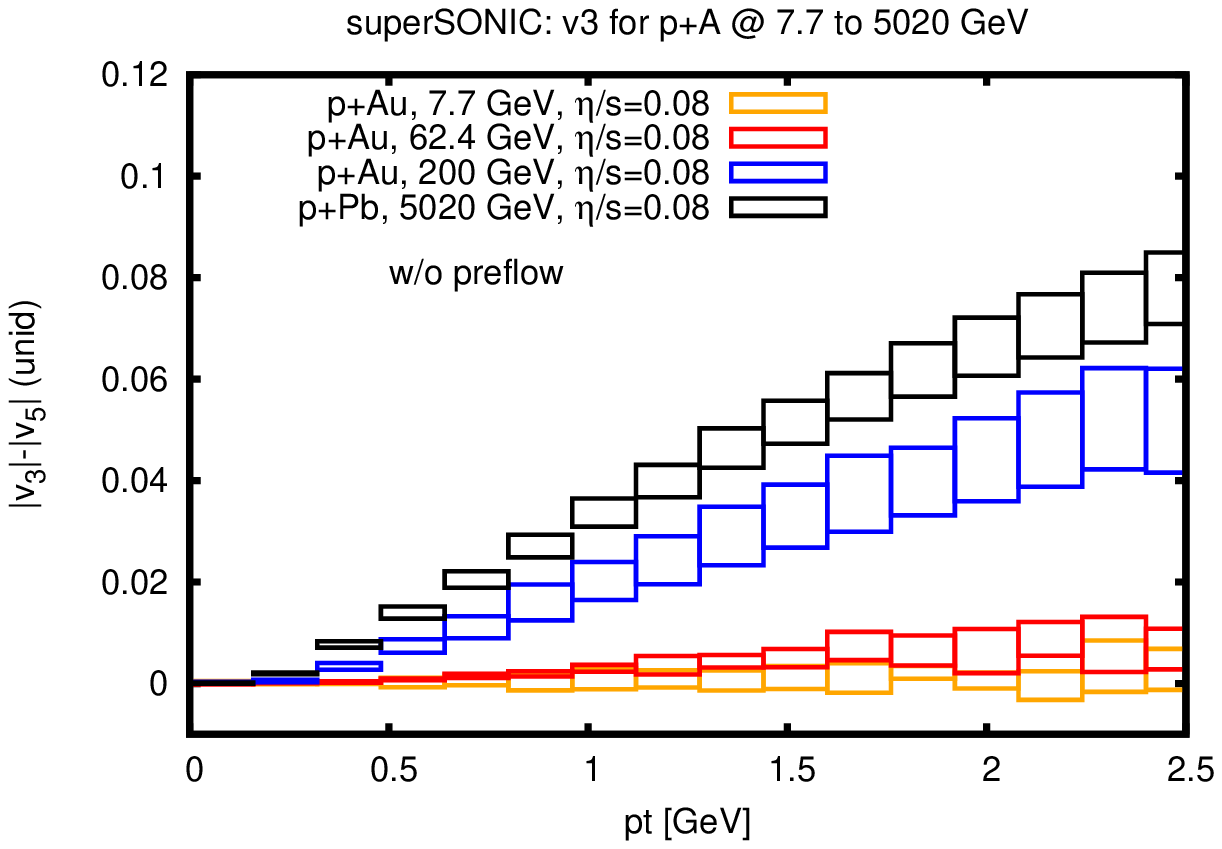}\hfill
\includegraphics[width=0.47\linewidth]{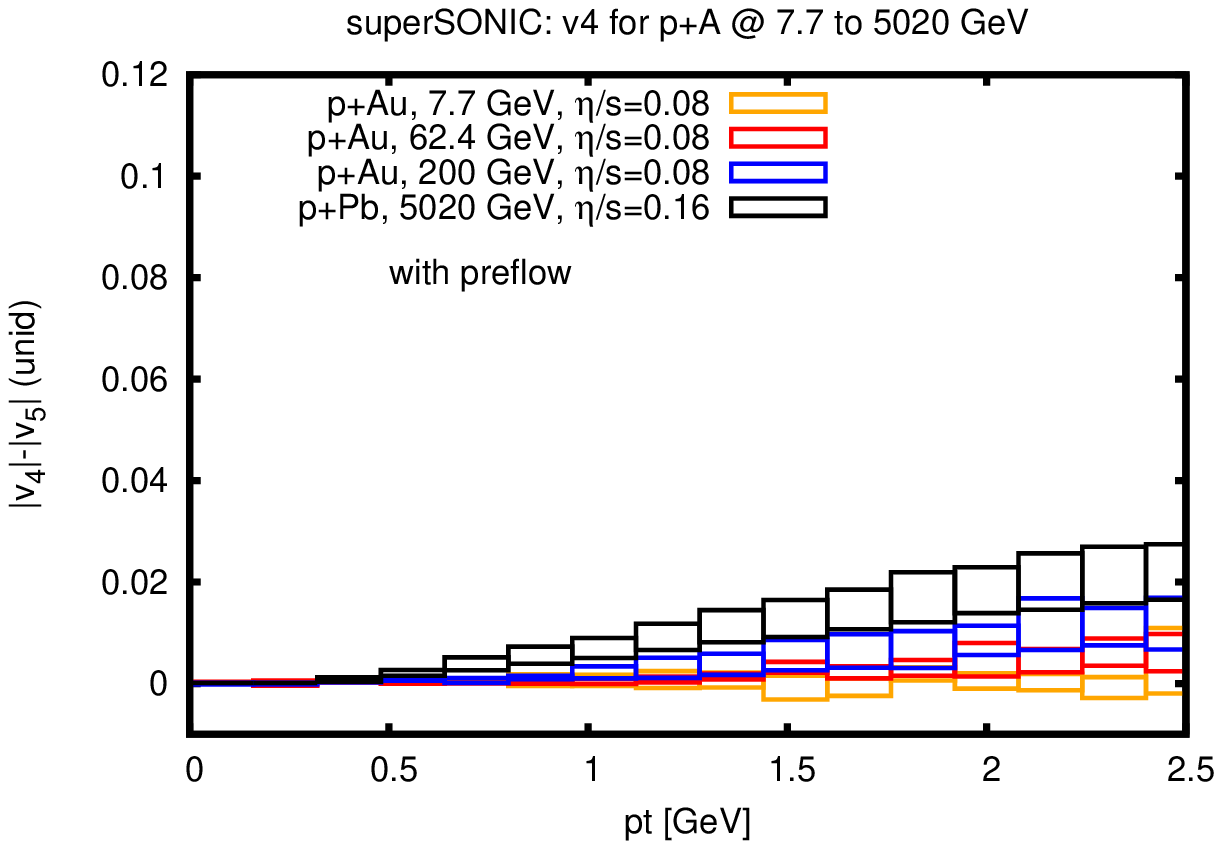}\hfill
\includegraphics[width=0.47\linewidth]{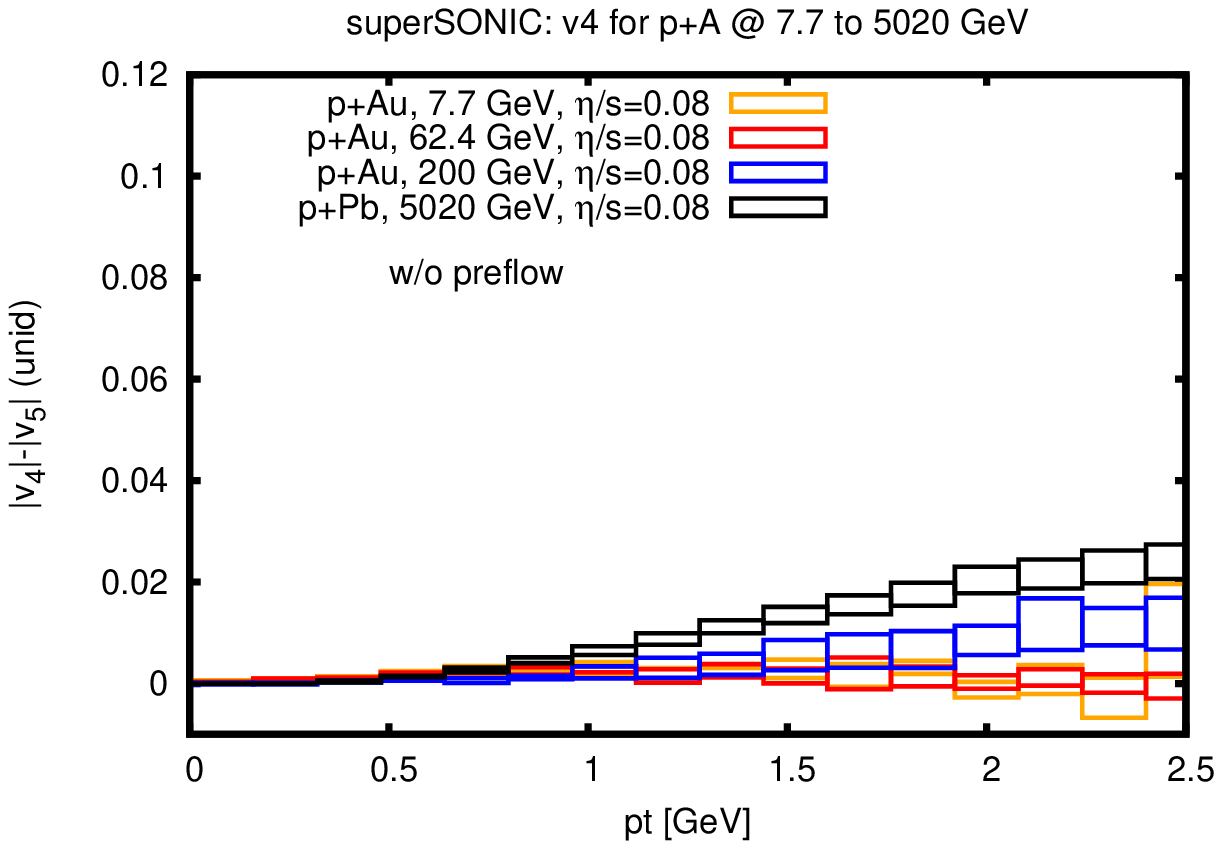}
\caption{\label{fig:allproton} 
Flow harmonics $v_n(p_T)-v_5(p_T)$ for unidentified charged particles (unid) for $n=2,3,4$ from superSONIC in proton-nucleus collisions, with and without pre-equilibrium flow. Boxes indicate combined statistic and estimated systematic error for hydrodynamics (latter from varying $C_\eta=2-3$). Y-axis is same scale on all plots. The \pPb system, which has the highest multiplicity (see. Tab.~\ref{tab:one}), shows non-vanishing flow components up to $n=4$, but as the multiplicity is decreased, first $v_4$, then $v_3$ and eventually also $v_2$ start to decrease and (in the case of $v_4$) eventually become consistent with zero.
}
\end{figure}

\begin{figure}[t]
\includegraphics[width=0.47\linewidth]{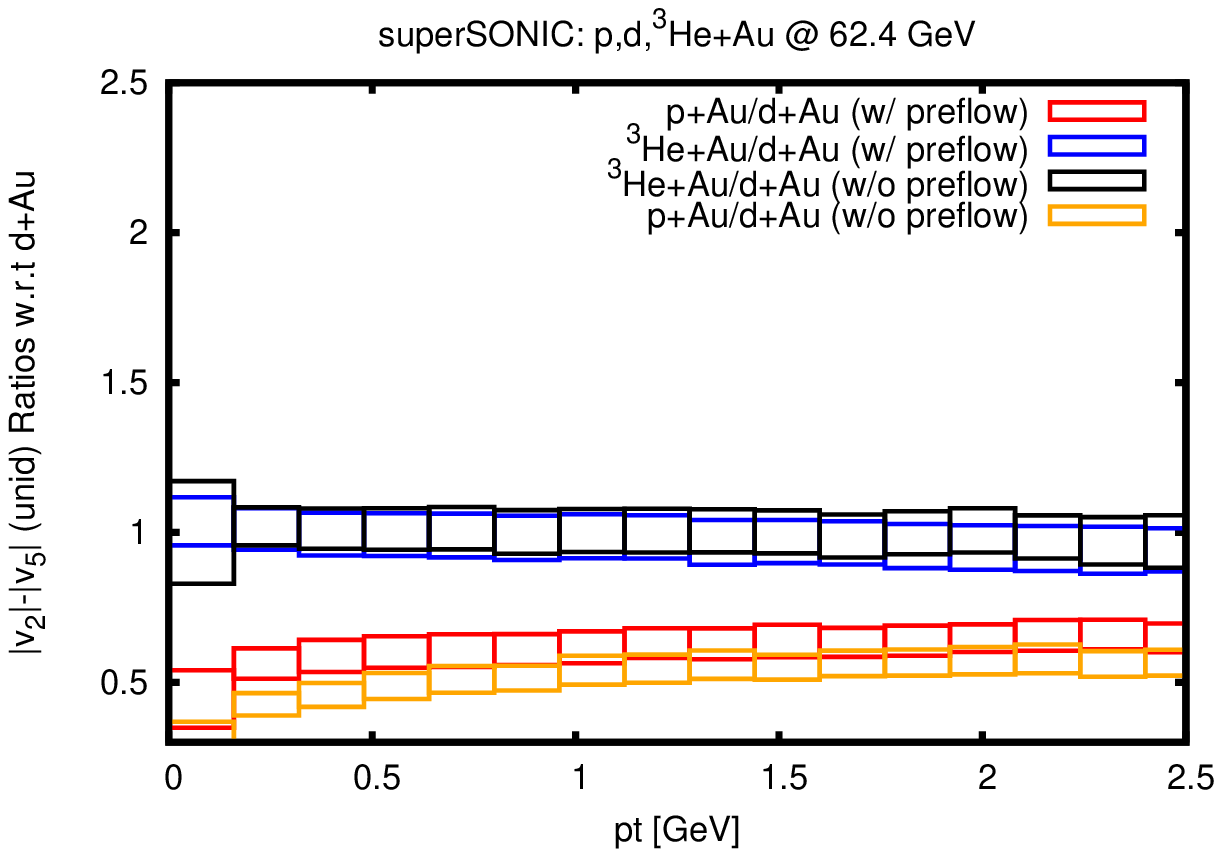}\hfill
\includegraphics[width=0.47\linewidth]{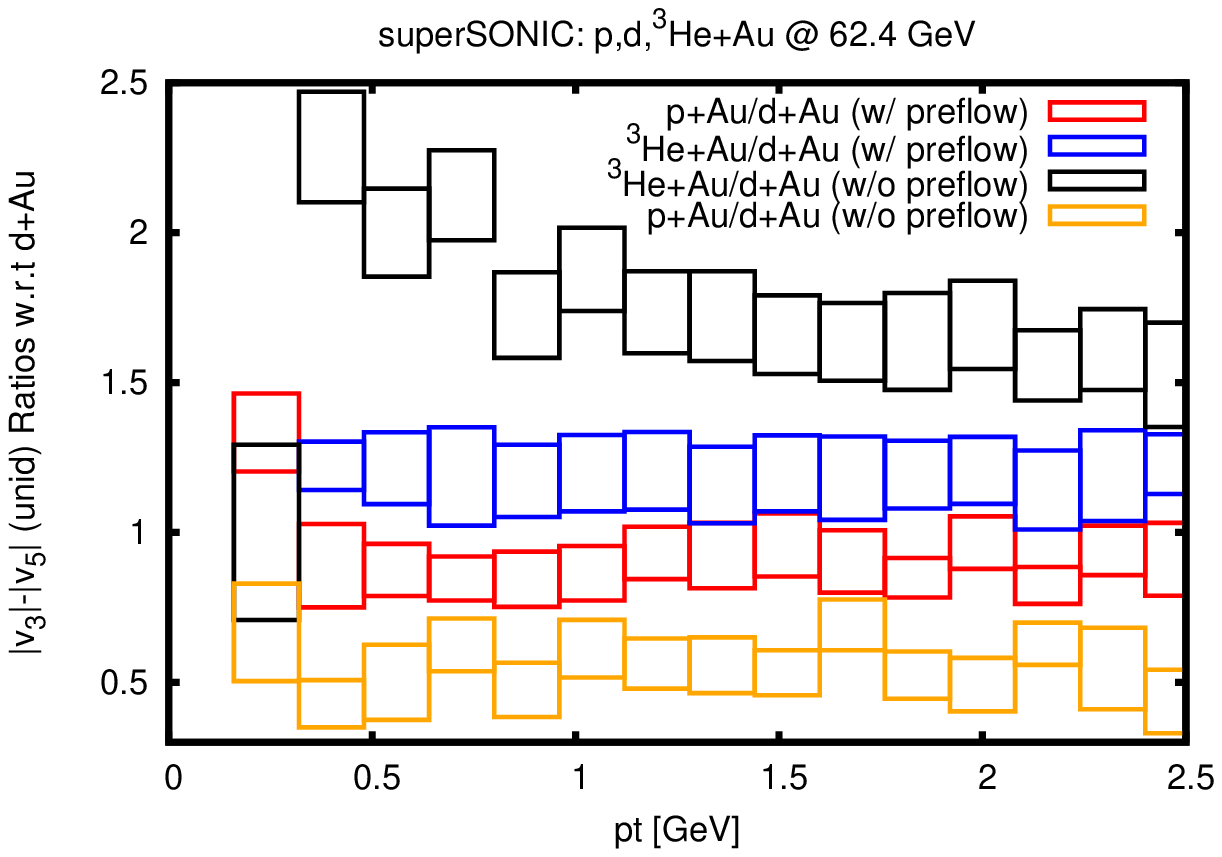}\hfill
\includegraphics[width=0.47\linewidth]{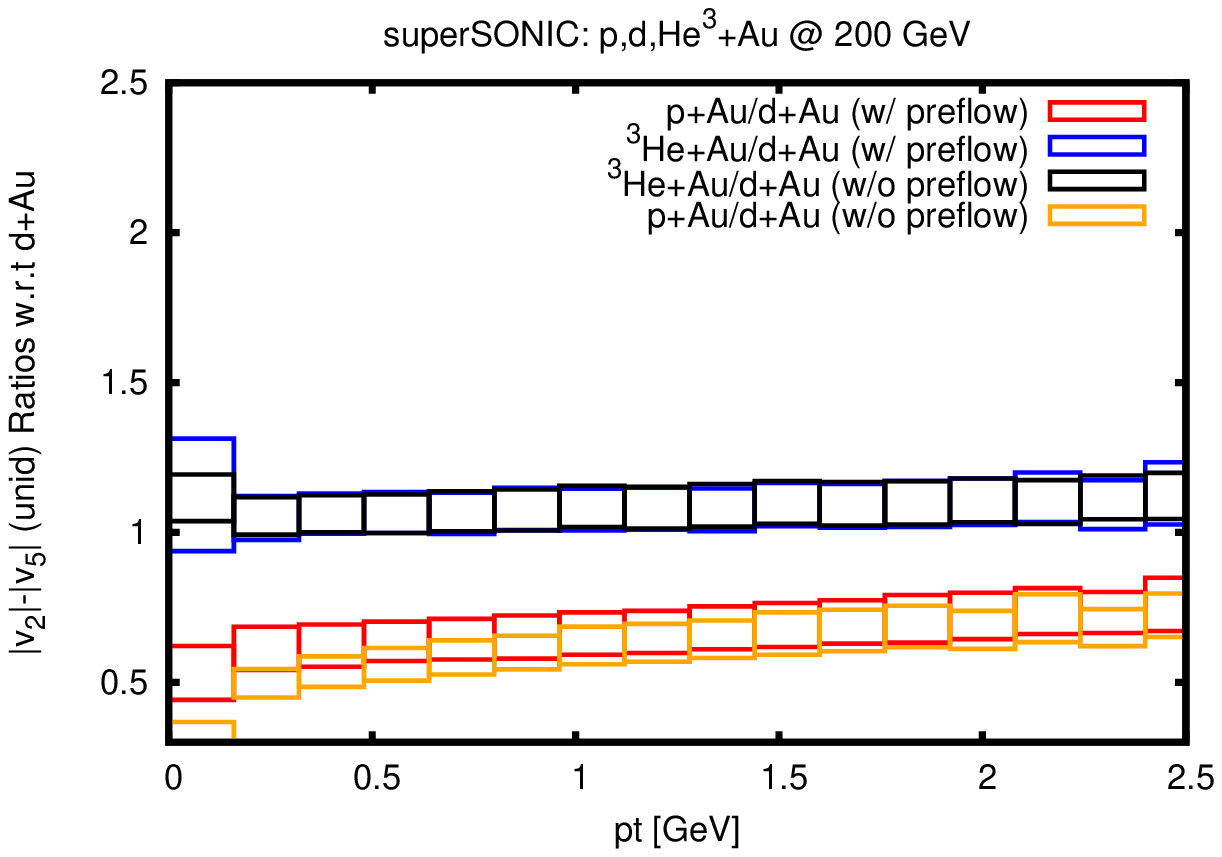}\hfill
\includegraphics[width=0.47\linewidth]{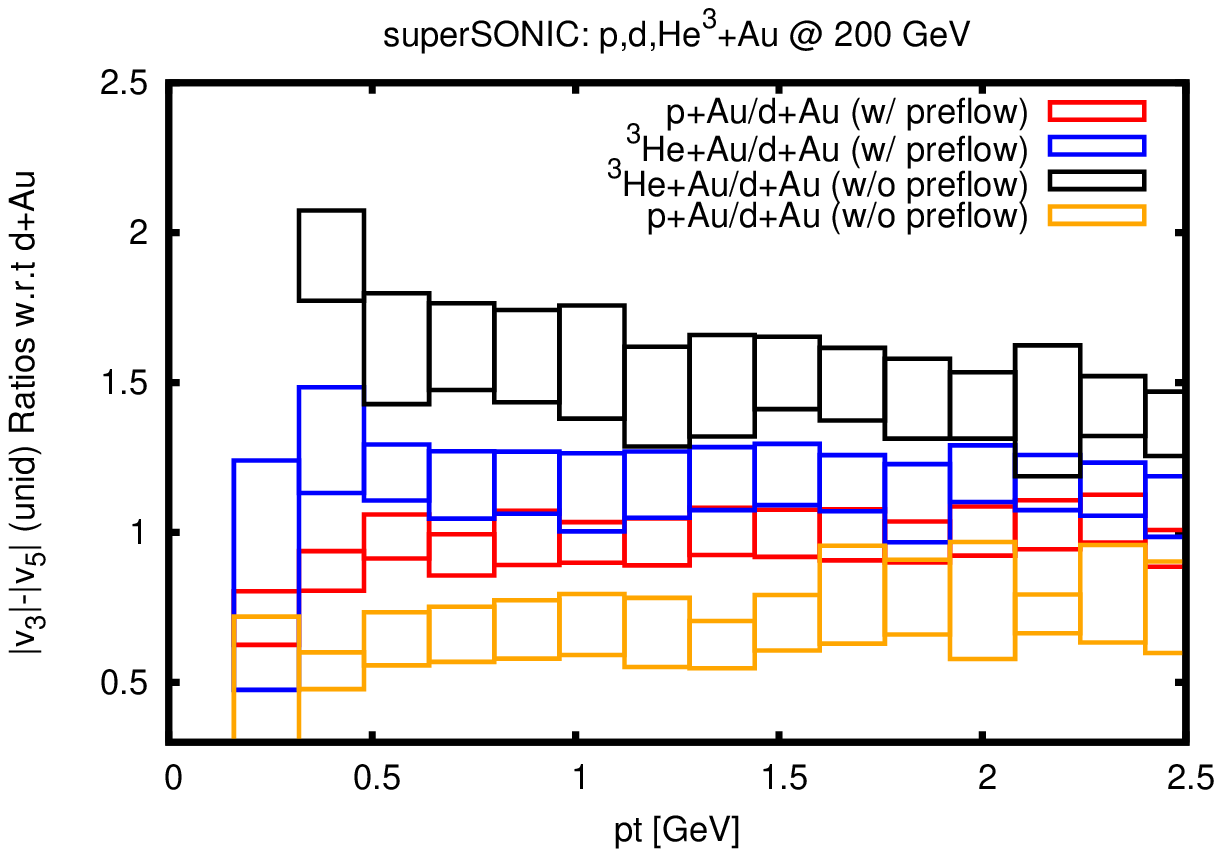}\hfill
\includegraphics[width=0.47\linewidth]{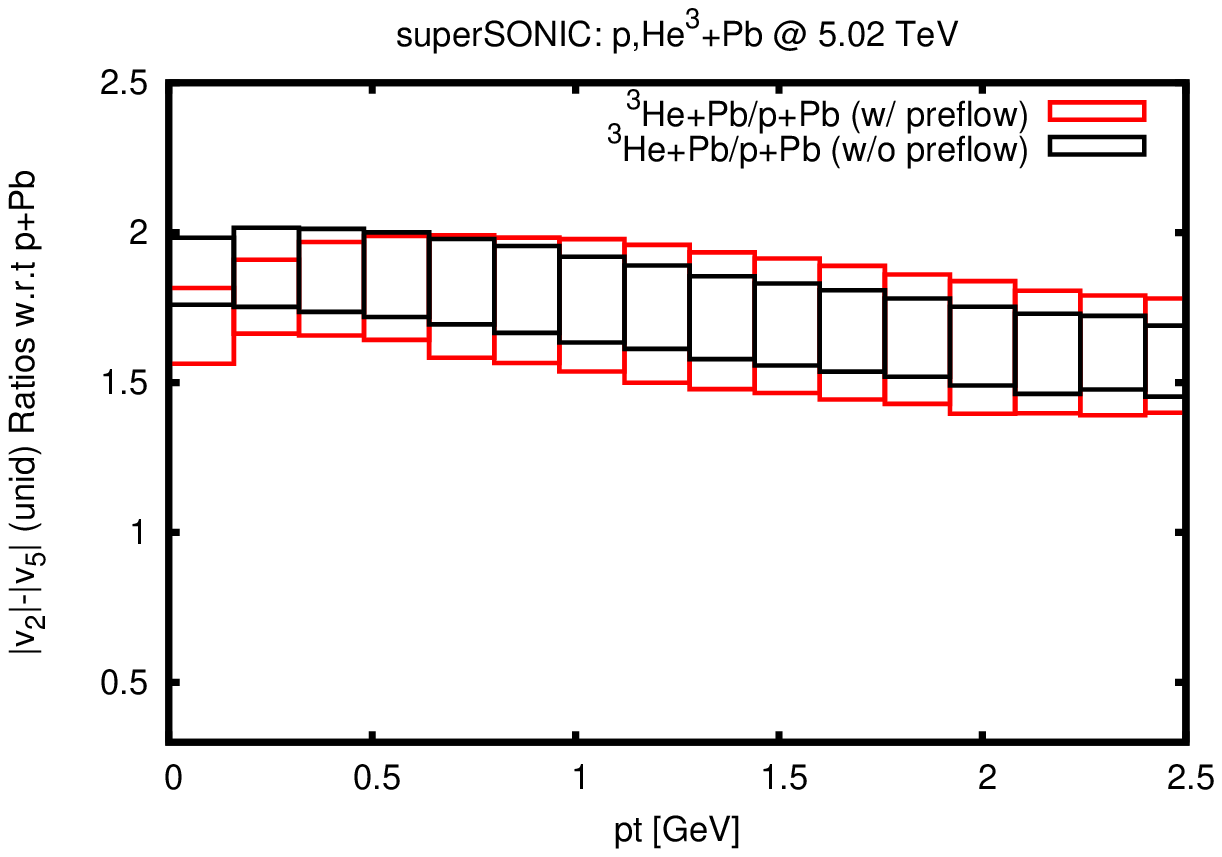}\hfill
\includegraphics[width=0.47\linewidth]{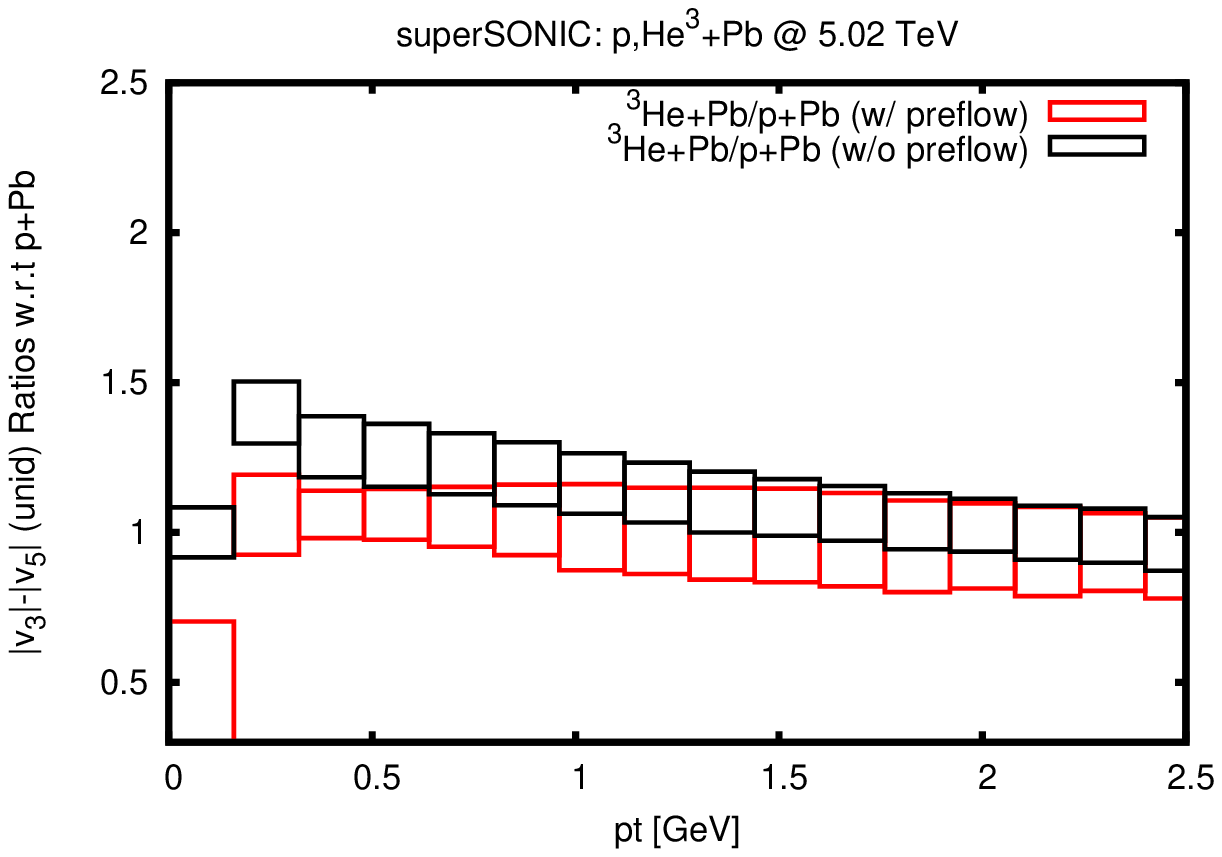}\hfill
\caption{\label{fig:ratios} Ratios of $v_2,v_3$ between \hAu,\pAu and \dAu (and \hPb and \pPb), with and without pre-equilibrium flow (``preflow'').  Boxes indicate combined statistic and estimated systematic error for hydrodynamics (latter from changing $C_\eta =2-3$). For $\sqrt{s}\leq 200$ GeV, the $v_3$ ratios between \hAu and \pAu to \dAu are rather sensitive to the presence of preflow. For $\sqrt{s}=5.02$ TeV, this difference is much smaller since the evolution is less sensitive to the presence of pre-equilibrium flow.}
\end{figure}

\begin{figure}[t]
\begin{center}
\begin{Large}
LHC, $\sqrt{s}=5.02$ TeV
\end{Large}
\end{center}
\includegraphics[width=0.45\linewidth]{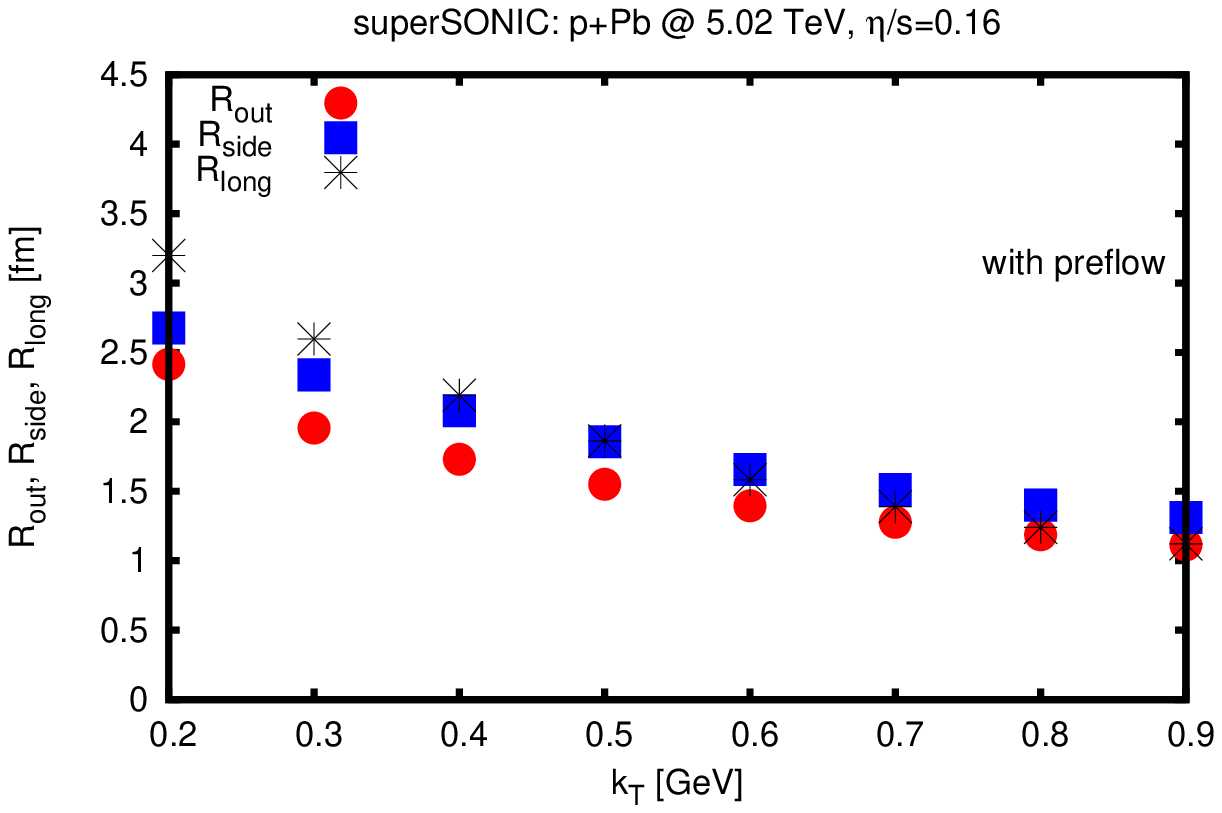}\hfill
\includegraphics[width=0.45\linewidth]{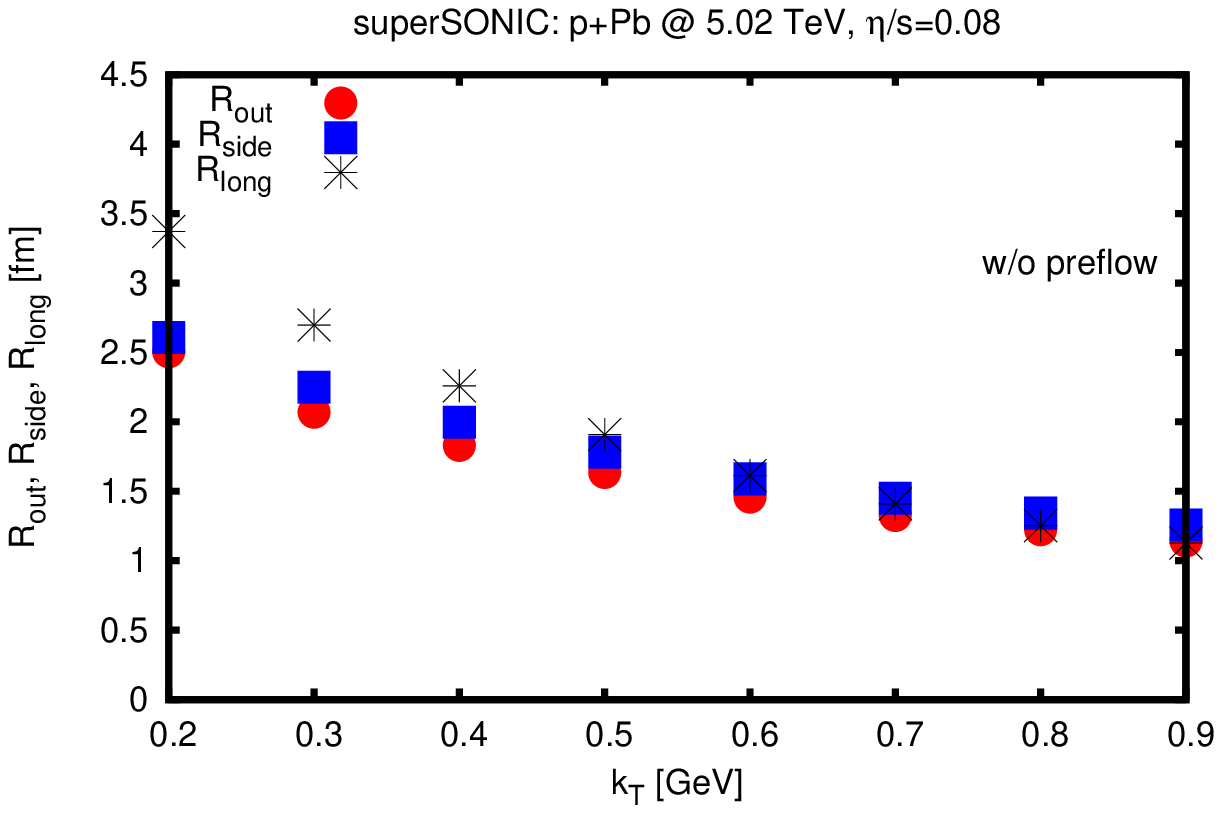}\hfill
\includegraphics[width=0.45\linewidth]{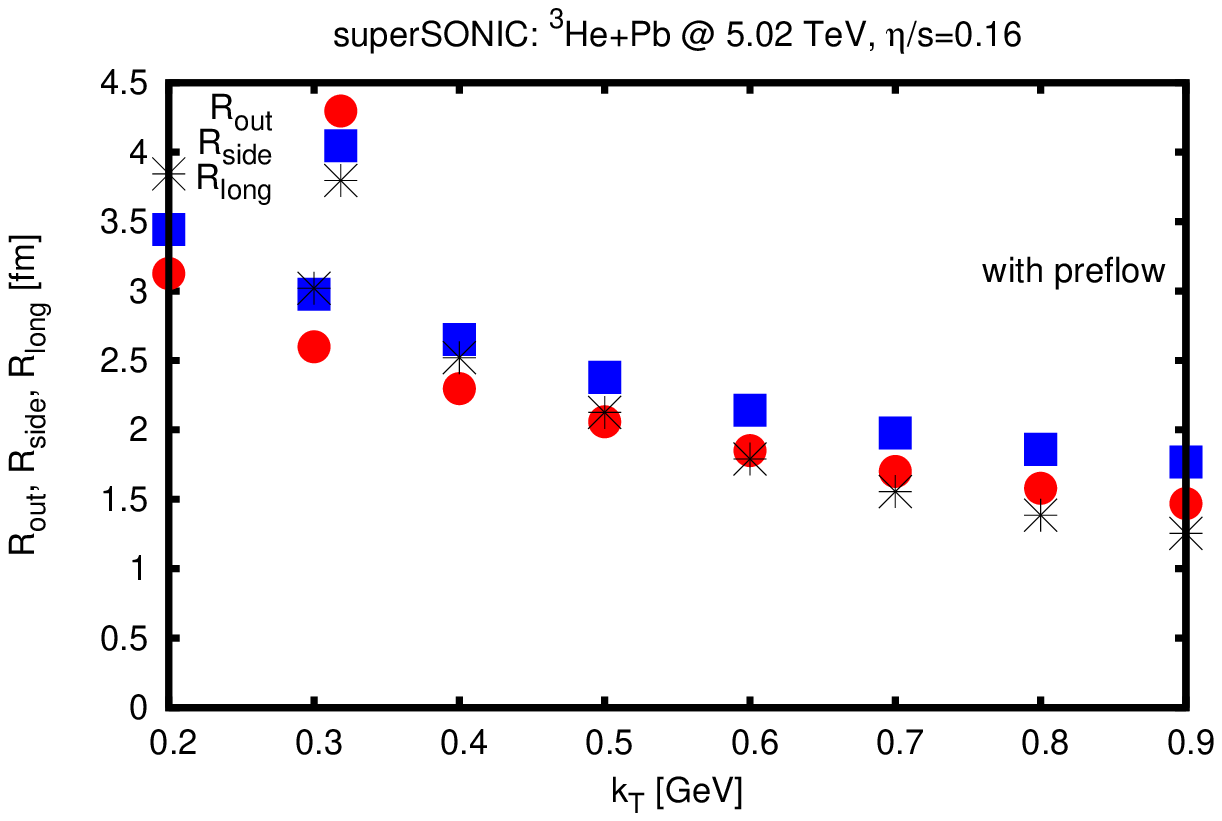}\hfill
\includegraphics[width=0.45\linewidth]{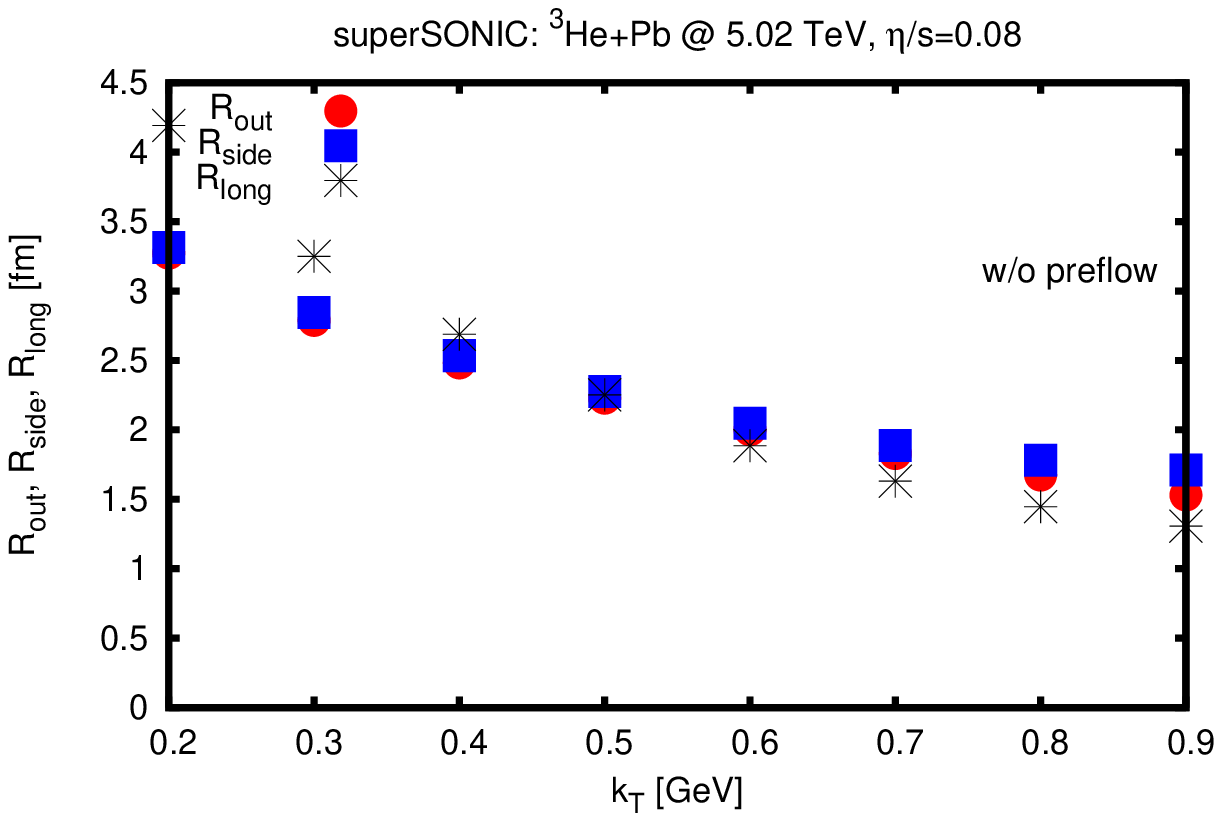}
\caption{\label{fig:all5000hbt}
HBT Radii for pions for \pPb and \hPb collisions at $\sqrt{s}=5.02$ TeV. Varying $C_\eta=2-3$ in hydrodynamics results in changes smaller than the symbol size shown.}
\end{figure}

\begin{figure}[t]
\begin{center}
\begin{Large}
RHIC, $\sqrt{s}=200$ GeV
\end{Large}
\end{center}
\includegraphics[width=0.45\linewidth]{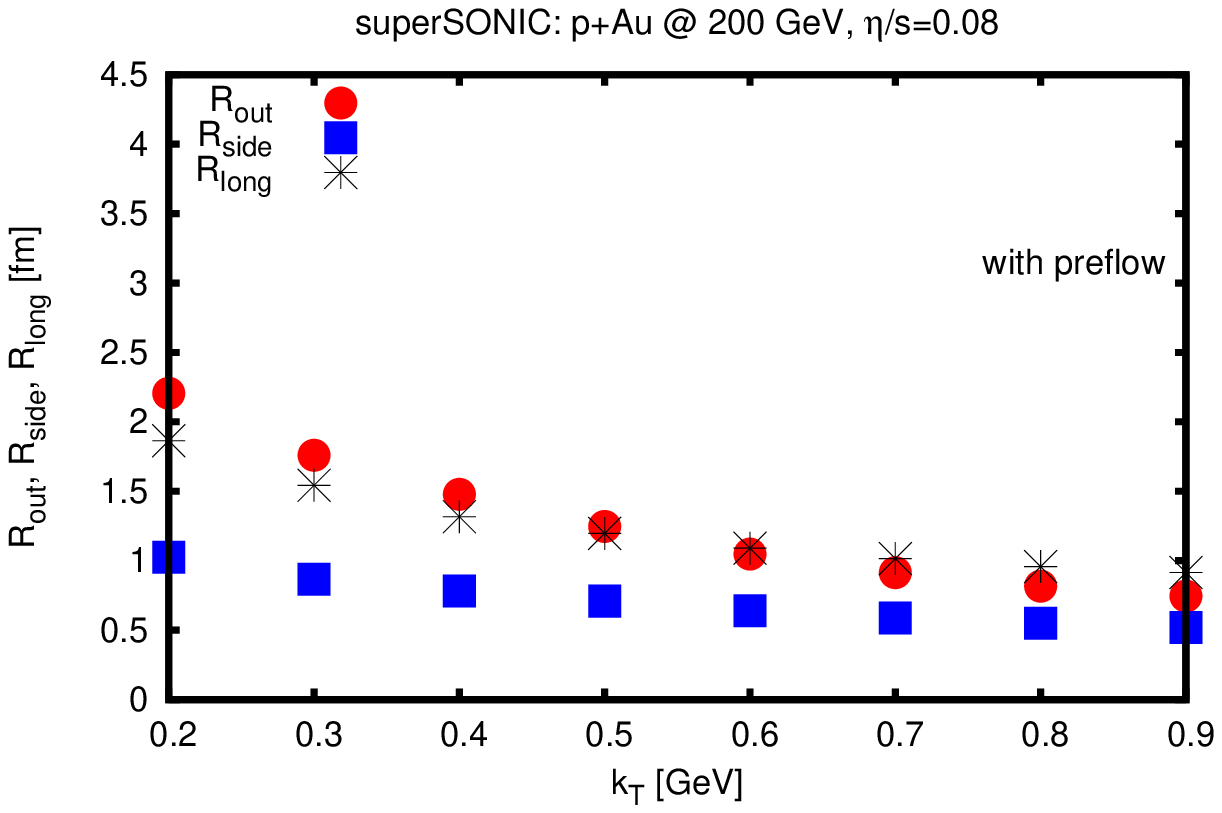}\hfill
\includegraphics[width=0.45\linewidth]{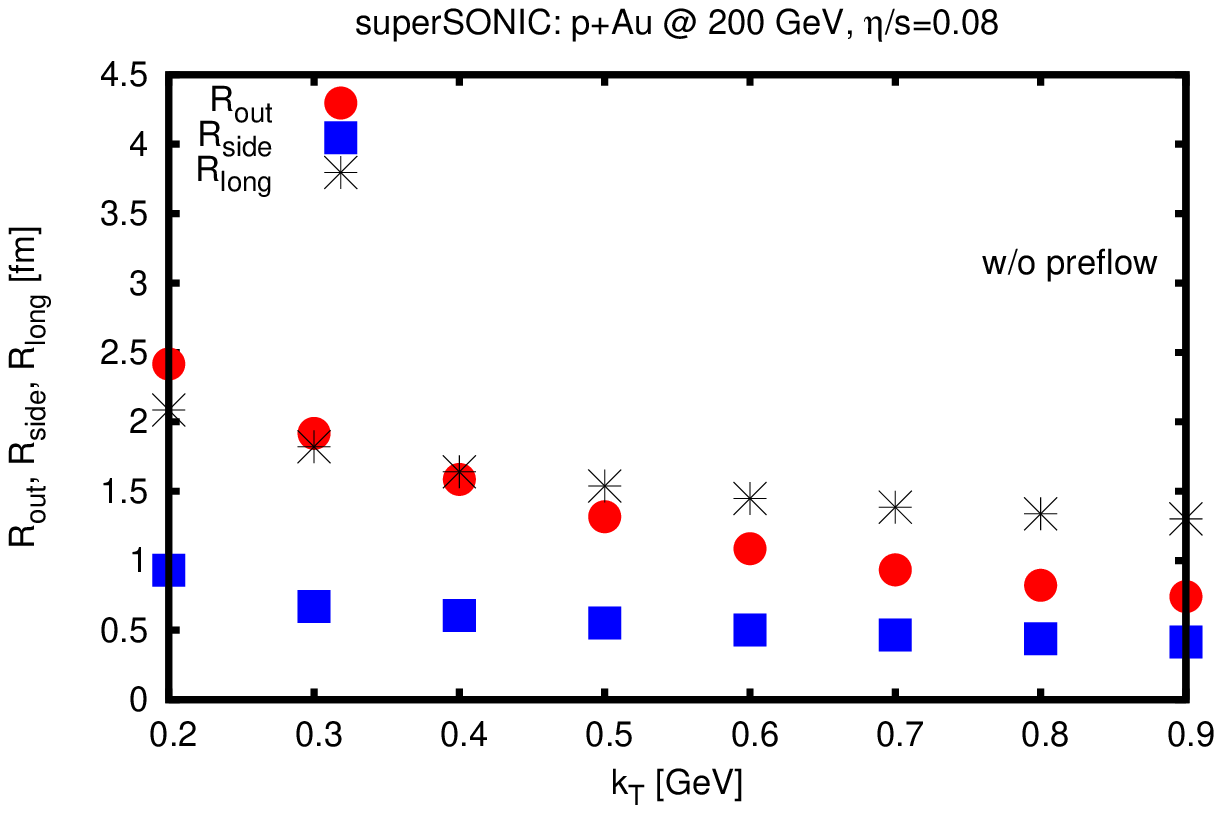}\hfill
\includegraphics[width=0.45\linewidth]{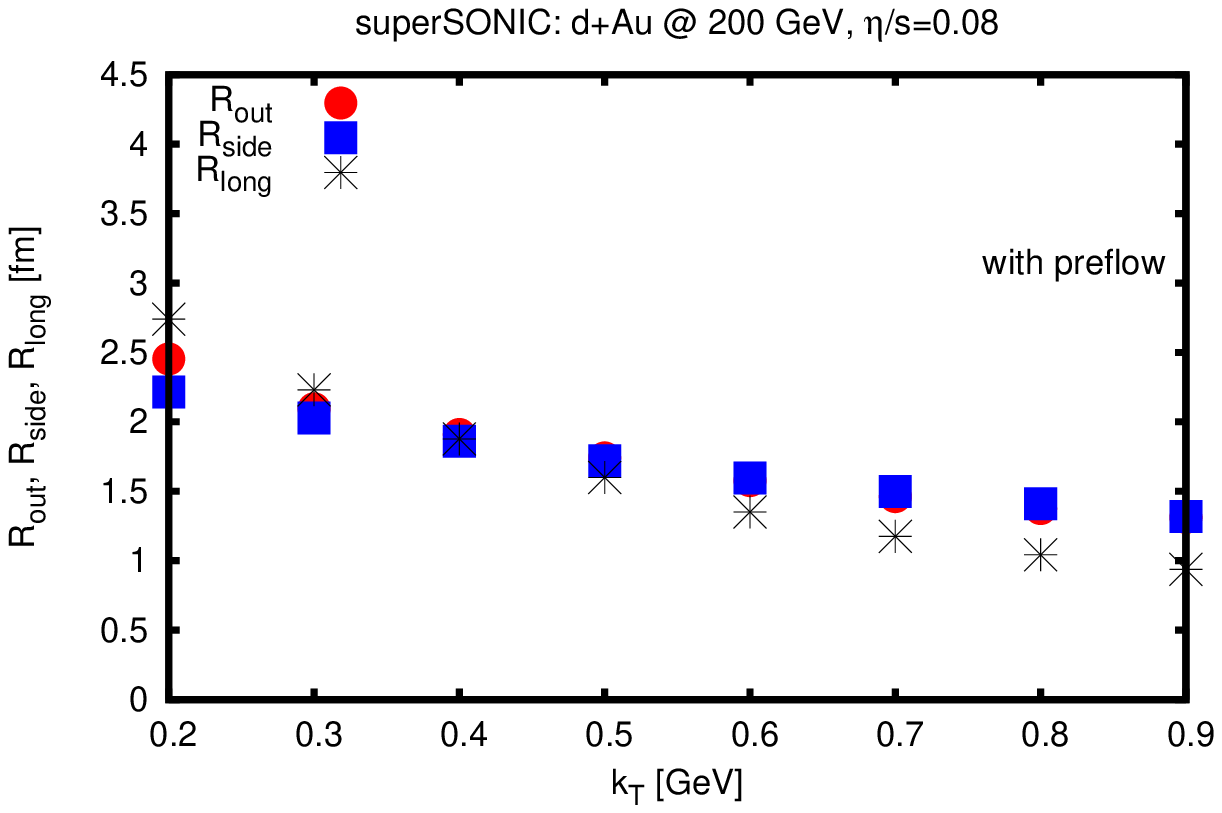}\hfill
\includegraphics[width=0.45\linewidth]{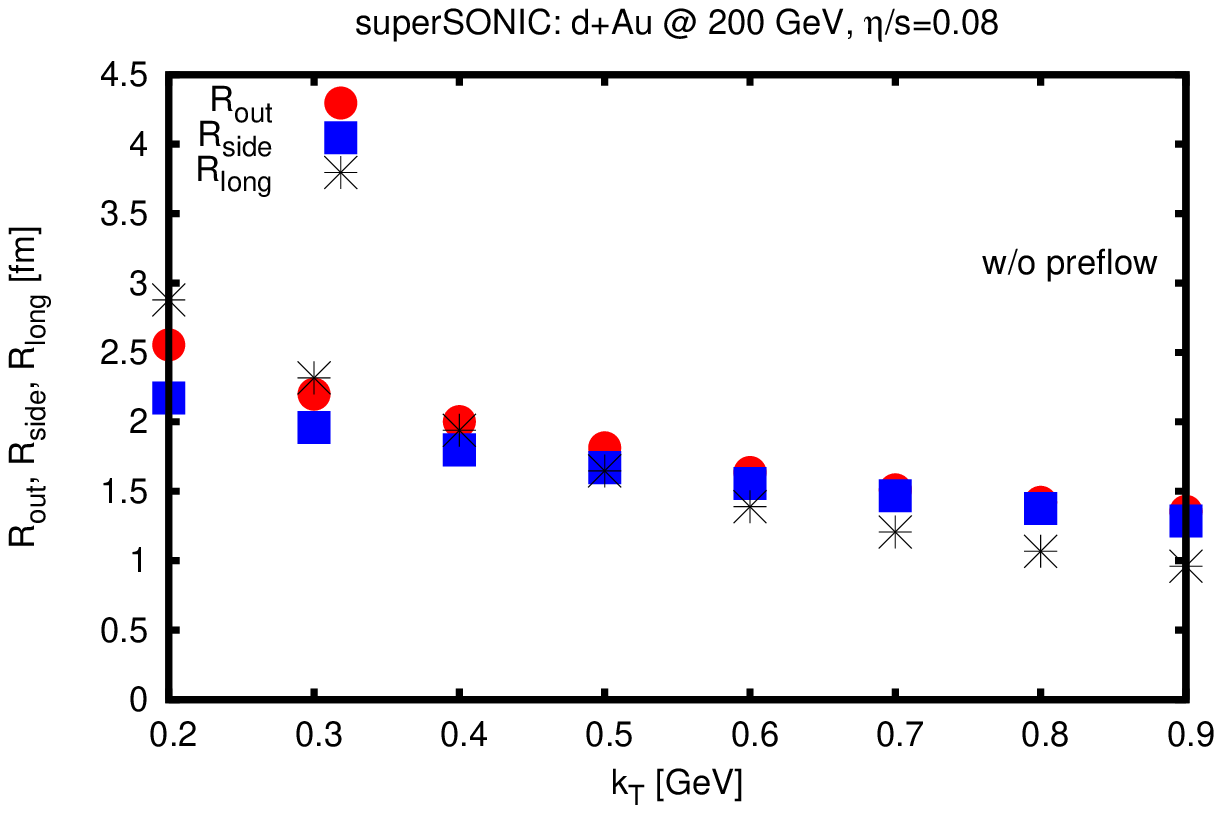}\hfill
\includegraphics[width=0.45\linewidth]{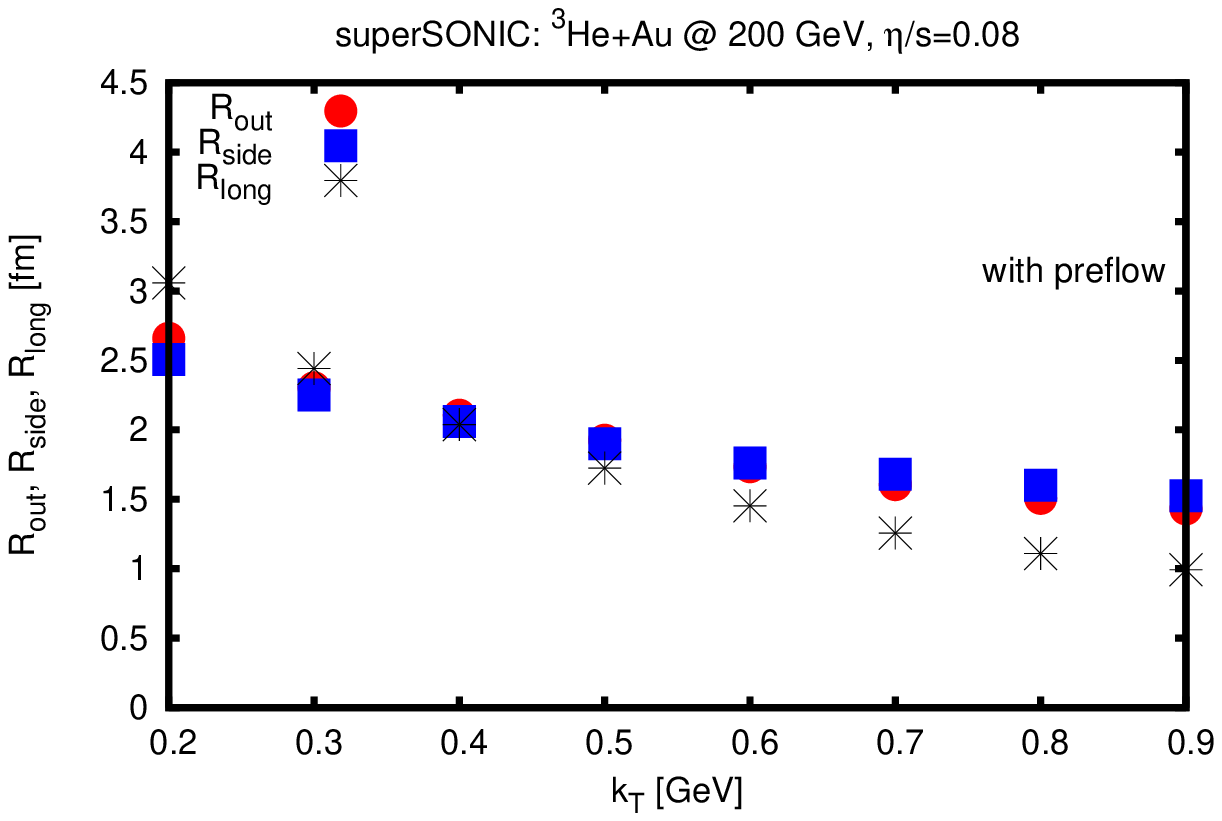}\hfill
\includegraphics[width=0.45\linewidth]{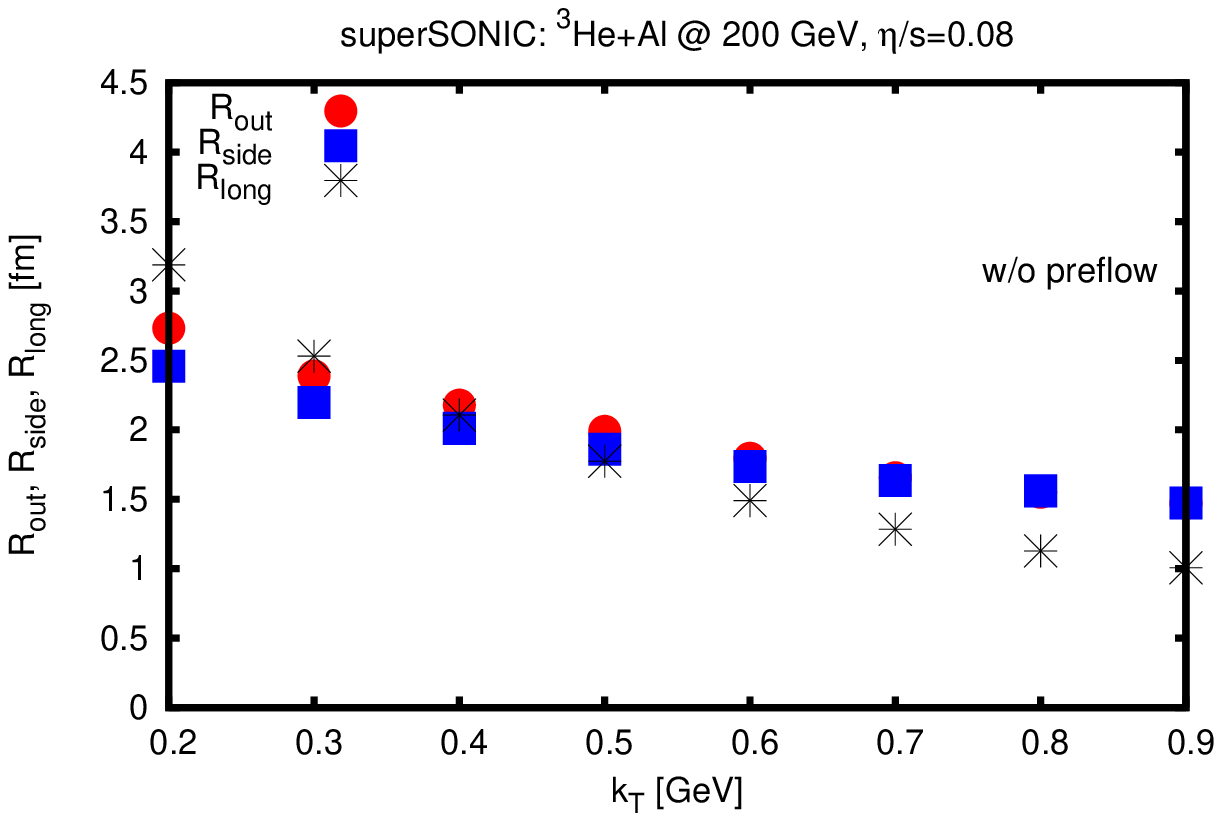}
\caption{\label{fig:all200hbt}
HBT Radii for pions for \pAu, \dAu and \hAu collisions at $\sqrt{s}=200$ GeV. Waring $C_\eta=2-3$ in hydrodynamics results in changes smaller than the symbol size shown.}
\end{figure}

\begin{figure}[t]
\begin{center}
\begin{Large}
RHIC, $\sqrt{s}=62.4$ GeV
\end{Large}
\end{center}
\includegraphics[width=0.45\linewidth]{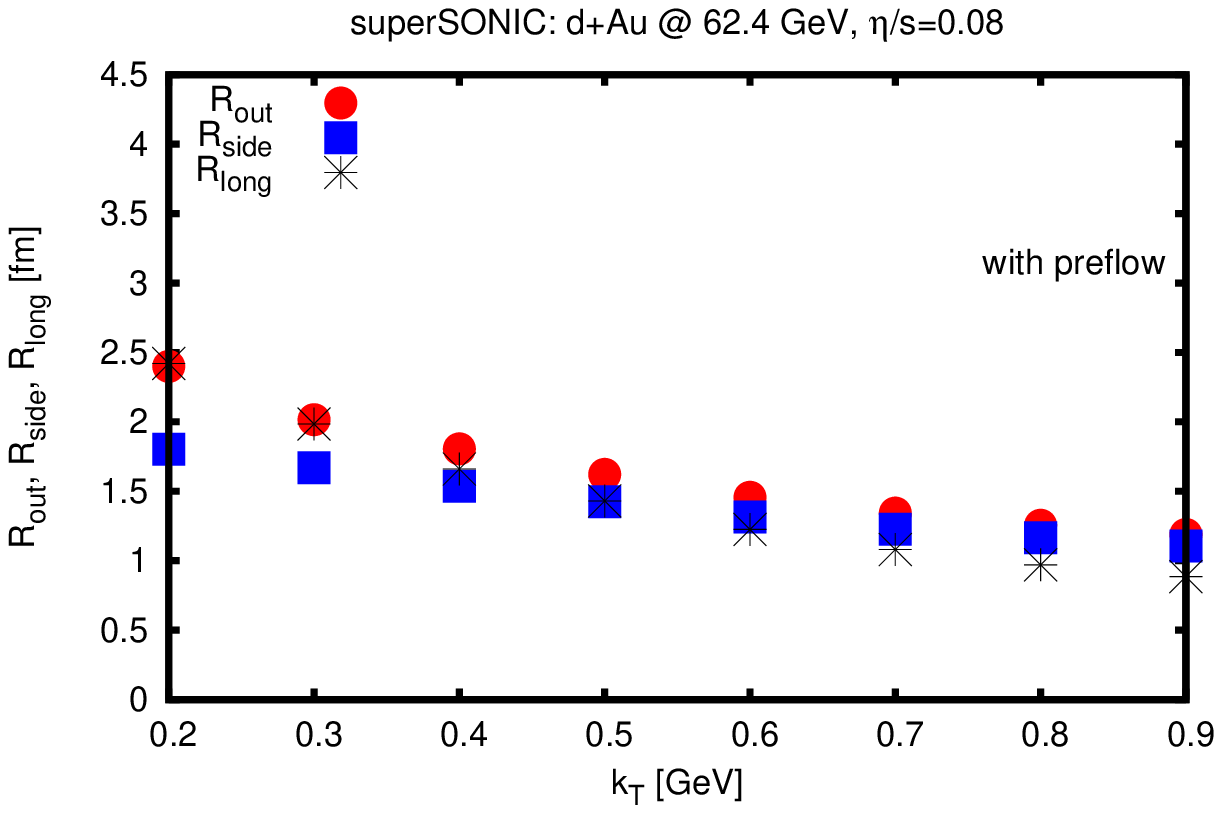}\hfill
\includegraphics[width=0.45\linewidth]{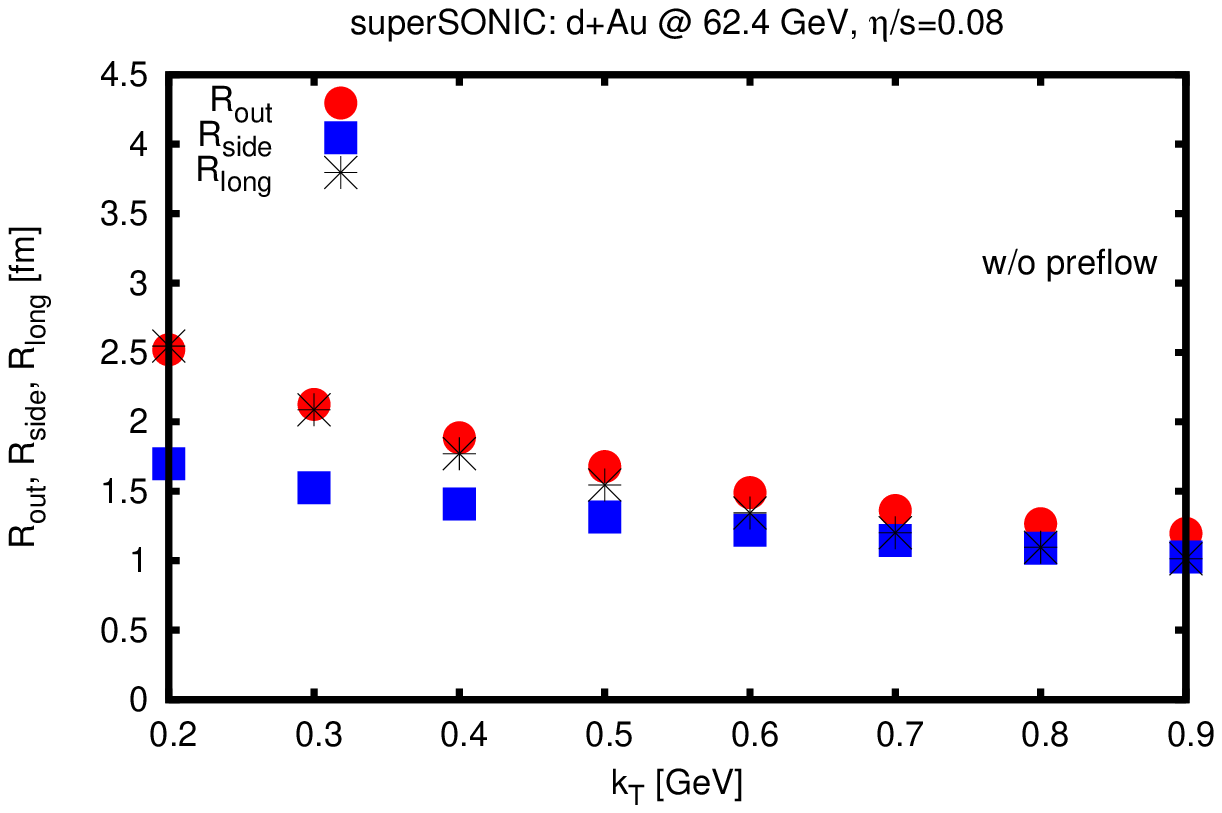}\hfill
\includegraphics[width=0.45\linewidth]{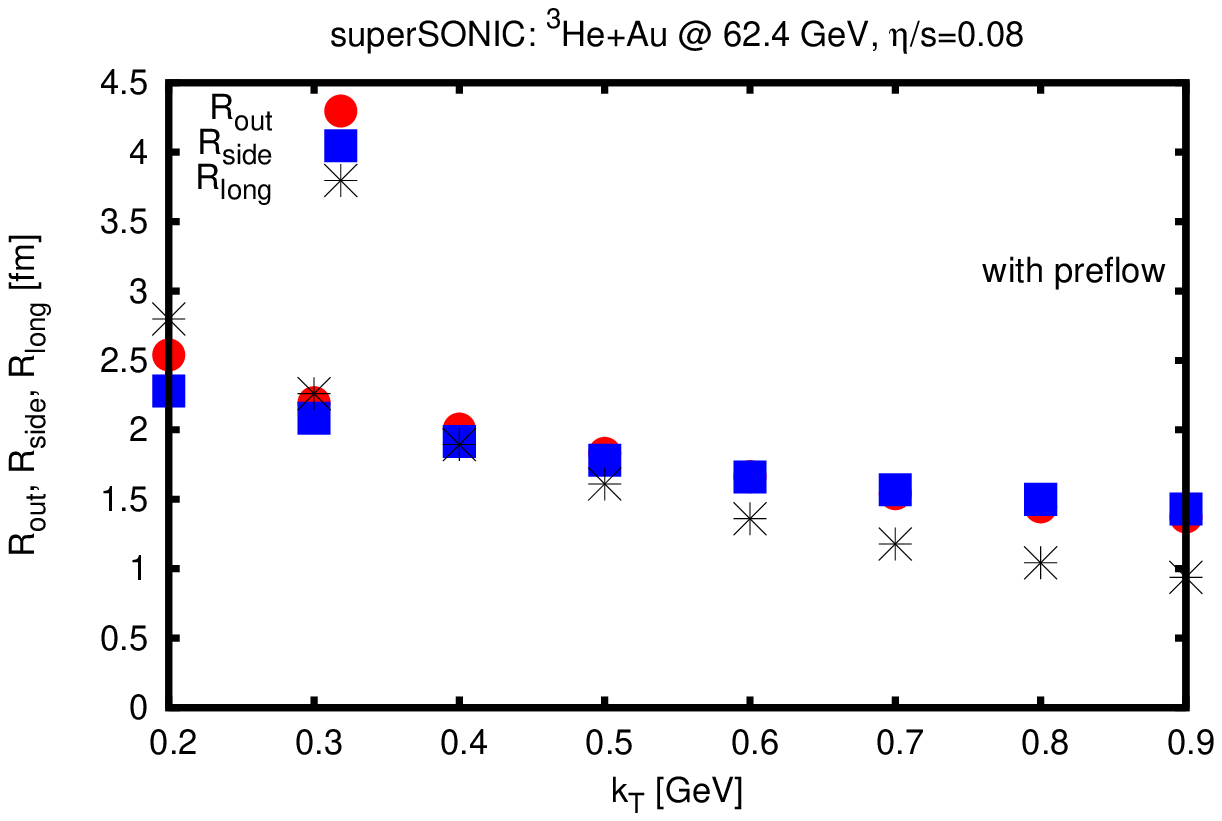}\hfill
\includegraphics[width=0.45\linewidth]{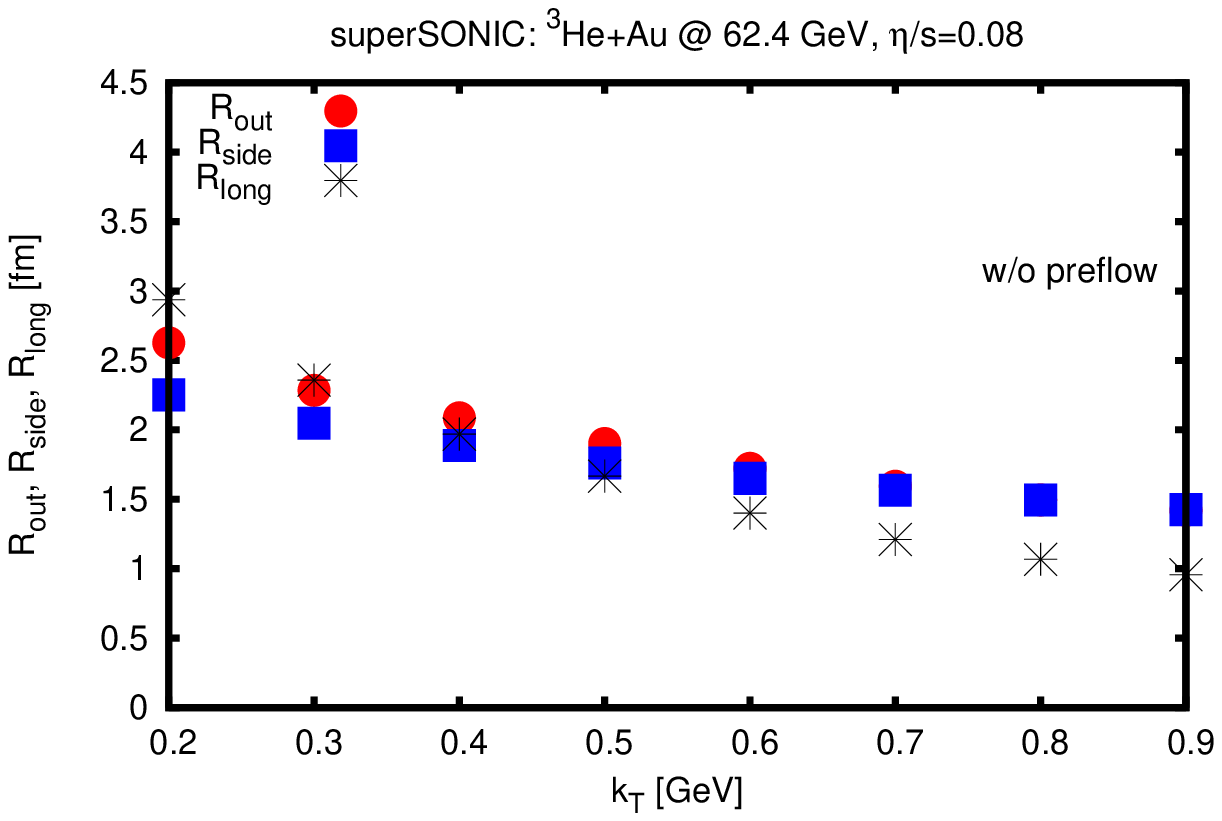}
\caption{\label{fig:all62hbt}
HBT Radii for pions for \dAu and \hAu collisions at $\sqrt{s}=62.4$ GeV. Waring $C_\eta=2-3$ in hydrodynamics results in changes smaller than the symbol size shown.}
\end{figure}
\end{appendix}

\end{document}